\documentclass[showpacs,preprintnumbers,amsmath,amssymb]{revtex4-2}

\usepackage{graphicx}
\usepackage{dcolumn}
\usepackage{bm}
\usepackage{graphicx,subfigure,color}

\newcommand{\bq}{\begin{equation}}
	\newcommand{\eq}{\end{equation}}
\newcommand{\bqa}{\begin{eqnarray}}
	\newcommand{\eqa}{\end{eqnarray}}
\newcommand{\nn}{\nonumber \\}

\def\be     {\begin{equation}}
	\def\ee     {\end{equation}}
\def\bea        {\begin{eqnarray}}
	\def\eea        {\end{eqnarray}}
\def\bnn    {\begin{eqnarray*}}
	\def\enn    {\end{eqnarray*}}

\begin{document}

\title{Renormalization group flow to effective quantum mechanics at IR in an emergent dual holographic description for spontaneous chiral symmetry breaking}
\author{Ki-Seok Kim$^{a,b}$, Mitsuhiro Nishida$^{a}$, and Yoonseok Choun$^{a,b}$}
\affiliation{$^{a}$Department of Physics, POSTECH, Pohang, Gyeongbuk 37673, Korea \\ $^{b}$Asia Pacific Center for Theoretical Physics (APCTP), Pohang, Gyeongbuk 37673, Korea}

\email[Ki-Seok Kim: ]{tkfkd@postech.ac.kr}
\email[Mitsuhiro Nishida: ]{mnishida124@gmail.com}
\email[Yoonseok Choun: ]{ychoun@gmail.com}
\date{\today}

\begin{abstract}
Implementing the Wilsonian renormalization group (RG) transformation in a nonperturbative way, we construct an effective holographic dual description with an emergent extradimension identified with an RG scale. Taking the large$-N$ limit, we obtain an equation of motion of an order-parameter field, here the chiral condensate for our explicit demonstration. In particular, an intertwined structure manifests between the first-order RG flow equations of renormalized coupling functions and the second-order differential equation of the order-parameter field, thus referred to as a nonperturbative RG-improved mean-field theory. Assuming translational symmetry as a vacuum state, we solve these nonlinear coupled mean-field equations based on a matching method between UV- and IR-regional solutions. As a result, we find an RG flow from a weakly-coupled chiral-symmetric UV fixed point to a strongly-correlated chiral-symmetry broken IR fixed point, where the renormalized velocity of Dirac fermions vanishes most rapidly and effective quantum mechanics appears at IR. Furthermore, we translate these RG flows of coupling functions into those of emergent metric tensors and extract out geometrical properties of the emergent holographic spacetime constructed from the UV- and IR-regional solutions. Surprisingly, we obtain the volume law of entanglement entropy in the Ryu-Takayanagi formula, which implies appearance of a black hole type solution in the limit of infinite cutoff even at zero temperature. We critically discuss our field theoretic interpretation for this solution in terms of potentially gapless multi-particle excitation spectra.
%
%
\end{abstract}


\maketitle

\section{Introduction}

Renormalization group (RG) improved mean-field theory (RG-MFT) is not only a natural but also a general framework at least in the conceptual aspect. Here, interaction vertices are renormalized in the one-loop level, or resummed in the context of the Bethe-Salpeter equation \cite{QFT_Textbook}. These renormalized coupling functions enter a mean-field equation for the description of a phase transition given by an order-parameter field. Unfortunately, this `perturbative' RG framework is not enough to understand dynamics of strongly coupled quantum field theories.

To generalize the Wilsonian perturbative RG transformation in a nonperturbative way, we introduce an energy-scale dependent order-parameter field and propose an intertwined RG structure between the RG flow equations of renormalized coupling functions and an extended mean-field equation of the order-parameter field at each energy scale. Suppose the Wilsonian RG transformation for a given quantum field theory at an energy scale $\Lambda$. Then, we solve the resulting effective interacting field theory in a mean-field fashion but with renormalized interaction vertices at the scale $\Lambda$. Usually, this is the end of the procedure, referred to as the RG-MFT mentioned above.

In this study, we extend this procedure in an iterative way. After performing the first iteration in the RG-MFT, we consider the second RG transformation for all dynamical variables including the order-parameter field. This second iteration is supported by the renormalized coupling functions and the mean-field `background' value of the order parameter at the scale $\Lambda$ in the first step of the RG-MFT. Again, we take the large$-N$ limit to perform the mean-field analysis with the updated renormalized interaction vertices after the second RG implementation. This completes the second iteration in the RG-MFT.

Repeating these RG iterations with the scale-dependent mean-field analysis in the large$-N$ limit and expressing the discrete variable of the RG iteration with a continuous `coordinate' $z$, we construct an effective field theory, where the RG transformation manifests with an emergent extradimension denoted by $z$ \cite{SungSik_Holography_I,SungSik_Holography_II,SungSik_Holography_III,Nonperturbative_Wilson_RG,Einstein_Klein_Gordon_RG_Kim,Einstein_Dirac_RG_Kim, RG_GR_Geometry_I_Kim,RG_GR_Geometry_II_Kim,Kondo_Holography_Kim,Kitaev_Entanglement_Entropy_Kim,RG_Holography_First_Kim}. Interestingly, we observe that this nonperturbative RG-MFT not only shares some characteristic features of the holographic dual effective field theory \cite{Holographic_Duality_I,Holographic_Duality_II,Holographic_Duality_III,Holographic_Duality_IV,Holographic_Duality_V,Holographic_Duality_VI,
Holographic_Duality_VII} but also modifies them in two ways. First, the background geometry in the dual holographic description corresponds to the RG flow equations of renormalized coupling functions in the nonperturbative RG-MFT \cite{Holographic_Duality_V,Holographic_Duality_VI,Holographic_Duality_VII}. Second, other fields besides the metric tensor describe the dynamics of `order parameters' or collective fields in a dual fashion \cite{Holographic_Duality_I,Holographic_Duality_II,Holographic_Duality_III,Holographic_Duality_IV}. For example, the scalar field in the Einstein-scalar field theory corresponds to the chiral condensate in the present description. These two aspects reinterpret the holographic dual effective field theory in terms of nonperturbative Wilsonian RG transformations. Third, the nonperturbative RG-MFT introduces RG $\beta-$functions of interaction vertices beyond the holographic dual effective field theory with gravity, where the RG flows for all dynamical variables appear naturally to show conformal invariance only in the low-energy limit \cite{Nonperturbative_Wilson_RG,Einstein_Klein_Gordon_RG_Kim,Einstein_Dirac_RG_Kim,RG_GR_Geometry_I_Kim,RG_GR_Geometry_II_Kim,Kondo_Holography_Kim,
Kitaev_Entanglement_Entropy_Kim,RG_Holography_First_Kim,Vasudev_Shyam_I,Vasudev_Shyam_II,Vasudev_Shyam_III,Vasudev_Shyam_IV}. Fourth, there is essential difference on how to assign UV and IR boundary conditions in the present description, where an effective on-shell (boundary) action determines both boundary conditions \cite{Nonperturbative_Wilson_RG,Einstein_Klein_Gordon_RG_Kim,Einstein_Dirac_RG_Kim,RG_GR_Geometry_I_Kim,RG_GR_Geometry_II_Kim,Kondo_Holography_Kim,
Kitaev_Entanglement_Entropy_Kim,RG_Holography_First_Kim}. Here, the effective on-shell (boundary) action may be regarded as a solution of the Hamilton-Jacobi equation \cite{Nonperturbative_Wilson_RG,Einstein_Klein_Gordon_RG_Kim}.

For concreteness, we consider spontaneous chiral symmetry breaking \cite{QFT_Textbook}. Here, we have three types of coupling functions, corresponding to the wave-function renormalization constant for Dirac fermions, the velocity of Dirac fermions, and their effective interactions for chiral symmetry breaking, respectively. In addition, we have one order parameter field to describe the chiral symmetry breaking. Taking the large$-N$ limit in the nonperturbative RG-MFT, where $N$ is the flavor number of Dirac fermions, we obtain four coupled differential equations, where three of them are given by the first order to describe the RG flows of the coupling functions and the last is the second-order differential equation for the order-parameter field, supported by renormalized coupling functions at a given energy scale. We verify that the wave-function renormalization constant (the velocity renormalization constant) in the nonperturbative RG-MFT corresponds to the space (time) component of the metric tensor in the holographic dual effective field theory.

%
%

We believe that it would not be an easy task to solve these four heavily intertwined differential equations. Here, we apply a matching method to solving these coupled differential equations \cite{Holography_Matching_Method}. First, we solve such coupled differential equations near both UV and IR boundary regions, where these equations become simplified. Second, we apply the UV (IR) boundary condition to the UV-regional (IR-) solution. Since the number of boundary conditions would be less than that of integration constants, some of the integration constants remain undetermined in both UV- and IR-regional solutions. Third, we require that the UV-regional solution should be smoothly connected to the IR-regional solution in the extradimensional space. Of course, there must be a certain condition for the existence of this matching solution, which will be clarified later. Based on this matching method, we find an RG flow from a weakly-coupled chiral-symmetric UV fixed point to a strongly-correlated chiral-symmetry broken IR fixed point, where the renormalized velocity of Dirac fermions vanishes most rapidly and effective quantum mechanics appears at IR. It is a feature of the nonperturbative RG-MFT the appearance of this local strong-coupling fixed point. Furthermore, we investigate several geometrical objects in the holographic spacetime constructed from the UV- and IR-regional solutions we found. In particular, we calculate the minimal surface for a single interval, which is essential for the holographic entanglement entropy formula \cite{Ryu:2006bv, Ryu:2006ef}. We uncovered the volume law of entanglement entropy at zero temperature, which implies appearance of a black hole type geometry at $\Lambda\to\infty$ to describe the strongly-correlated chiral-symmetry broken IR fixed point. We critically discuss our field theoretic interpretation that strong correlations may allow gapless multi-particle spectra between the single-particle excitation gap due to spontaneous chiral symmetry breaking.

This paper is organized as follows. First, we discuss a general formulation of emergent dual holography in section \ref{General_Formulation}. Here, we introduce a general framework for the RG transformation and clarify how the present construction implements the RG transformation in a nonperturbaive way. In addition, we introduce an emergent dual holographic description given by gravity with an extradimension and argue that our nonperturbative RG framework may be regarded as an effective holographic dual description. Second, we take the large$-N$ limit in this effective field theory and obtain the equation of motion for an order-parameter field, here the chiral condensate, in section \ref{General_Formulation}. Here, we find an intertwined structure for the RG flows between the renormalized coupling vertices and the order-parameter field. Third, we try to solve these coupled mean-field equations based on a matching method between UV- and IR-regional solutions in section \ref{Mean_Field_Theory}, assuming translational symmetry as a vacuum solution. In particular, we could reveal some analytic behaviors in the low-energy limit and find a local strong-coupling fixed point. In section \ref{HologarphicSpacetime}, we compute the Ricci scalar curvature and the minimal surface on the three-dimensional holographic spacetime. Then, we summarize our main results and discuss several unresolved issues here in the last section.

\section{A nonperturbative approach of the Wilsonian renormalization group theory for spontaneous chiral symmetry breaking} \label{General_Formulation}

\subsection{A general framework for the renormalization group transformation}

We start our discussions, reviewing a general framework of the RG transformation \cite{QFT_Textbook}. Here, we consider interacting Dirac fermions, described by
\bqa && Z = \int D \psi_{b \sigma}(x) \exp\Big[ - \int d^{D} x \Big\{ \bar{\psi}_{b \sigma}(x) \gamma^{\mu} \partial_{\mu} \psi_{b \sigma}(x) + \frac{\lambda_{b \chi}}{2 N} \bar{\psi}_{b \sigma}(x) \psi_{b \sigma}(x) \bar{\psi}_{b \sigma'}(x) \psi_{b \sigma'}(x) \Big\} \Big] . \eqa
$\psi_{b \sigma}(x)$ is a Dirac spinor at $x$ in $D$ spacetime dimensions, where $\sigma$ is a flavor index from $1$ to $N$ and $b$ is `bare' or `unrenormalized'. $\bar{\psi}_{b \sigma}(x) = \psi_{b \sigma}^{\dagger}(x) \gamma^{0}$ is the canonical conjugate variable to $\psi_{b \sigma}(x)$. $\gamma^{\mu}$ with $\mu = 0, ..., D-1$ is the Dirac matrix to satisfy the Clifford algebra in $D$ spacetime dimensions. $\lambda_{b \chi}$ is an interaction coefficient for chiral symmetry breaking, where $b$ also denotes `bare' or `unrenormalized'.

Performing the Hubbard-Stratonovich transformation for the chiral symmetry breaking channel, we introduce an order-parameter field $\varphi_{b}(x)$ as follows
\bqa && Z = \int D \psi_{b \sigma}(x) D \varphi_{b}(x) \exp\Big[ - \int d^{D} x \Big\{ \bar{\psi}_{b \sigma}(x) \Big(\gamma^{\tau} \partial_{\tau} - i v_{b \psi} \gamma^{i} \partial_{i}\Big) \psi_{b \sigma}(x) - i \lambda_{b \chi} \varphi_{b}(x) \bar{\psi}_{b \sigma}(x) \psi_{b \sigma}(x) + \frac{N \lambda_{b \chi}}{2} \varphi_{b}^{2}(x) \Big\} \Big] . \nn \eqa
Here, we introduced $v_{b \psi}$ as the `bare' velocity of Dirac fermions. To describe quantum fluctuations of chiral symmetry breaking, we integrate over short-distance fluctuations of Dirac fermions and obtain
\bqa && Z = \int D \psi_{b \sigma}(x) D \varphi_{b}(x) \exp\Big[ - \int d^{D} x \Big\{ \bar{\psi}_{b \sigma}(x) \Big(\gamma^{\tau} \partial_{\tau} - i v_{b \psi} \gamma^{i} \partial_{i}\Big) \psi_{b \sigma}(x) - i \lambda_{b \chi} \varphi_{b}(x) \bar{\psi}_{b \sigma}(x) \psi_{b \sigma}(x) \nn && + N \varphi_{b}(x) \Big( - \partial_{\tau}^{2} - v_{b \varphi}^{2} \partial_{i}^{2} + m_{b \varphi}^{2} \Big) \varphi_{b}(x) \Big\} \Big] . \eqa
Here, the kinetic energy of the order-parameter field results from the polarization bubble of high-energy quantum fluctuations of Dirac fermions, denoted by $\Pi(x-x') = \lambda_{b \chi}^{2} \Big\langle \bar{\psi}_{b \sigma}(x) \psi_{b \sigma}(x) \bar{\psi}_{b \sigma'}(x') \psi_{b \sigma'}(x') \Big\rangle$, where $\langle ... \rangle$ is an ensemble average. Performing the Fourier transformation and expanding it up to the second order with respect to the frequency and momentum, we obtain $\Pi(i\Omega,q) = c_{1} N \frac{\lambda_{b \chi}^{2}}{\Lambda_{f}^{2}}(\Omega^{2} + c_{2} v_{b \psi}^{2} q^{2}) + \Pi(0,0)$, where $c_{1}$ and $c_{2}$ are numerical constants. $\Lambda_{f}$ is the short-distance cutoff to control high-energy fluctuations. Rescaling coordinates, fields, and coupling constants appropriately, we obtain the above expression for the kinetic energy of the order-parameter field, where $v_{b \varphi}$ and $m_{b \varphi}$ are the `bare' velocity and `bare' mass of the collective field $\varphi_{b}(x)$.

Now, we consider the RG transformation. First, the coordinate transforms as
\bqa && x \rightarrow \mu x , \eqa
where $\mu$ is the scaling parameter. To make the time-derivative term be invariant under this coordinate transformation, both fermion and boson fields have to transform as follows
\bqa && \psi_{b \sigma}(x) = \mu^{- \frac{D-1}{2}} Z_{\psi}^{\frac{1}{2}} \psi_{r \sigma}(x) , ~~~~~ \varphi_{b} = \mu^{- \frac{D-2}{2}} Z_{\varphi}^{\frac{1}{2}} \varphi_{r} . \eqa
Here, $Z_{\psi}$ and $Z_{\varphi}$ are wave-function renormalization constants to take divergent contributions from quantum corrections and to relate fermion and boson bare fields with their renormalized ones denoted by the subscript $r$. Accordingly, fermion and boson velocities transform as
\bqa && v_{b \psi} = Z_{\psi}^{-1} Z_{v_{\psi}} v_{r \psi} , ~~~~~ v_{b \varphi}^{2} = Z_{\varphi}^{-1} Z_{v_{\varphi}^{2}} v_{r \varphi}^{2} , \eqa
which lead the space-derivative term to be invariant under this scaling transformation. $Z_{v_{\psi}}$ and $Z_{v_{\varphi}^{2}}$ are velocity renormalization constants for fermions and bosons, respectively. The boson mass term remains invariant if the mass parameter transforms as \bqa && m_{b \varphi}^{2} = \mu^{2} Z_{\varphi}^{-1} Z_{m_{\varphi}^{2}} m_{r \varphi}^{2} , \eqa where $Z_{m_{\varphi}^{2}}$ is the mass renormalization constant. Finally, the Yukawa coupling term is invariant if the Yukawa interaction vertex transforms as \bqa && \lambda_{b \chi} = \mu^{- \frac{D-4}{2}} Z_{\psi}^{-1} Z_{\varphi}^{-\frac{1}{2}} Z_{\lambda_{\chi}} \lambda_{r \chi} , \eqa where $Z_{\lambda_{\chi}}$ is the interaction renormalization constant.
%
%

Replacing all bare fields and coupling constants with renormalized ones, we obtain the following effective field theory
\bqa && Z = (\mu^{- (2 D - 3)} Z_{\psi}^{N} Z_{\varphi})^{V_{D}} \int D \psi_{r \sigma}(x) D \varphi_{r}(x) \exp\Big[ - \int d^{D} x \Big\{ \bar{\psi}_{r \sigma}(x) \Big(Z_{\psi} \gamma^{\tau} \partial_{\tau} - i Z_{v_{\psi}} v_{r \psi} \gamma^{i} \partial_{i}\Big) \psi_{r \sigma}(x) \nn && - i Z_{\lambda_{\chi}} \lambda_{r \chi} \varphi_{r}(x) \bar{\psi}_{r \sigma}(x) \psi_{r \sigma}(x) + N \varphi_{r}(x) \Big( - Z_{\varphi} \partial_{\tau}^{2} - Z_{v_{\varphi}^{2}} v_{r \varphi}^{2} \partial_{i}^{2} + Z_{m_{\varphi}^{2}} m_{r \varphi}^{2} \Big) \varphi_{r}(x) \Big\} \Big] , \label{RG_EFT_General} \eqa
where $V_{D}$ is the volume factor of the $D-$dimensional spacetime. Below, we find all these renormalization constants in the nonperturbative RG-MFT.

\subsection{Nonperturbative implementation of the Wilsonian renormalization group transformation}

To implement the Wilsonian RG transformation in a nonperturbative way, we introduce $w_{\psi}$ with an order-parameter field as follows
\bqa && Z = \int D \psi_{\sigma}(x) D \varphi(x) \exp\Big[ - \int d^{D} x \Big\{ \bar{\psi}_{\sigma}(x) \Big(w_{\psi} \gamma^{\tau} \partial_{\tau} - i v_{\psi} \gamma^{i} \partial_{i}\Big) \psi_{\sigma}(x) - i \lambda_{\chi} \varphi(x) \bar{\psi}_{\sigma}(x) \psi_{\sigma}(x) + \frac{N \lambda_{\chi}}{2} \varphi^{2}(x) \Big\} \Big] . \nn \eqa
As expected, $w_{\psi}$ will play the role of the wave-function renormalization constant for Dirac fermions.

Now, we perform the RG transformation. First, we separate fast and slow degrees of freedom from all dynamical fields at a given energy scale $\Lambda$, here Dirac fermions $\psi_{\sigma}(x)$ and chiral symmetry breaking fluctuations $\varphi(x)$. Second, we integrate over short-distance fluctuations for $\varphi(x)$ and obtain newly-generated effective interactions between Dirac fermions. Third, we perform the Hubbard-Stratonovich transformation for such RG-transformation generated interactions and have an additional collective field variable, saying $\varphi^{(1)}(x)$ to decompose RG-generated effective interactions. Calling the previous low-energy mode $\varphi^{(0)}(x)$ and shifting $\varphi^{(1)}(x)$ to $\varphi^{(1)}(x) - \varphi^{(0)}(x)$, we finish the first RG transformation for chiral symmetry breaking fluctuations. Fourth, we perform the path integral for short-distance fluctuations of Dirac fermions. Actually, this RG transformation gives rise to genuine renormalization effects for all coupling functions introduced in the above UV effective field theory. Although any concrete procedures would depend on regularization, we have the following structure in the Wilsonian RG transformation. The path integral for high-energy quantum fluctuations of Dirac fermions generates an effective potential in terms of unrenormalized coupling functions and $\varphi^{(1)}(x)$. Fifth, taking a functional derivative of the RG-generated effective potential with respect to each coupling function and $\varphi^{(1)}(x)$, we obtain the so called RG $\beta-$function to encode how the coupling function evolves through this RG transformation. Calling the previous unrenormalized coupling function that with a superscript $(0)$, we update it to a renormalized one with a superscript $(1)$. This completes the first iteration of the RG transformation.

To proceed the second iteration of the RG transformation, we consider quantum fluctuations of $\psi_{\sigma}(x)$ and $\varphi^{(1)}(x)$. Here, $\varphi^{(0)}(x)$ is determined by free-energy minimization as usual. In this respect $\varphi^{(0)}(x)$ is an energy-scale dependent order-parameter field as discussed in the introduction. Now, it is straightforward to perform the RG transformation in a recursive way. Each coupling function with a superscript $(k-1)$ renormalizes into that with a superscript $(k)$ by its RG$-\beta$ function. Within these background renormalized coupling functions, the energy-dependent order parameter field is determined by minimization of the free energy functional. To make the resulting effective field theory be more tractable, we introduce a continuum variable $z$ to replace the discrete step $(k)$ and rewrite the RG transformation from $(k-1)$ to $(k)$ in a form of the $z-$derivative. As a result, the RG flows of the coupling functions and the RG evolution of the order-parameter field become manifested through the emergent extradimensional space denoted by the coordinate $z$. This completes our nonperturbative RG-MFT \cite{Nonperturbative_Wilson_RG,Einstein_Klein_Gordon_RG_Kim,Einstein_Dirac_RG_Kim,RG_GR_Geometry_I_Kim,RG_GR_Geometry_II_Kim,Kondo_Holography_Kim,
Kitaev_Entanglement_Entropy_Kim,RG_Holography_First_Kim}.

%
%

All these discussions can be summarized by the following effective field theory
\bqa && Z = \int D \psi_{\sigma}(x) D \varphi(x,z) D \pi_{\varphi}(x,z) D w_{\psi}(x,z) D \pi_{w_{\psi}}(x,z) D v_{\psi}(x,z) D \pi_{v_{\psi}}(x,z) D \lambda_{\chi}(x,z) D \pi_{\lambda_{\chi}}(x,z) \nn && \exp\Big[ - \int d^{D} x \Big\{ \bar{\psi}_{\sigma}(x) \Big(w_{\psi}(x,z_{f}) \gamma^{\tau} \partial_{\tau} - i v_{\psi}(x,z_{f}) \gamma^{i} \partial_{i}\Big) \psi_{\sigma}(x) - i \lambda_{\chi}(x,z_{f}) \varphi(x,z_{f}) \bar{\psi}_{\sigma}(x) \psi_{\sigma}(x) + \frac{N \lambda_{\chi}(x,0)}{2} \varphi^{2}(x,0) \Big\} \nn && - N \int_{0}^{z_{f}} d z \int d^{D} x \Big\{ \pi_{\varphi}(x,z) \Big( \partial_{z} \varphi(x,z) - \beta_{\varphi}[\varphi(x,z),w_{\psi}(x,z),v_{\psi}(x,z),\lambda_{\chi}(x,z)] \Big) - \frac{1}{2 \lambda_{\chi}(x,z)} \pi_{\varphi}^{2}(x,z) \nn && + \pi_{w_{\psi}}(x,z) \Big( \partial_{z} w_{\psi}(x,z) - \beta_{w_{\psi}}[\varphi(x,z),w_{\psi}(x,z),v_{\psi}(x,z),\lambda_{\chi}(x,z)] \Big) \nn && + \pi_{v_{\psi}}(x,z) \Big( \partial_{z} v_{\psi}(x,z) - \beta_{v_{\psi}}[\varphi(x,z),w_{\psi}(x,z),v_{\psi}(x,z),\lambda_{\chi}(x,z)] \Big) \nn && + \pi_{\lambda_{\chi}}(x,z) \Big( \partial_{z} \lambda_{\chi}(x,z) - \beta_{\lambda_{\chi}}[\varphi(x,z),w_{\psi}(x,z),v_{\psi}(x,z),\lambda_{\chi}(x,z)] \Big) \nn && + \mathcal{V}_{eff}[\varphi(x,z),w_{\psi}(x,z),v_{\psi}(x,z),\lambda_{\chi}(x,z)] \Big\} \Big] . \label{RG_MFT_General} \eqa
Although this expression looks rather complicated, we explain several characteristic features of this effective field theory, discussed above.

$w_{\psi}(x,z)$, $v_{\psi}(x,z)$, and $\lambda_{\chi}(x,z)$ are renormalized coupling functions. In particular, $w_{\psi}(x,z)$ is related with the field renormalization to be clarified in the next subsection. $\pi_{w_{\psi}}(x,z)$, $\pi_{v_{\psi}}(x,z)$, and $\pi_{\lambda_{\chi}}(x,z)$ are their canonical conjugate fields, respectively. Integrating over these canonical fields, we obtain the RG$-$flow equations for these coupling functions as follows
\bqa \partial_{z} w_{\psi}(x,z) &=& \beta_{w_{\psi}}[\varphi(x,z),w_{\psi}(x,z),v_{\psi}(x,z),\lambda_{\chi}(x,z)] , \\ \partial_{z} v_{\psi}(x,z) &=& \beta_{v_{\psi}}[\varphi(x,z),w_{\psi}(x,z),v_{\psi}(x,z),\lambda_{\chi}(x,z)] , \\ \partial_{z} \lambda_{\chi}(x,z) &=& \beta_{\lambda_{\chi}}[\varphi(x,z),w_{\psi}(x,z),v_{\psi}(x,z),\lambda_{\chi}(x,z)] . \eqa
Physical meaning of these first-order differential equations is clear. The real question is how to find these RG $\beta-$functions in a nonperturbative way, to be clarified below.

$\varphi(x,z)$ is an energy-dependent chiral symmetry breaking order-parameter field, whose dynamics is determined by minimization of the bulk effective action in the large$-N$ limit, given by
\bqa && \mathcal{S}_{bulk} = N \int_{0}^{z_{f}} d z \int d^{D} x \Big\{ \pi_{\varphi}(x,z) \Big( \partial_{z} \varphi(x,z) - \beta_{\varphi}[\varphi(x,z),w_{\psi}(x,z),v_{\psi}(x,z),\lambda_{\chi}(x,z)] \Big) - \frac{1}{2 \lambda_{\chi}(x,z)} \pi_{\varphi}^{2}(x,z) \nn && + \mathcal{V}_{eff}[\varphi(x,z),w_{\psi}(x,z),v_{\psi}(x,z),\lambda_{\chi}(x,z)] \Big\} . \eqa
$\pi_{\varphi}(x,z)$ is the canonical conjugate field to $\varphi(x,z)$. In this respect this effective bulk action is written in the Hamiltonian formulation. The essential ingredient is an effective potential $\mathcal{V}_{eff}[\varphi(x,z),w_{\psi}(x,z),v_{\psi}(x,z),\lambda_{\chi}(x,z)]$, generated by the RG transformation for Dirac fermions and given by
\bqa && \mathcal{V}_{eff}[\varphi(x,z),w_{\psi}(x,z),v_{\psi}(x,z),\lambda_{\chi}(x,z)] \nn && = - \frac{1}{N} \ln \int_{\Lambda(z)} D \psi_{\sigma}(x) \exp\Big[ - \int d^{D} x \Big\{ \bar{\psi}_{\sigma}(x) \Big(w_{\psi}(x,z) \gamma^{\tau} \partial_{\tau} - i v_{\psi}(x,z) \gamma^{i} \partial_{i}\Big) \psi_{\sigma}(x) - i \lambda_{\chi}(x,z) \varphi(x,z) \bar{\psi}_{\sigma}(x) \psi_{\sigma}(x) \Big\} \Big] . \nn \label{Effective_Potential_RG_MFT} \eqa
Here, $\int_{\Lambda(z)} D \psi_{\sigma}(x)$ means that the fermion path integral is performed at a given energy scale $\Lambda(z)$, which will be clarified in the next section. Accordingly, the RG $\beta-$function of the chiral condensate is given by the functional derivative of this effective potential with respect to the order-parameter field \cite{QFT_Textbook} as follows
\bqa \beta_{\varphi}[\varphi(x,z),w_{\psi}(x,z),v_{\psi}(x,z),\lambda_{\chi}(x,z)] &=& - \delta_{\varphi} \mathcal{V}_{eff}[\varphi(x,z),w_{\psi}(x,z),v_{\psi}(x,z),\lambda_{\chi}(x,z)] = \frac{i}{N} \Big\langle \lambda_{\chi}(x,z) \bar{\psi}_{\sigma}(x) \psi_{\sigma}(x) \Big\rangle . \nn \eqa
Here, $\langle ... \rangle$ is an ensemble average with respect to the effective action functional
\bqa && \mathcal{S}_{\Lambda(z)} = \int d^{D} x \Big\{ \bar{\psi}_{\sigma}(x) \Big(w_{\psi}(x,z) \gamma^{\tau} \partial_{\tau} - i v_{\psi}(x,z) \gamma^{i} \partial_{i}\Big) \psi_{\sigma}(x) - i \lambda_{\chi}(x,z) \varphi(x,z) \bar{\psi}_{\sigma}(x) \psi_{\sigma}(x) \Big\} . \eqa

Three other RG $\beta-$functions are given by functional derivatives of this effective potential with respect to the corresponding coupling function in the following way
\bqa \beta_{w_{\psi}}[\varphi(x,z),w_{\psi}(x,z),v_{\psi}(x,z),\lambda_{\chi}(x,z)] &=& - \delta_{w_{\psi}} \mathcal{V}_{eff}[\varphi(x,z),w_{\psi}(x,z),v_{\psi}(x,z),\lambda_{\chi}(x,z)] = - \frac{1}{N} \Big\langle \bar{\psi}_{\sigma}(x) \gamma^{\tau} \partial_{\tau} \psi_{\sigma}(x) \Big\rangle , \nn \\ \beta_{v_{\psi}}[\varphi(x,z),w_{\psi}(x,z),v_{\psi}(x,z),\lambda_{\chi}(x,z)] &=& - \delta_{v_{\psi}} \mathcal{V}_{eff}[\varphi(x,z),w_{\psi}(x,z),v_{\psi}(x,z),\lambda_{\chi}(x,z)] = \frac{i}{N} \Big\langle \bar{\psi}_{\sigma}(x) \gamma^{x} \partial_{x} \psi_{\sigma}(x) \Big\rangle , \nn \\ \beta_{\lambda_{\chi}}[\varphi(x,z),w_{\psi}(x,z),v_{\psi}(x,z),\lambda_{\chi}(x,z)] &=& - \delta_{\lambda_{\chi}} \mathcal{V}_{eff}[\varphi(x,z),w_{\psi}(x,z),v_{\psi}(x,z),\lambda_{\chi}(x,z)] = \frac{i}{N} \Big\langle \varphi(x,z) \bar{\psi}_{\sigma}(x) \psi_{\sigma}(x) \Big\rangle . \nn \eqa
This procedure is completely consistent with what we learnt in the RG transformation \cite{QFT_Textbook}.

It is straightforward to find the Hamiltonian equation of motion in the large$-N$ limit, given by
\bqa && \pi_{\varphi}(x,z) = \lambda_{\chi}(x,z)\Big( \partial_{z} \varphi(x,z) - \beta_{\varphi}[\varphi(x,z),w_{\psi}(x,z),v_{\psi}(x,z),\lambda_{\chi}(x,z)] \Big) \eqa
and
\bqa && \partial_{z} \pi_{\varphi}(x,z) = - \beta_{\varphi}[\varphi(x,z),w_{\psi}(x,z),v_{\psi}(x,z),\lambda_{\chi}(x,z)] - \pi_{\varphi}(x,z) \delta_{\varphi} \beta_{\varphi}[\varphi(x,z),w_{\psi}(x,z),v_{\psi}(x,z),\lambda_{\chi}(x,z)] , \eqa
respectively. UV and IR boundary conditions are determined by the following boundary effective action
\bqa && \mathcal{S}_{b.c.} = N \int d^{D} x \Big\{ \mathcal{V}_{eff}[\varphi(x,z_{f}),w_{\psi}(x,z_{f}),v_{\psi}(x,z_{f}),\lambda_{\chi}(x,z_{f})] + \pi_{\varphi}(x,z_{f}) \varphi(x,z_{f}) + \frac{\lambda_{\chi}(x,0)}{2} \varphi^{2}(x,0) - \pi_{\varphi}(x,0) \varphi(x,0) \Big\} , \nn \eqa
where renormalized Dirac fields $\psi_{\sigma}(x)$ living in the $D-$dimensional spacetime IR boundary $z = z_{f}$ were integrated out to give $\mathcal{V}_{eff}[\varphi(x,z_{f}),w_{\psi}(x,z_{f}),v_{\psi}(x,z_{f}),\lambda_{\chi}(x,z_{f})]$. In addition, $\pi_{\varphi}(x,z_{f}) \varphi(x,z_{f}) - \pi_{\varphi}(x,0) \varphi(x,0)$ results from the integration by part for $\int_{0}^{z_{f}} d z \pi_{\varphi}(x,z) \partial_{z} \varphi(x,z)$ in the bulk effective action, where equations of motion for all other fields have been utilized. As a result, the UV boundary condition is
\bqa && \lambda_{\chi}(x,0) \varphi(x,0) = \pi_{\varphi}(x,0) , \eqa
and the IR one is
\bqa && \pi_{\varphi}(x,z_{f}) = \beta_{\varphi}[\varphi(x,z_{f}),w_{\psi}(x,z_{f}),v_{\psi}(x,z_{f}),\lambda_{\chi}(x,z_{f})] . \eqa

For completeness, we would like to point out the Lagrangian formulation. Integrating over renormalized Dirac fermions, we obtain the following partition function
\bqa && Z = \int D \varphi(x,z) \exp\Big[ - N \int_{0}^{z_{f}} d z \int d^{D} x \Big\{ \frac{\lambda_{\chi}(x,z)}{2} \Big( \partial_{z} \varphi(x,z) - \beta_{\varphi}[\varphi(x,z),w_{\psi}(x,z),v_{\psi}(x,z),\lambda_{\chi}(x,z)] \Big)^{2} \nn && + \mathcal{V}_{eff}[\varphi(x,z),w_{\psi}(x,z),v_{\psi}(x,z),\lambda_{\chi}(x,z)] \Big\} \nn && - N \int d^{D} x \Big\{ \mathcal{V}_{eff}[\varphi(x,z_{f}),w_{\psi}(x,z_{f}),v_{\psi}(x,z_{f}),\lambda_{\chi}(x,z_{f})] + \frac{\lambda_{\chi}(x,0)}{2} \varphi^{2}(x,0) \Big\} \Big] , \eqa
supported by the RG-flow equations of three coupling functions. Then, the Lagrange equation of motion is given by
\bqa && \partial_{z}^{2} \varphi(x,z) + \frac{\beta_{\lambda_{\chi}}(x,z)}{\lambda_{\chi}(x,z)} \partial_{z} \varphi(x,z) \nn && = - \Big\{\frac{1 - \beta_{\lambda_{\chi}}(x,z)}{\lambda_{\chi}(x,z)} - \Big(\beta_{\varphi}(x,z) \delta_{\varphi} + \beta_{w_{\psi}}(x,z) \delta_{w_{\psi}} + \beta_{v_{\psi}}(x,z) \delta_{v_{\psi}} + \beta_{\lambda_{\chi}}(x,z) \delta_{\lambda_{\chi}} \Big) \Big\} \beta_{\varphi}(x,z) , \eqa
supported by the IR boundary condition
\bqa && \partial_{z_{f}} \varphi(x,z_{f}) = \Big(1 + \frac{1}{\lambda_{\chi}(x,z_{f})}\Big) \beta_{\varphi}[\varphi(x,z_{f}),w_{\psi}(x,z_{f}),v_{\psi}(x,z_{f}),\lambda_{\chi}(x,z_{f})] \eqa
and the UV one
\bqa && \partial_{z} \varphi(x,z) \Big|_{z = 0} = \varphi(x,0) + \beta_{\varphi}[\varphi(x,0),w_{\psi}(x,0),v_{\psi}(x,0),\lambda_{\chi}(x,0)] . \eqa

Now, we have a complete intertwined structure between the RG flows of the coupling functions and the equation of motion of the chiral condensate. It is interesting to notice that the mean-field equation of the order-parameter field is given by the second-order differential equation instead of an algebraic equation. These renormalized coupling functions and the order-parameter field appear in the IR boundary action functional as
\bqa && \mathcal{S}_{IR} = \int d^{D} x \Big\{ \bar{\psi}_{\sigma}(x) \Big(w_{\psi}(x,z_{f}) \gamma^{\tau} \partial_{\tau} - i v_{\psi}(x,z_{f}) \gamma^{i} \partial_{i}\Big) \psi_{\sigma}(x) - i \lambda_{\chi}(x,z_{f}) \varphi(x,z_{f}) \bar{\psi}_{\sigma}(x) \psi_{\sigma}(x) \Big\} . \label{On-Shell_Action_IR} \eqa
This discussion implies self-consistency of this framework, which means that everything is determined by the effective field theory itself as it should. Actually, this point can be more clarified based on the Hamilton-Jacobi formulation \cite{Nonperturbative_Wilson_RG}, which will not be further discussed in this study.

Finally, we obtain an effective on-shell free-energy functional, given by
\bqa && F = \frac{N}{\beta} \int d^{D} x \Big\{ \mathcal{V}_{eff}[\varphi(x,z_{f}),w_{\psi}(x,z_{f}),v_{\psi}(x,z_{f}),\lambda_{\chi}(x,z_{f})] \nn && + \beta_{\varphi}(x,z_{f}) \varphi(x,z_{f}) + \beta_{w_{\psi}}(x,z_{f}) w_{\psi}(x,z_{f}) + \beta_{v_{\psi}}(x,z_{f}) v_{\psi}(x,z_{f}) + \beta_{\lambda_{\chi}}(x,z_{f}) \lambda_{\chi}(x,z_{f}) \nn && - \frac{\lambda_{\chi}(x,0)}{2} \varphi^{2}(x,0) \Big\} . \label{On-Shell_Free_Energy} \eqa
Here, the renormalized Dirac fermions were integrated out to give the boundary effective potential expressed by $\mathcal{V}_{eff}[\varphi(x,z_{f}),w_{\psi}(x,z_{f}),v_{\psi}(x,z_{f}),\lambda_{\chi}(x,z_{f})]$. Both the equation of motion for the chiral condensate and the RG$-$flow equations for the coupling functions have been used to kill most parts of the bulk action. Both UV and IR boundary conditions were also incorporated. One may regard this expression as the Legendre transformation for the effective potential \cite{QFT_Textbook,QFT_Finite_T}.

\subsection{Comparison between the renormalized effective field theory Eq. (\ref{RG_EFT_General}) and the nonperturbative RG-improved mean-field theory Eq. (\ref{RG_MFT_General})}

We are ready to read renormalization constants in a nonperturbative way from the nonperturbative RG-MFT Eq. (\ref{RG_MFT_General}). Comparing the renormalized effective field theory Eq. (\ref{RG_EFT_General}) with the nonperturbative RG-MFT Eq. (\ref{RG_MFT_General}) or the on-shell effective free energy Eq. (\ref{On-Shell_Free_Energy}) with the on-shell IR effective action Eq. (\ref{On-Shell_Action_IR}), we obtain
\bqa && w_{\psi}(x,z_{f}) = Z_{\psi} , ~~~~~ v_{\psi}(x,z_{f}) = Z_{v_{\psi}} v_{r \psi} , ~~~~~ \lambda_{\chi}(x,z_{f}) = Z_{\lambda_{\chi}} \lambda_{r \chi} \eqa
for renormalized coupling functions. It is natural to consider the following identification
\bqa && \varphi(x,z_{f}) = \varphi_{r}(x) . \eqa
Then, the comparison between the renormalized effective field theory Eq. (\ref{RG_EFT_General}) and the on-shell effective free energy Eq. (\ref{On-Shell_Free_Energy}) with the on-shell IR effective action Eq. (\ref{On-Shell_Action_IR}) gives rise to
\bqa && Z_{m_{\varphi}^{2}} m_{r \varphi}^{2} = \varphi^{- 2}(x,z_{f}) \Big( \ln w_{\psi}(x,z_{f}) + \beta_{\varphi}(x,z_{f}) \varphi(x,z_{f}) + \beta_{w_{\psi}}(x,z_{f}) w_{\psi}(x,z_{f}) \nn && + \beta_{v_{\psi}}(x,z_{f}) v_{\psi}(x,z_{f}) + \beta_{\lambda_{\chi}}(x,z_{f}) \lambda_{\chi}(x,z_{f}) - \frac{\lambda_{\chi}(x,0)}{2} \varphi^{2}(x,0) \Big) . \eqa
Here, we neglected the kinetic energy term (spatial and temporal fluctuations) in Eq. (\ref{RG_EFT_General}). As a result, the renormalization constants of $Z_{\psi}$, $Z_{v_{\psi}}$, $Z_{\lambda_{\chi}}$, and $Z_{m_{\varphi}^{2}}$ are determined by the nonperturbative RG-MFT Eq. (\ref{RG_MFT_General}).

\subsection{A holographic dual effective field theory in a gravity formulation}

Our previous discussions show that the nonperturbative RG-MFT not only shares several characteristic features of the holographic dual effective field theory but also extends the holographic duality conjecture to that away from conformal invariance, as pointed out in the introduction. In this subsection, we clarify this correspondence more explicitly. To pursue this conceptual direction, we need to consider the RG transformation in a gravitational background. We start from the following effective field theory \cite{GR_Textbook}
\bqa && Z = \int D \psi_{\sigma}(x) D e_{a}^{\mu}(x) \delta\Big(e_{a}^{\mu}(x) - \delta_{a}^{\mu}\Big) \exp\Big[ - \int d^{D} x \sqrt{g(x)} \Big\{ \bar{\psi}_{\sigma}(x) \gamma^{a} e_{a}^{\tau}(x) \Big( \partial_{\tau} - \frac{i}{4} \omega_{\tau}^{a'b'}(x) \sigma_{a'b'} \Big) \psi_{\sigma}(x) \nn && - i \bar{\psi}_{\sigma}(x) \gamma^{a} e_{a}^{i}(x) \Big( \partial_{i} - \frac{i}{4} \omega_{i}^{a'b'}(x) \sigma_{a'b'} \Big) \psi_{\sigma}(x) + \frac{\lambda_{\chi}^{g}}{2 N} \bar{\psi}_{\sigma}(x) \psi_{\sigma}(x) \bar{\psi}_{\sigma'}(x) \psi_{\sigma'}(x) \Big\} \Big] . \eqa
Here, $\psi_{\sigma}(x)$ is a Dirac spinor at $x$ in $D$ spacetime dimensions. $\sigma$ runs from $1$ to $N$, denoting the flavor degeneracy of Dirac fermions. $\gamma^{a}$ is a Dirac $\gamma$ matrix, defined in a local rest frame at $x$ and satisfying the Clifford algebra $\{\gamma^{a}, \gamma^{b}\} = 2 \delta^{ab}$ with the Euclidean signature. $e_{a}^{\mu}(x)$ defines the local rest frame given by the tangent manifold at $x$, called vierbein. The corresponding background metric is given by the vierbein as follows $g_{\mu\nu} = e_{\mu}^{a} e_{\nu}^{b} \delta_{ab}$. $\omega_{\mu}^{ab} = e_{\nu}^{a} \partial_{\mu} e^{\nu b} + e_{\nu}^{a} \Gamma^{\nu}_{\sigma\mu} e^{\sigma b}$ is a background spin connection and $\sigma_{ab} = \frac{i}{2} [\gamma^{a}, \gamma^{b}]$ is a commutator of Dirac gamma matrixes in the local rest frame. Here, $\Gamma_{\sigma\mu}^{\nu} = \frac{1}{2} g^{\nu\delta} (\partial_{\sigma} g_{\delta\mu} + \partial_{\mu} g_{\sigma\delta} - \partial_{\delta} g_{\sigma\mu})$ is the Christoffel symbol. $\lambda_{\chi}^{g}$ is the coupling constant of an effective interaction term for dynamical mass generation. We point out that this curved background geometry is introduced just formally, denoted by $\delta\Big(e_{a}^{\mu}(x) - \delta_{a}^{\mu}\Big)$.

Following our previous works \cite{Nonperturbative_Wilson_RG,Einstein_Klein_Gordon_RG_Kim,Einstein_Dirac_RG_Kim,RG_GR_Geometry_I_Kim,RG_GR_Geometry_II_Kim}, the RG transformation of which is essentially the same as that of the previous subsection, we obtain
\bqa && Z = \int D \psi_{\sigma}(x) D \varphi_{g}(x,z) D \pi_{\varphi}^{g}(x,z) D g_{\mu\nu}(x,z) D \pi^{\mu\nu}(x,z) D \lambda_{\chi}^{g}(x,z) D \pi_{\lambda_{\chi}^{g}}(x,z) \delta\Big(e_{a}^{\mu}(x,0) - \delta_{a}^{\mu}\Big) \nn && \exp\Bigg[ - \int d^{D} x \Bigg\{ \sqrt{g(x,z_{f})} \Bigg( \bar{\psi}_{\sigma}(x) \gamma^{a} e_{a}^{\tau}(x,z_{f}) \Big( \partial_{\tau} - \frac{i}{4} \omega_{\tau}^{a'b'}(x,z_{f}) \sigma_{a'b'} \Big) \psi_{\sigma}(x) \nn && - i \bar{\psi}_{\sigma}(x) \gamma^{a} e_{a}^{i}(x,z_{f}) \Big( \partial_{i} - \frac{i}{4} \omega_{i}^{a'b'}(x,z_{f}) \sigma_{a'b'} \Big) \psi_{\sigma}(x) - i \lambda_{\chi}^{g}(x,z_{f}) \varphi_{g}(x,z_{f}) \bar{\psi}_{\sigma}(x) \psi_{\sigma}(x) \Bigg) \nn && + N \sqrt{g(x,0)} \frac{\lambda_{\chi}^{g}(x,0)}{2} \varphi_{g}^{2}(x,0) \Bigg\} - N \int_{0}^{z_{f}} d z \int d^{D} x \Bigg\{ \pi_{\varphi}^{g}(x,z) \Big( \partial_{z} \varphi_{g}(x,z) - \beta_{\varphi}^{g}[\varphi_{g}(x,z),g_{\mu\nu}(x,z),\lambda_{\chi}^{g}(x,z)] \Big) \nn && - \frac{1}{2 \lambda_{\chi}^{g}(x,z) \sqrt{g(x,z)}} \pi_{\varphi}^{{g} 2}(x,z) + \pi^{\mu\nu}(x,z) \Big( \partial_{z} g_{\mu\nu}(x,z) - \beta_{\mu\nu}^{g}[\varphi_{g}(x,z),g_{\mu\nu}(x,z),\lambda_{\chi}^{g}(x,z)] \Big) \nn && + \pi_{\lambda_{\chi}^{g}}(x,z) \Big( \partial_{z} \lambda_{\chi}^{g}(x,z) - \beta_{\lambda_{\chi}^{g}}[\varphi_{g}(x,z),g_{\mu\nu}(x,z),\lambda_{\chi}^{g}(x,z)] \Big) + \mathcal{V}_{eff}[\varphi_{g}(x,z),g_{\mu\nu}(x,z),\lambda_{\chi}^{g}(x,z)] \Bigg\} \Bigg] . \eqa
Although this expression also looks quite complicated, its structure is essentially the same as what we have discussed in the previous subsection. Here, we consider the gaussian normal coordinate system $d s^{2} = d z^{2} + g_{\mu\nu}(x,z) d x^{\mu} d x^{\nu}$ in the Arnowitt-Deser-Misner (ADM) decomposition \cite{ADM_Hamiltonian_Formulation}, given by
\bqa && d s^{2} = \Big( \mathcal{N}^{2}(x,z) + \mathcal{N}_{\mu}(x,z) \mathcal{N}^{\mu}(x,z) \Big) d z^{2} + 2 \mathcal{N}_{\mu}(x,z) d x^{\mu} d z + g_{\mu\nu}(x,z) d x^{\mu} d x^{\nu} , \eqa
where the gauge fixing condition for the lapse function $\mathcal{N}(x,z) = 1$ and the shift vector $\mathcal{N}_{\mu}(x,z) = 0$ is taken into account.

The Hamiltonian equation of motion for the chiral condensate can be found from the effective bulk action, given by
\bqa && \mathcal{S}_{bulk} = N \int_{0}^{z_{f}} d z \int d^{D} x \Bigg\{ \pi_{\varphi}^{g}(x,z) \Big( \partial_{z} \varphi_{g}(x,z) - \beta_{\varphi}^{g}[\varphi_{g}(x,z),g_{\mu\nu}(x,z),\lambda_{\chi}^{g}(x,z)] \Big) - \frac{1}{2 \lambda_{\chi}^{g}(x,z) \sqrt{g(x,z)}} \pi_{\varphi}^{{g} 2}(x,z) \nn && + \mathcal{V}_{eff}[\varphi_{g}(x,z),g_{\mu\nu}(x,z),\lambda_{\chi}^{g}(x,z)] \Bigg\} . \eqa
Here, we do not repeat to show it. Again, the effective potential is generated from the RG transformation for Dirac fermions as follows
\bqa && N \mathcal{V}_{eff}[\varphi_{g}(x,z),g_{\mu\nu}(x,z),\lambda_{\chi}^{g}(x,z)] \nn && = - \ln \int_{\Lambda(z)} D \psi_{\sigma}(x) \exp\Big[ - \int d^{D} x \sqrt{g(x,z)} \Big\{ \bar{\psi}_{\sigma}(x) \gamma^{a} e_{a}^{\tau}(x,z) \Big( \partial_{\tau} - \frac{i}{4} \omega_{\tau}^{a'b'}(x,z) \sigma_{a'b'} \Big) \psi_{\sigma}(x) \nn && - i \bar{\psi}_{\sigma}(x) \gamma^{a} e_{a}^{i}(x,z) \Big( \partial_{i} - \frac{i}{4} \omega_{i}^{a'b'}(x,z) \sigma_{a'b'} \Big) \psi_{\sigma}(x) - i \lambda_{\chi}^{g}(x,z) \varphi_{g}(x,z) \bar{\psi}_{\sigma}(x) \psi_{\sigma}(x) \Big\} \Big] . \label{Effective_Potential_HDEFT}  \eqa
Then, the RG $\beta-$function for the chiral condensate is given by
\bqa N \beta_{\varphi}^{g}[\varphi_{g}(x,z),g_{\mu\nu}(x,z),\lambda_{\chi}^{g}(x,z)] &=& - \frac{1}{\sqrt{g(x,z)}} \delta_{\varphi} N \mathcal{V}_{eff}[\varphi_{g}(x,z),g_{\mu\nu}(x,z),\lambda_{\chi}^{g}(x,z)] \equiv \frac{i \lambda_{\chi}^{g}(x,z)}{\sqrt{g(x,z)}} \Big\langle \bar{\psi}_{\sigma}(x) \psi_{\sigma}(x) \Big\rangle , \nn \eqa
where the ensemble average $\langle ... \rangle$ is taken into account based on
\bqa && \mathcal{S}_{\Lambda(z)} = \int d^{D} x \sqrt{g(x,z)} \Big\{ \bar{\psi}_{\sigma}(x) \gamma^{a} e_{a}^{\tau}(x,z) \Big( \partial_{\tau} - \frac{i}{4} \omega_{\tau}^{a'b'}(x,z) \sigma_{a'b'} \Big) \psi_{\sigma}(x) \nn && - i \bar{\psi}_{\sigma}(x) \gamma^{a} e_{a}^{i}(x,z) \Big( \partial_{i} - \frac{i}{4} \omega_{i}^{a'b'}(x,z) \sigma_{a'b'} \Big) \psi_{\sigma}(x) - i \lambda_{\chi}^{g}(x,z) \varphi_{g}(x,z) \bar{\psi}_{\sigma}(x) \psi_{\sigma}(x) \Big\} . \eqa

As a side remark, the gradient expansion of the RG-generated effective potential for both the metric tensor and the chiral condensate order-parameter field with the effective interaction parameter gives rise to
\bqa && \mathcal{V}_{eff}[\varphi_{g}(x,z),g_{\mu\nu}(x,z),\lambda_{\chi}^{g}(x,z)] \approx \sqrt{g(x,z)} \Big\{ \frac{1}{2 \kappa_{g}} \Big(R(x,z) - 2 \Lambda_{g}\Big) \nn && + \frac{\mathcal{C}_{\varphi}^{g}}{2} g^{\mu\nu}(x,z) \Big(\partial_{\mu} [\lambda_{\chi}^{g}(x,z) \varphi_{g}(x,z)] \Big) \Big(\partial_{\nu} [\lambda_{\chi}^{g}(x,z) \varphi_{g}(x,z)] \Big) + \xi_{\varphi}^{g} R(x,z) [\lambda_{\chi}^{g}(x,z) \varphi_{g}(x,z)]^{2} \Big\} . \eqa
The first Einstein-Hilbert action with D$-$dimensional Ricci scalar $R(x,z)$ is referred to as induced gravity \cite{Gradient_Expansion_Gravity_I,Gradient_Expansion_Gravity_II}, where higher curvature terms \cite{Higher_Curvature_GR} are not taken into account. Here, both the cosmological constant $\Lambda_{g}$ and the effective gravitational one $\kappa_{g}$ can be determined in principle by performing the gradient expansion on a general curved spacetime manifold explicitly. However, it is demanding in practice due to renormalization effects. In this respect we regard them as some phenomenological parameters. The second term counts the D$-$dimensional kinetic energy of the order-parameter field at a given energy scale $z$. The last term describes curvature-induced effective mass for scalar-field fluctuations.

$g_{\mu\nu}(x,z)$ plays the role of coupling functions such as the wave-function renormalization and the velocity renormalization, which will be discussed below more clearly. $\pi^{\mu\nu}(x,z)$ is the canonical conjugate field to $g_{\mu\nu}(x,z)$. Integrating over $\pi^{\mu\nu}(x,z)$, we obtain the RG-flow equation for the metric tensor, given by
\bqa && \partial_{z} g_{\mu\nu}(x,z) = \beta_{\mu\nu}^{g}[\varphi_{g}(x,z),g_{\mu\nu}(x,z),\lambda_{\chi}^{g}(x,z)] , \eqa
where the RG$-\beta$ function for the metric tensor is
\bqa N \beta_{\mu\nu}^{g}[\varphi_{g}(x,z),g_{\mu\nu}(x,z),\lambda_{\chi}^{g}(x,z)] &=& - \frac{2}{\sqrt{g(x,z)}} \mathcal{G}_{\mu\nu\rho\gamma}(x,z) \delta_{g_{\rho\gamma}} N \mathcal{V}_{eff}[\varphi_{g}(x,z),g_{\mu\nu}(x,z),\lambda_{\chi}^{g}(x,z)] \nn &\equiv& - \mathcal{G}_{\mu\nu\rho\gamma}(x,z) \Big\langle T^{\rho\gamma}(x,z) \Big\rangle . \eqa
Here, $\mathcal{G}_{\mu\nu\rho\gamma}(x,z) \equiv g_{\mu\rho}(x,z) g_{\nu\gamma}(x,z) - \frac{1}{D-1} g_{\mu\nu}(x,z) g_{\rho\gamma}(x,z)$ is de Witt supermetric \cite{DeWitt_Metric}, taking into account transverseness. $T^{\rho\gamma}(x,z)$ is the energy-momentum tensor for Dirac fermions.

One may point out why $\frac{\lambda}{2} \frac{1}{\sqrt{g(x,z)}} \pi^{\mu\nu}(x,z) \mathcal{G}_{\mu\nu\rho\gamma}(x,z) \pi^{\rho\gamma}(x,z)$ does not appear in this effective field theory. We recall that $\frac{1}{2 \lambda_{\chi}^{g}(x,z) \sqrt{g(x,z)}} \pi_{\varphi}^{{g} 2}(x,z)$ in the bulk effective action for the chiral condensate originates from the mass term $\sqrt{g(x)} \frac{\lambda_{\chi}^{g}}{2} \varphi_{g}^{2}(x)$ in the RG transformation. To have $\frac{\lambda}{2} \frac{1}{\sqrt{g(x,z)}} \pi^{\mu\nu}(x,z) \mathcal{G}_{\mu\nu\rho\gamma}(x,z) \pi^{\rho\gamma}(x,z)$ in the bulk effective action for the metric tensor, we need a mass term for gravitational fluctuations in the RG transformation. These massive gravitational fluctuations can be generated by the so called $T \bar{T}$ deformation \cite{TTbar_Deformation}, here $\sim T^{\mu\nu}(x) \mathcal{G}_{\mu\nu\rho\gamma}(x) T^{\rho\gamma}(x)$, where $T^{\mu\nu}(x)$ is the energy-momentum tensor of Dirac fermions. We refer this aspect to our recent studies \cite{Nonperturbative_Wilson_RG,Einstein_Klein_Gordon_RG_Kim,Einstein_Dirac_RG_Kim}. Since we do not take into account such effective interactions from the beginning, we do not have the second-order differential equation for the RG evolution of the metric tensor as the chiral condensate. We will discuss this point in the last section.

The RG flow of the interaction vertex for chiral symmetry breaking is given by
\bqa && \partial_{z} \lambda_{\chi}^{g}(x,z) = \beta_{\lambda_{\chi}^{g}}[\varphi_{g}(x,z),g_{\mu\nu}(x,z),\lambda_{\chi}^{g}(x,z)] , \eqa
where the RG$-\beta$ function is
\bqa N \beta_{\lambda_{\chi}^{g}}[\varphi_{g}(x,z),g_{\mu\nu}(x,z),\lambda_{\chi}^{g}(x,z)] &=& - \frac{1}{\sqrt{g(x,z)}} \delta_{\lambda_{\chi}^{g}} N \mathcal{V}_{eff}[\varphi_{g}(x,z),g_{\mu\nu}(x,z),\lambda_{\chi}^{g}(x,z)] \equiv \frac{i \varphi_{g}(x,z)}{\sqrt{g(x,z)}} \Big\langle \bar{\psi}_{\sigma}(x) \psi_{\sigma}(x) \Big\rangle . \nn \eqa

All these renormalized coupling functions and the order-parameter field appear in the IR boundary effective action as follows
\bqa && \mathcal{S}_{IR} = \int d^{D} x \sqrt{g(x,z_{f})} \Big\{ \bar{\psi}_{\sigma}(x) \gamma^{a} e_{a}^{\tau}(x,z_{f}) \Big( \partial_{\tau} - \frac{i}{4} \omega_{\tau}^{a'b'}(x,z_{f}) \sigma_{a'b'} \Big) \psi_{\sigma}(x) \nn && - i \bar{\psi}_{\sigma}(x) \gamma^{a} e_{a}^{i}(x,z_{f}) \Big( \partial_{i} - \frac{i}{4} \omega_{i}^{a'b'}(x,z_{f}) \sigma_{a'b'} \Big) \psi_{\sigma}(x) - i \lambda_{\chi}^{g}(x,z_{f}) \varphi_{g}(x,z_{f}) \bar{\psi}_{\sigma}(x) \psi_{\sigma}(x) \Big\} . \eqa
This effective action defines the IR boundary condition unambiguously.

%
%

Our final question in the formulation perspective is how the nonperturbative RG-MFT in the previous subsection can be linked to the emergent holographic dual effective field theory in the present subsection. Comparing the effective potential Eq. (\ref{Effective_Potential_RG_MFT}) in the RG-MFT with that Eq. (\ref{Effective_Potential_HDEFT}) in the emergent holographic dual effective field theory, we identify the field renormalization and the velocity one with the time and spatial component of the vierbein tensor, respectively, as follows
\bqa && w_{\psi}(x,z) = \sqrt{g(x,z)} e_{a}^{\tau}(x,z) , ~~~~~ v_{\psi}(x,z) = \sqrt{g(x,z)} e_{a}^{i}(x,z) . \label{ivb}\eqa
The chiral condensate order-parameter field matches as
\bqa && \lambda_{\chi}(x,z) \varphi(x,z) = \sqrt{g(x,z)} \lambda_{\chi}^{g}(x,z) \varphi_{g}(x,z) . \eqa
Based on this identification, we obtain equivalence between two effective potentials
%
%
\bqa && \mathcal{V}_{eff}[\varphi(x,z),w_{\psi}(x,z),v_{\psi}(x,z),\lambda_{\chi}(x,z)] = \mathcal{V}_{eff}[\varphi_{g}(x,z),g_{\mu\nu}(x,z),\lambda_{\chi}^{g}(x,z)] . \eqa
Here, we did not take into account the spin connection in the gravity formulation. Frankly speaking, we do not find the corresponding observable to the spin connection in the RG-MFT framework. Although we do not have a definite answer for this discrepancy, we suspect that diffeomorphism invariance may not be taken into account carefully in the RG-MFT framework. One may regard that this correspondence holds in two spacetime dimensions, where the spin connection vanishes \cite{GR_Textbook}.

%
%
%
%

\section{A vacuum solution of the holographic dual effective field theory} \label{Mean_Field_Theory}

\subsection{Renormalized partition function with translational symmetry}

In the previous section, we obtained heavily intertwined nonlinear first- and second-order differential equations between renormalized coupling functions and order parameter fields in the large$-N$ limit. To solve these coupled differential equations, we take the `perturbative' approach as usual. First, we consider a vacuum solution, characterized by translational symmetry. Second, we introduce small fluctuations around this translational invariant solution and linearize all coupled nonlinear differential equations based on this vacuum solution. Solving such linearized coupled differential equations, we find corresponding collective-mode spectra from the vacuum state in the holographic dual description \cite{Holographic_Liquid_Son_I,Holographic_Liquid_Son_II,Holographic_Liquid_Son_III,Holographic_Liquid_Son_IV}.

Now, we enforce the translational symmetry for all the coupling functions and the chiral symmetry breaking field. Then, we obtain the following partition function
\bqa && Z = \int D \psi_{\sigma}(x) D \varphi(z) D \pi_{\varphi}(z) D w_{\psi}(z) D \pi_{w_{\psi}}(z) D v_{\psi}(z) D \pi_{v_{\psi}}(z) D \lambda_{\chi}(z) D \pi_{\lambda_{\chi}}(z) \nn && \exp\Big[ - \int d^{2} x \Big\{ \bar{\psi}_{\sigma}(x) \Big(w_{\psi}(z_{f}) \gamma^{\tau} \partial_{\tau} - i v_{\psi}(z_{f}) \gamma^{x} \partial_{x}\Big) \psi_{\sigma}(x) - i \lambda_{\chi}(z_{f}) \varphi(z_{f}) \bar{\psi}_{\sigma}(x) \psi_{\sigma}(x) + \frac{N \lambda_{\chi}(0)}{2} \varphi^{2}(0) \Big\} \nn && - \beta L N \int_{0}^{z_{f}} d z \Big\{ \pi_{\varphi}(z) \Big( \partial_{z} \varphi(z) - \beta_{\varphi}[\varphi(z),w_{\psi}(z),v_{\psi}(z),\lambda_{\chi}(z)] \Big) - \frac{1}{2 \lambda_{\chi}(z)} \pi_{\varphi}^{2}(z) \nn && + \pi_{w_{\psi}}(z) \Big( \partial_{z} w_{\psi}(z) - \beta_{w_{\psi}}[\varphi(z),w_{\psi}(z),v_{\psi}(z),\lambda_{\chi}(z)] \Big) + \pi_{v_{\psi}}(z) \Big( \partial_{z} v_{\psi}(z) - \beta_{v_{\psi}}[\varphi(z),w_{\psi}(z),v_{\psi}(z),\lambda_{\chi}(z)] \Big) \nn && + \pi_{\lambda_{\chi}}(z) \Big( \partial_{z} \lambda_{\chi}(z) - \beta_{\lambda_{\chi}}[\varphi(z),w_{\psi}(z),v_{\psi}(z),\lambda_{\chi}(z)] \Big) + \mathcal{V}_{eff}[\varphi(z),w_{\psi}(z),v_{\psi}(z),\lambda_{\chi}(z)] \Big\} \Big] , \eqa
where the spacetime dependence disappears in all the dynamical fields except for the Dirac-spinor field. Although one may consider general spacetime dimensions, we focus on two spacetime dimensions here just for simplicity, which will be more clarified below.

The presence of the translational symmetry allows us to perform the path integral for Dirac fermions in the energy-momentum space. As a result, we obtain the following effective potential
\bqa && \mathcal{V}_{eff}[\varphi(z),w_{\psi}(z),v_{\psi}(z),\lambda_{\chi}(z)] \nn && = - \frac{1}{N} \ln \int_{\Lambda(z)} D \psi_{\sigma}(x) \exp\Big[ - \int d^{2} x \Big\{ \bar{\psi}_{\sigma}(x) \Big(w_{\psi}(z) \gamma^{\tau} \partial_{\tau} - i v_{\psi}(z) \gamma^{x} \partial_{x}\Big) \psi_{\sigma}(x) - i \lambda_{\chi}(z) \varphi(z) \bar{\psi}_{\sigma}(x) \psi_{\sigma}(x) \Big\} \Big] \nn && = - \frac{2}{\beta} \sum_{k = \Lambda(z)} \ln \Big\{ 2 w_{\psi}(z) \cosh \Big( \frac{\beta}{2 w_{\psi}(z)} \sqrt{v_{\psi}^{2}(z) k^{2} - \lambda_{\chi}^{2}(z) \varphi^{2}(z)} \Big) \Big\} \eqa after the Matsubara frequency summation at finite temperatures \cite{QFT_Finite_T}. $\beta = T^{-1}$ is inverse temperature. Here, the path integral for Dirac fermions has been performed at a given energy scale $\Lambda(z)$. In this respect the momentum $k$ is determined by $\Lambda(z)$, which will be clarified below.

Taking functional derivatives for this effective potential with respect to the coupling functions and the order-parameter field, we obtain RG $\beta-$functions as follows
\bqa && \beta_{\varphi}[\varphi(z),w_{\psi}(z),v_{\psi}(z),\lambda_{\chi}(z)] = - \frac{1}{w_{\psi}(z)} \sum_{k = \Lambda(z)} \frac{ \lambda_{\chi}^{2}(z) \varphi(z)}{\sqrt{v_{\psi}^{2}(z) k^{2} - \lambda_{\chi}^{2}(z) \varphi^{2}(z)}} \tanh \Big( \frac{\beta}{2 w_{\psi}(z)} \sqrt{v_{\psi}^{2}(z) k^{2} - \lambda_{\chi}^{2}(z) \varphi^{2}(z)} \Big) , \nn \\ && \beta_{w_{\psi}}[\varphi(z),w_{\psi}(z),v_{\psi}(z),\lambda_{\chi}(z)] \nn && = \sum_{k = \Lambda(z)} \Big\{ \frac{2}{\beta} \frac{1}{w_{\psi}(z)} - \frac{1}{w_{\psi}^{2}(z)} \sqrt{v_{\psi}^{2}(z) k^{2} - \lambda_{\chi}^{2}(z) \varphi^{2}(z)} \tanh \Big( \frac{\beta}{2 w_{\psi}(z)} \sqrt{v_{\psi}^{2}(z) k^{2} - \lambda_{\chi}^{2}(z) \varphi^{2}(z)} \Big) \Big\} , \\ && \beta_{v_{\psi}}[\varphi(z),w_{\psi}(z),v_{\psi}(z),\lambda_{\chi}(z)] = - \frac{1}{w_{\psi}(z)} \sum_{k = \Lambda(z)} \frac{ v_{\psi}(z) k^{2} }{\sqrt{v_{\psi}^{2}(z) k^{2} - \lambda_{\chi}^{2}(z) \varphi^{2}(z)}} \tanh \Big( \frac{\beta}{2 w_{\psi}(z)} \sqrt{v_{\psi}^{2}(z) k^{2} - \lambda_{\chi}^{2}(z) \varphi^{2}(z)} \Big) , \nn \\ && \beta_{\lambda_{\chi}}[\varphi(z),w_{\psi}(z),v_{\psi}(z),\lambda_{\chi}(z)] = \frac{1}{w_{\psi}(z)} \sum_{k = \Lambda(z)} \frac{ \lambda_{\chi}(z) \varphi^{2}(z)}{\sqrt{v_{\psi}^{2}(z) k^{2} - \lambda_{\chi}^{2}(z) \varphi^{2}(z)}} \tanh \Big( \frac{\beta}{2 w_{\psi}(z)} \sqrt{v_{\psi}^{2}(z) k^{2} - \lambda_{\chi}^{2}(z) \varphi^{2}(z)} \Big) . \nn \eqa

Introducing the effective potential and all these RG $\beta-$functions into the partition function, we obtain the following expression
\bqa && Z = \int D \varphi(z) D \pi_{\varphi}(z) D w_{\psi}(z) D \pi_{w_{\psi}}(z) D v_{\psi}(z) D \pi_{v_{\psi}}(z) D \lambda_{\chi}(z) D \pi_{\lambda_{\chi}}(z) \nn && \exp\Bigg[ - \beta L N \Bigg( - \frac{2}{\beta} \ln \Big\{ 2 w_{\psi}(z_{f}) \cosh \Big( \frac{\beta}{2 w_{\psi}(z_{f})} \sqrt{v_{\psi}^{2}(z_{f}) [\Lambda - z_{f}]^{2} - \lambda_{\chi}^{2}(z_{f}) \varphi^{2}(z_{f})} \Big) \Big\}  + \frac{\lambda_{\chi}(0)}{2} \varphi^{2}(0) \Bigg) \nn && - \beta L N \int_{0}^{z_{f}} d z \Bigg\{ \pi_{\varphi}(z) \Bigg( \partial_{z} \varphi(z) + \frac{1}{w_{\psi}(z)} \frac{ \lambda_{\chi}^{2}(z) \varphi(z)}{\sqrt{v_{\psi}^{2}(z) [\Lambda - z]^{2} - \lambda_{\chi}^{2}(z) \varphi^{2}(z)}} \tanh \Big( \frac{\beta}{2 w_{\psi}(z)} \sqrt{v_{\psi}^{2}(z) [\Lambda - z]^{2} - \lambda_{\chi}^{2}(z) \varphi^{2}(z)} \Big) \Bigg) \nn && - \frac{1}{2 \lambda_{\chi}(z)} \pi_{\varphi}^{2}(z) \nn && + \pi_{w_{\psi}}(z) \Bigg( \partial_{z} w_{\psi}(z) - \Big\{ \frac{2}{\beta} \frac{1}{w_{\psi}(z)} - \frac{1}{w_{\psi}^{2}(z)} \sqrt{v_{\psi}^{2}(z) [\Lambda - z]^{2} - \lambda_{\chi}^{2}(z) \varphi^{2}(z)} \tanh \Big( \frac{\beta}{2 w_{\psi}(z)} \sqrt{v_{\psi}^{2}(z) [\Lambda - z]^{2} - \lambda_{\chi}^{2}(z) \varphi^{2}(z)} \Big) \Big\} \Bigg) \nn && + \pi_{v_{\psi}}(z) \Bigg( \partial_{z} v_{\psi}(z) + \frac{1}{w_{\psi}(z)} \frac{ v_{\psi}(z) [\Lambda - z]^{2} }{\sqrt{v_{\psi}^{2}(z) [\Lambda - z]^{2} - \lambda_{\chi}^{2}(z) \varphi^{2}(z)}} \tanh \Big( \frac{\beta}{2 w_{\psi}(z)} \sqrt{v_{\psi}^{2}(z) [\Lambda - z]^{2} - \lambda_{\chi}^{2}(z) \varphi^{2}(z)} \Big) \Bigg) \nn && + \pi_{\lambda_{\chi}}(z) \Bigg( \partial_{z} \lambda_{\chi}(z) - \frac{1}{w_{\psi}(z)} \frac{ \lambda_{\chi}(z) \varphi^{2}(z)}{\sqrt{v_{\psi}^{2}(z) [\Lambda - z]^{2} - \lambda_{\chi}^{2}(z) \varphi^{2}(z)}} \tanh \Big( \frac{\beta}{2 w_{\psi}(z)} \sqrt{v_{\psi}^{2}(z) [\Lambda - z]^{2} - \lambda_{\chi}^{2}(z) \varphi^{2}(z)} \Big) \Bigg) \nn && - \frac{2}{\beta} \ln \Big\{ 2 w_{\psi}(z) \cosh \Big( \frac{\beta}{2 w_{\psi}(z)} \sqrt{v_{\psi}^{2}(z) [\Lambda - z]^{2} - \lambda_{\chi}^{2}(z) \varphi^{2}(z)} \Big) \Big\} \Bigg\} \Bigg] . \eqa
Here, $L$ is the size of our one-dimensional system. $\Lambda$ is the UV cutoff, where our interacting field theory is introduced. $z_{f}$ represents the final recursion step of the RG transformation, given by $z_{f} = f d z$, where $f$ counts the total number of RG transformations and $d z$ is an RG scale \cite{Einstein_Klein_Gordon_RG_Kim,Einstein_Dirac_RG_Kim}. When $z_{f} = \Lambda$ is reached, whole Dirac fermions are integrated out to give the bulk effective action except for the IR boundary condition. One more important point is that the momentum $k$ is identified with an energy scale of $\Lambda(z)$ as $k = z$. We believe that this is our natural choice in a dimensional ground, regarded to be consistent gauge fixing for the coordinate of the extradimensional space.

Recalling the on-shell free-energy $F = \frac{N}{\beta} \int d^{D} x \Big\{ \mathcal{V}_{eff}(x,z_{f}) + \beta_{\varphi}(x,z_{f}) \varphi(x,z_{f}) + \beta_{w_{\psi}}(x,z_{f}) w_{\psi}(x,z_{f}) + \beta_{v_{\psi}}(x,z_{f}) v_{\psi}(x,z_{f}) + \beta_{\lambda_{\chi}}(x,z_{f}) \lambda_{\chi}(x,z_{f}) - \frac{\lambda_{\chi}(x,0)}{2} \varphi^{2}(x,0) \Big\}$ of Eq. (\ref{On-Shell_Free_Energy}), we obtain
\bqa && \frac{1}{L N} F(z_{f}) = - \frac{2}{\beta} \ln \Big\{ 2 w_{\psi}(z_{f}) \cosh \Big( \frac{\beta}{2 w_{\psi}(z_{f})} \sqrt{v_{\psi}^{2}(z_{f}) [\Lambda - z_{f}]^{2} - \lambda_{\chi}^{2}(z_{f}) \varphi^{2}(z_{f})} \Big) \Big\} - \frac{\lambda_{\chi}(0)}{2} \varphi^{2}(0) \nn && - \frac{1}{w_{\psi}(z_{f})} \frac{ \lambda_{\chi}^{2}(z_{f}) \varphi(z_{f})}{\sqrt{v_{\psi}^{2}(z_{f}) [\Lambda - z_{f}]^{2} - \lambda_{\chi}^{2}(z_{f}) \varphi^{2}(z_{f})}} \tanh \Big( \frac{\beta}{2 w_{\psi}(z_{f})} \sqrt{v_{\psi}^{2}(z_{f}) [\Lambda - z_{f}]^{2} - \lambda_{\chi}^{2}(z_{f}) \varphi^{2}(z_{f})} \Big) \varphi(z_{f}) \nn && + \Bigg( \frac{2}{\beta} \frac{1}{w_{\psi}(z_{f})} - \frac{1}{w_{\psi}^{2}(z_{f})} \sqrt{v_{\psi}^{2}(z_{f}) [\Lambda - z_{f}]^{2} - \lambda_{\chi}^{2}(z_{f}) \varphi^{2}(z_{f})} \tanh \Big( \frac{\beta}{2 w_{\psi}(z_{f})} \sqrt{v_{\psi}^{2}(z_{f}) [\Lambda - z_{f}]^{2} - \lambda_{\chi}^{2}(z_{f}) \varphi^{2}(z_{f})} \Big) \Bigg) w_{\psi}(z_{f}) \nn && - \frac{1}{w_{\psi}(z_{f})} \frac{ v_{\psi}(z_{f}) [\Lambda - z_{f}]^{2} }{\sqrt{v_{\psi}^{2}(z_{f}) [\Lambda - z_{f}]^{2} - \lambda_{\chi}^{2}(z_{f}) \varphi^{2}(z_{f})}} \tanh \Big( \frac{\beta}{2 w_{\psi}(z_{f})} \sqrt{v_{\psi}^{2}(z_{f}) [\Lambda - z_{f}]^{2} - \lambda_{\chi}^{2}(z_{f}) \varphi^{2}(z_{f})} \Big) v_{\psi}(z_{f}) \nn && + \frac{1}{w_{\psi}(z_{f})} \frac{ \lambda_{\chi}(z_{f}) \varphi^{2}(z_{f})}{\sqrt{v_{\psi}^{2}(z_{f}) [\Lambda - z_{f}]^{2} - \lambda_{\chi}^{2}(z_{f}) \varphi^{2}(z_{f})}} \tanh \Big( \frac{\beta}{2 w_{\psi}(z_{f})} \sqrt{v_{\psi}^{2}(z_{f}) [\Lambda - z_{f}]^{2} - \lambda_{\chi}^{2}(z_{f}) \varphi^{2}(z_{f})} \Big) \lambda_{\chi}(z_{f}) . \eqa
Taking the zero-temperature limit of $\beta \rightarrow \infty$, we obtain
\bqa && \frac{1}{L N} E_{G}(z_{f}) = - \frac{3}{w_{\psi}(z_{f})} \sqrt{v_{\psi}^{2}(z_{f}) [\Lambda - z_{f}]^{2} - \lambda_{\chi}^{2}(z_{f}) \varphi^{2}(z_{f})} - \frac{1}{w_{\psi}(z_{f})} \frac{ \lambda_{\chi}^{2}(z_{f}) \varphi^{2}(z_{f})}{\sqrt{v_{\psi}^{2}(z_{f}) [\Lambda - z_{f}]^{2} - \lambda_{\chi}^{2}(z_{f}) \varphi^{2}(z_{f})}} - \frac{\lambda_{\chi}(0)}{2} \varphi^{2}(0) . \nn \eqa
If we perform the RG transformation completely, we are allowed to set $z_{f} = \Lambda$. Then, we obtain
\bqa && \frac{1}{L N} E_{G}(\Lambda) = - 2 \frac{\lambda_{\chi}(\Lambda) \varphi(\Lambda)}{w_{\psi}(\Lambda)} + \frac{\lambda_{\chi}(0)}{2} \varphi^{2}(0) , \eqa
where $\varphi(0) \rightarrow i \varphi(0)$ and $\varphi(z_{f}) \rightarrow i \varphi(z_{f})$ have been taken into account. Here, we assumed that both $\lambda_{\chi}(\Lambda)$ and $\varphi(\Lambda)$ are positive. The negative sign in $- 2 \frac{\lambda_{\chi}(\Lambda) \varphi(\Lambda)}{w_{\psi}(\Lambda)}$ gives the possibility of spontaneous chiral symmetry breaking, the solution of which will be discussed below.

Considering the high-temperature limit of $\beta \rightarrow 0$, we obtain
\bqa && \frac{1}{L N} F(z_{f}) = \frac{2}{\beta} \Big( 1 - \ln 2 w_{\psi}(z_{f}) \Big) - \beta \frac{1}{4 w_{\psi}^{2}(z_{f})} \Big(5 v_{\psi}^{2}(z_{f}) [\Lambda - z_{f}]^{2} - 3 \lambda_{\chi}^{2}(z_{f}) \varphi^{2}(z_{f})\Big) - \frac{\lambda_{\chi}(0)}{2} \varphi^{2}(0) . \eqa
Taking $z_{f} = \Lambda$, we have
\bqa && \frac{1}{L N} F(z_{f}) = \frac{2}{\beta} \Big( 1 - \ln 2 w_{\psi}(z_{f}) \Big) - \beta \frac{3 \lambda_{\chi}^{2}(z_{f}) \varphi^{2}(z_{f})}{4 w_{\psi}^{2}(z_{f})} + \frac{\lambda_{\chi}(0)}{2} \varphi^{2}(0) , \eqa
where $\varphi(0) \rightarrow i \varphi(0)$ and $\varphi(z_{f}) \rightarrow i \varphi(z_{f})$ have been taken into account.

Before going further, we discuss what happens above two spacetime dimensions. Suppose
\bqa && \int d \tau \int d^{D-1} x = \int d \tau \int \frac{d^{D-1} k}{(2 \pi)^{D-1}} = \int d \tau \int \frac{d^{D-2} \Omega}{(2 \pi)^{D-2}} \int \frac{d k}{2 \pi} k^{D-2} \equiv \int d \tau \int d z \rho(z) , \eqa
where $\int \frac{d^{D-2} \Omega}{(2 \pi)^{D-2}}$ represents $D-2$ dimensional solid-angle integration. The last equality results in
\bqa && \rho(z) \propto z^{D-2} \eqa
on a dimensional ground. In this respect a pseudogap behavior appears in the density of states, where the integral is more `regularized' at IR above two spacetime dimensions.

\subsection{Renormalization group flows for all the coupling functions and the chiral condensate}

RG-flow equations for three coupling functions are given by
\bqa \partial_{z} w_{\psi}(z) &=& \frac{2}{\beta} \frac{1}{w_{\psi}(z)} - \frac{1}{w_{\psi}^{2}(z)} \sqrt{v_{\psi}^{2}(z) [\Lambda - z]^{2} - \lambda_{\chi}^{2}(z) \varphi^{2}(z)} \tanh \Big( \frac{\beta}{2 w_{\psi}(z)} \sqrt{v_{\psi}^{2}(z) [\Lambda - z]^{2} - \lambda_{\chi}^{2}(z) \varphi^{2}(z)} \Big) , \nn \partial_{z} v_{\psi}(z) &=& - \frac{1}{w_{\psi}(z)} \frac{ v_{\psi}(z) [\Lambda - z]^{2} }{\sqrt{v_{\psi}^{2}(z) [\Lambda - z]^{2} - \lambda_{\chi}^{2}(z) \varphi^{2}(z)}} \tanh \Big( \frac{\beta}{2 w_{\psi}(z)} \sqrt{v_{\psi}^{2}(z) [\Lambda - z]^{2} - \lambda_{\chi}^{2}(z) \varphi^{2}(z)} \Big) , \nn \partial_{z} \lambda_{\chi}(z) &=& \frac{1}{w_{\psi}(z)} \frac{ \lambda_{\chi}(z) \varphi^{2}(z)}{\sqrt{v_{\psi}^{2}(z) [\Lambda - z]^{2} - \lambda_{\chi}^{2}(z) \varphi^{2}(z)}} \tanh \Big( \frac{\beta}{2 w_{\psi}(z)} \sqrt{v_{\psi}^{2}(z) [\Lambda - z]^{2} - \lambda_{\chi}^{2}(z) \varphi^{2}(z)} \Big) . \eqa
%
%
These first-order differential equations are supported by their UV boundary conditions, i.e., their initial values at a given UV cutoff $\Lambda$.

The Hamiltonian equation of motion for the chiral condensate is given by
\bqa && \pi_{\varphi}(z) = \lambda_{\chi}(z) \Bigg( \partial_{z} \varphi(z) + \frac{1}{w_{\psi}(z)} \frac{ \lambda_{\chi}^{2}(z) \varphi(z)}{\sqrt{v_{\psi}^{2}(z) [\Lambda - z]^{2} - \lambda_{\chi}^{2}(z) \varphi^{2}(z)}} \tanh \Big( \frac{\beta}{2 w_{\psi}(z)} \sqrt{v_{\psi}^{2}(z) [\Lambda - z]^{2} - \lambda_{\chi}^{2}(z) \varphi^{2}(z)} \Big) \Bigg) , \nn \\ && \partial_{z} \pi_{\varphi}(z) = \frac{1}{w_{\psi}(z)} \frac{ \lambda_{\chi}^{2}(z) \varphi(z)}{\sqrt{v_{\psi}^{2}(z) [\Lambda - z]^{2} - \lambda_{\chi}^{2}(z) \varphi^{2}(z)}} \tanh \Big( \frac{\beta}{2 w_{\psi}(z)} \sqrt{v_{\psi}^{2}(z) [\Lambda - z]^{2} - \lambda_{\chi}^{2}(z) \varphi^{2}(z)} \Big) \nn && + \pi_{\varphi}(z) \frac{1}{w_{\psi}(z)} \frac{ \lambda_{\chi}^{2}(z) }{\sqrt{v_{\psi}^{2}(z) [\Lambda - z]^{2} - \lambda_{\chi}^{2}(z) \varphi^{2}(z)}} \tanh \Big( \frac{\beta}{2 w_{\psi}(z)} \sqrt{v_{\psi}^{2}(z) [\Lambda - z]^{2} - \lambda_{\chi}^{2}(z) \varphi^{2}(z)} \Big) \nn && + \pi_{\varphi}(z) \frac{1}{w_{\psi}(z)} \frac{ \lambda_{\chi}^{4}(z) \varphi^{2}(z)}{\Big(v_{\psi}^{2}(z) [\Lambda - z]^{2} - \lambda_{\chi}^{2}(z) \varphi^{2}(z)\Big)^{\frac{3}{2}}} \tanh \Big( \frac{\beta}{2 w_{\psi}(z)} \sqrt{v_{\psi}^{2}(z) [\Lambda - z]^{2} - \lambda_{\chi}^{2}(z) \varphi^{2}(z)} \Big) \nn && - \beta \pi_{\varphi}(z) \frac{1}{2 w_{\psi}^{2}(z)} \frac{ \lambda_{\chi}^{4}(z) \varphi^{2}(z)}{v_{\psi}^{2}(z) [\Lambda - z]^{2} - \lambda_{\chi}^{2}(z) \varphi^{2}(z)} \mbox{sech}^{2} \Big( \frac{\beta}{2 w_{\psi}(z)} \sqrt{v_{\psi}^{2}(z) [\Lambda - z]^{2} - \lambda_{\chi}^{2}(z) \varphi^{2}(z)} \Big) . \eqa
These coupled first-order differential equations or the corresponding second-order differential equation needs two boundary conditions, here IR and UV ones
\bqa && \partial_{z_{f}} \varphi(z_{f}) = - \Big(1 + \frac{1}{\lambda_{\chi}(z_{f})}\Big) \frac{1}{w_{\psi}(z_{f})} \frac{ \lambda_{\chi}^{2}(z_{f}) \varphi(z_{f})}{\sqrt{v_{\psi}^{2}(z_{f}) [\Lambda - z_{f}]^{2} - \lambda_{\chi}^{2}(z_{f}) \varphi^{2}(z_{f})}} \tanh \Big( \frac{\beta}{2 w_{\psi}(z_{f})} \sqrt{v_{\psi}^{2}(z_{f}) [\Lambda - z_{f}]^{2} - \lambda_{\chi}^{2}(z_{f}) \varphi^{2}(z_{f})} \Big) , \nn \\ && \partial_{z} \varphi(z) \Big|_{z = 0} = \varphi(0) - \frac{1}{w_{\psi}(0)} \frac{ \lambda_{\chi}^{2}(0) \varphi(0)}{\sqrt{v_{\psi}^{2}(0) \Lambda^{2} - \lambda_{\chi}^{2}(0) \varphi^{2}(0)}} \tanh \Big( \frac{\beta}{2 w_{\psi}(0)} \sqrt{v_{\psi}^{2}(0) \Lambda^{2} - \lambda_{\chi}^{2}(0) \varphi^{2}(0)} \Big) . \eqa

\subsection{Possible vacuum solutions at high temperatures ($\beta \rightarrow 0$)}

First, we solve all these coupled equations in the high-temperature limit of $\beta \rightarrow 0$. The RG-flow equations of the coupling functions are simplified as
\bqa && \partial_{z} w_{\psi}(z) = \frac{2}{\beta} \frac{1}{w_{\psi}(z)} - \frac{\beta}{2 w_{\psi}^{3}(z)} \Big( v_{\psi}^{2}(z) [\Lambda - z]^{2} - \lambda_{\chi}^{2}(z) \varphi^{2}(z) \Big) \longrightarrow \frac{2}{\beta} \frac{1}{w_{\psi}(z)} , \nn && \partial_{z} v_{\psi}(z) = - \beta \frac{v_{\psi}(z) [\Lambda - z]^{2}}{2 w_{\psi}^{2}(z)} \longrightarrow 0 , ~~~~~ \partial_{z} \lambda_{\chi}(z) = \beta \frac{\lambda_{\chi}(z) \varphi^{2}(z)}{2 w_{\psi}^{2}(z)} \longrightarrow 0 \eqa
up to the first order in $\beta$. If we focus on the zeroth order in $\beta$, we obtain vanishing RG $\beta-$functions, denoted by the long right arrow. These vanishing RG $\beta-$functions define the UV fixed point that we start from. On the other hand, the RG $\beta-$function of $w_{\psi}^{2}(z)$ shows the divergence of $w_{\psi}^{2}(z)$. This implies the classical nature of the dynamics at the UV fixed point, where the time-derivative term may be neglected safely.

The Hamiltonian equation of motion is also reduced as
\bqa && \pi_{\varphi}(z) = \lambda_{\chi}(z) \Big( \partial_{z} \varphi(z) + \beta \frac{\lambda_{\chi}^{2}(z) \varphi(z)}{2 w_{\psi}^{2}(z)} \Big) \longrightarrow \lambda_{\chi}(z) \partial_{z} \varphi(z) , \nn && \partial_{z} \pi_{\varphi}(z) = \beta \frac{\lambda_{\chi}^{2}(z) \varphi(z)}{2 w_{\psi}^{2}(z)} + \beta \pi_{\varphi}(z) \frac{\lambda_{\chi}^{2}(z)}{2 w_{\psi}^{2}(z)} \longrightarrow 0 . \eqa
Then, the Lagrange equation of motion is given by
\bqa && \partial_{z}^{2} \varphi(z) - \frac{2 \lambda_{\chi}^{2}(z) \varphi(z)}{w_{\psi}^{4}(z)} + \beta \frac{\varphi^{2}(z)}{2 w_{\psi}^{2}(z)} [\partial_{z} \varphi(z)] - \beta \frac{\lambda_{\chi}(z) \varphi(z)}{2 w_{\psi}^{2}(z)} = 0 \longrightarrow \partial_{z}^{2} \varphi(z) - \frac{2 \lambda_{\chi}^{2}(z) \varphi(z)}{w_{\psi}^{4}(z)} = 0 . \eqa
This equation is supported by the UV boundary condition
\bqa && \partial_{z} \varphi(z) \Big|_{z = 0} = \varphi(0) - \beta \frac{\lambda_{\chi}^{2}(0) \varphi(0)}{2 w_{\psi}^{2}(0)} \longrightarrow \varphi(0) . \eqa

Considering the zeroth order in $\beta$ and introducing the UV boundary condition into this equation, we obtain
\bqa && \partial_{z}^{2} \varphi(z) = \frac{2 \lambda_{\chi}^{2}(z)}{w_{\psi}^{4}(z)} \partial_{z} \varphi(z) \longrightarrow \varphi(z) = \varphi(0) \frac{w_{\psi}^{4}(0)}{2 \lambda_{\chi}^{2}(0)} \Big\{ \exp\Big( \frac{2 \lambda_{\chi}^{2}(0)}{w_{\psi}^{4}(0)} z \Big) - 1 \Big\} + \varphi(0) \eqa
for the chiral condensate. $\varphi(0)$ is the UV value of the chiral symmetry breaking. Here, we do not discuss this high-temperature physics further.

%
%

\subsection{Possible vacuum solutions at zero temperature ($\beta \rightarrow \infty$)}

%
%

\subsubsection{Matching method}

	We enforce the reality condition for the chiral condensate as $\varphi(x) \longrightarrow i \varphi(x)$ and $\pi_{\varphi}(x) \longrightarrow i \pi_{\varphi}(x)$, where $x=\frac{z}{\Lambda}$ is a `normalized' coordinate of the extradimension. Taking the zero-temperature limit of $\beta \rightarrow \infty$, we have the RG flow equations for the coupling functions as
	\bqa \frac{1}{\Lambda}\partial_{x} w_{\psi}(x) &=& - \frac{1}{w_{\psi}^{2}(x)} \sqrt{\Lambda^2 (1 - x)^{2}v_{\psi}^{2}(x)+ \lambda_{\chi}^{2}(x) \varphi^{2}(x)} , 	\label{eq:1}\\
	\frac{1}{\Lambda} \partial_{x} v_{\psi}(x) &=& - \frac{1}{w_{\psi}(x)} \frac{\Lambda^2 (1 - x)^{2} v_{\psi}(x)  }{\sqrt{\Lambda^2 (1 - x)^{2} v_{\psi}^{2}(x)  + \lambda_{\chi}^{2}(x) \varphi^{2}(x)}} , \label{eq:2}\\
	\frac{1}{\Lambda}  \partial_{x} \lambda_{\chi}(x) &=& -\frac{1}{w_{\psi}(x)} \frac{ \lambda_{\chi}(x) \varphi^{2}(x)}{\sqrt{\Lambda^2 (1 - x)^{2} v_{\psi}^{2}(x)  + \lambda_{\chi}^{2}(x) \varphi^{2}(x)}} , \label{eq:3} \eqa
	and the Hamiltonian equation of motion for the chiral condensate as
	\bqa && \pi_{\varphi}(x) = \lambda_{\chi}(x) \Big( \frac{1}{\Lambda}\partial_{x} \varphi(x) + \frac{1}{w_{\psi}(x)} \frac{ \lambda_{\chi}^{2}(x) \varphi(x)}{\sqrt{\Lambda^2 (1 - x)^{2} v_{\psi}^{2}(x)+ \lambda_{\chi}^{2}(x) \varphi^{2}(x)}} \Big) ,\label{eq:4} \\
	&& \frac{1}{\Lambda} \partial_{x} \pi_{\varphi}(x) = \frac{1}{w_{\psi}(x)} \frac{ \lambda_{\chi}^{2}(x) \varphi(x)}{\sqrt{\Lambda^2 (1 - x)^{2}v_{\psi}^{2}(x) + \lambda_{\chi}^{2}(x) \varphi^{2}(x)}} + \pi_{\varphi}(x) \frac{1}{w_{\psi}(x)} \frac{ \lambda_{\chi}^{2}(x) }{\sqrt{\Lambda^2 (1 - x)^{2} v_{\psi}^{2}(x)+ \lambda_{\chi}^{2}(x) \varphi^{2}(x)}} \nn && - \pi_{\varphi}(x) \frac{1}{w_{\psi}(x)} \frac{ \lambda_{\chi}^{4}(x) \varphi^{2}(x)}{ \Big(\Lambda^2 (1 - x)^{2}v_{\psi}^{2}(x) + \lambda_{\chi}^{2}(x) \varphi^{2}(x)\Big)^{\frac{3}{2}}} .\label{eq:5} \eqa

	Both IR and UV boundary conditions in the zero-temperature limit are given by
	\bqa &&\frac{1}{\Lambda} \partial_{x} \varphi(x) \Big|_{x = 1} = - \Big(1 + \frac{1}{\lambda_{\chi}(1)}\Big) \frac{\lambda_{\chi} (1)}{w_{\psi}(1)} , \label{eq:6} \\
	&& \frac{1}{\Lambda}\partial_{x} \varphi(x) \Big|_{x = 0} = \varphi(0) - \frac{1}{w_{\psi}(0)} \frac{ \lambda_{\chi}^{2}(0) \varphi(0)}{\sqrt{v_{\psi}^{2}(0) \Lambda^{2} + \lambda_{\chi}^{2}(0) \varphi^{2}(0)}} , \label{eq:7} \eqa
supporting the Hamiltonian equation of motion for the chiral condensate. Here, we introduce the following notations
	\bqa  w_{\psi}(0)=w_0,\; v_{\psi}(0)=v_0,\;  \lambda_{\chi}(0)=\lambda_0,\; \varphi(0)= \varphi_0 \label{eq:8} \eqa
	for the UV boundary values and
	\bqa  w_{\psi}(1)=w_f,\; v_{\psi}(1)=v_f,\;  \lambda_{\chi}(1)=\lambda_f,\; \varphi(1)= \varphi_f \label{eq:9} \eqa
    for the IR boundary values. $w_0$, $v_0$, and $\lambda_0$ are given at UV while $w_f$, $v_f$, and $\lambda_f$ are unknown to be determined by the matching method. In addition, $\varphi_0$ and $\varphi_f$ are also unknown to be determined by the matching method.

%
%
	
%
%
	
	To solve these coupled nonlinear differential equations, we consider the matching method as follows:
	\begin{enumerate}
		\item
		$\lambda_{\chi}(x)$, $w_{\psi}(x)$, and $v_{\psi}(x)$: We say $\lambda_{\chi}(x)$ ($w_{\psi}(x)$ and $v_{\psi}(x)$) near UV boundary as $\lambda_{uv}(x)$ ($w_{uv}(x)$ and $v_{uv}(x)$) and $\lambda_{\chi}(x)$ ($w_{\psi}(x)$ and $v_{\psi}(x)$) near IR boundary as $\lambda_{ir}(x)$ ($w_{ir}(x)$ and $v_{ir}(x)$). Then, we consider the matching condition at $x_{1}$ ($x_{2}$ and $x_{3}$), given by $\lambda_{uv}(x_1)=\lambda_{ir}(x_1)$ ($w_{uv}(x_2) = w_{ir}(x_2)$ and $v_{uv}(x_3) = v_{ir}(x_3)$) and $\partial_{x}\lambda_{uv}(x) \Big|_{x = x_1} = \partial_{x}\lambda_{ir}(x) \Big|_{x = x_1}$ ($\partial_{x}w_{uv}(x) \Big|_{x = x_2} = \partial_{x}w_{ir}(x) \Big|_{x = x_2}$ and $\partial_{x}v_{uv}(x) \Big|_{x = x_3} = \partial_{x}v_{ir}(x) \Big|_{x = x_3}$). Solving these two equations, we obtain $x_1$ ($x_2$ and $x_3$) and determine $\lambda_f$ ($w_f$ and $v_f$) in a self-consistent way.
%
%
		\item
		$\varphi(x)$: Unlike the above three cases, $\varphi(x)$ satisfies the second order nonlinear differential equation. Then, we consider the following matching equation of (1) $\varphi_{uv}(x_4)=\varphi_{ir}(x_4)$, (2) $\partial_{x}\varphi_{uv}(x) \Big|_{x = x_4} =\partial_{x}\varphi_{ir}(x) \Big|_{x = x_4}$, and (3) $ \partial_{x}^2\varphi_{uv}(x) \Big|_{x = x_4} =\partial_{x}^2\varphi_{ir}(x) \Big|_{x = x_4}$. Solving these three equations, we determine $x_4$, $\varphi_0$, and $\varphi_f$.
	\end{enumerate}

	\subsubsection{$w_{\psi}(x)$, $v_{\psi}(x)$, $\lambda_{\chi}(x)$, and $\varphi(x) $ near IR boundary}

	Based on three RG flow equations of (\ref{eq:1})--(\ref{eq:3}), we obtain
	\bqa \varphi(x)= \Lambda  (1-x) \sqrt{\frac{  v_{\psi}(x)\; \partial_{x} \lambda_{\chi}(x) }{ \lambda_{\chi}(x)  \; \partial_{x} v_{\psi}(x) }}. \label{eq:10} \eqa
	Substituting Eq. (\ref{eq:1})--Eq. (\ref{eq:4}) and Eq. (\ref{eq:10}) into Eq. (\ref{eq:5}), we find the following second order nonlinear bulk equation for $\varphi(x)$
	\bqa  A(x)\;  \varphi''(x)+B(x)\;   \varphi'(x)+\mathcal{C}(x)\; \varphi (x)=0 , \label{eq:11} \eqa
	where
	\bqa A(x)= \frac{\lambda_{\chi} (x)\; w_{\psi}(x)^{12}\;  w_{\psi}'(x)  ^5}{\Lambda^5} , \label{eq:12} \eqa
	\bqa B(x)= \frac{ \lambda_{\chi} '(x)\; w_{\psi}(x)^{12}\;  w_{\psi}'(x) ^5}{\Lambda^5} , \label{eq:13} \eqa
	and
	\begin{small}
		\begin{eqnarray} C(x)&=&(x-1)^2 \Lambda ^3 \lambda_{\chi} (x)^2 v_{\psi} (x)^2 \left( -\frac{\Lambda ^2 (x-1)^2 \lambda_{\chi} (x)^2 \lambda_{\chi} '(x)^2 \left(w_{\psi} (x) \left(\Lambda -2 \lambda '(x)\right)+\lambda_{\chi} (x) w_{\psi} '(x)\right)}{v_{\psi} '(x)^2} \right. \nonumber\\
			&&-\Lambda ^2 (x-1)^2 v_{\psi} (x)^2 w_{\psi} (x) \left(\Lambda -3 \lambda_{\chi} '(x)\right)+\lambda_{\chi} (x)^3 w_{\psi} (x)^2 w_{\psi} '(x)  \label{eq:14}  \\
			&&+\Lambda ^2 (1-x) \lambda_{\chi} (x) v_{\psi}(x) \left((x-1) w_{\psi} (x) v_{\psi}'(x)+v_{\psi}(x) \left((x-1) w_{\psi} '(x)+w_{\psi} (x)\right)\right)\nonumber\\
			&&-\left.\frac{\Lambda ^2 (x-1) \lambda_{\chi} (x) \lambda_{\chi} '(x) \left(v_{\psi} (x) \left(w_{\psi} (x) \left((x-1) \left(2 \Lambda -5 \lambda_{\chi} '(x)\right)+\lambda_{\chi} (x)\right)+2 (x-1) \lambda_{\chi} (x) w_{\psi} '(x)\right)+(x-1) \lambda_{\chi} (x) w_{\psi} (x) v_{\psi} '(x)\right)}{v_{\psi} '(x)}
			\right) . \nonumber
		\end{eqnarray}
	\end{small}
Here, the $'$ symbol denotes a derivative with resect to the argument of the corresponding function.

	Taking the $x\rightarrow 1$ limit (IR boundary), we observe that Eq. (\ref{eq:11}) becomes simplified as
	\bqa
	\lambda_{\chi} '(x) \varphi '(x)+\lambda_{\chi} (x) \varphi ''(x)=0. \label{eq:15}
	\eqa
	Its solution is given by
	\bqa
	\varphi '(x)=-\frac{k}{\lambda_{\chi} (x)}, \label{eq:16}
	\eqa
	where $k$ is an integration constant.
	
	Again, as $x\rightarrow 1$, Eq. (\ref{eq:1})--Eq. (\ref{eq:3}) become
	\bqa
	\frac{1}{\Lambda}\partial_{x} w_{\psi}(x) &=& - \frac{\lambda_{\chi}(x) \varphi(x)}{w_{\psi}^{2}(x)}, 	\label{eq:17}\\
	\frac{1}{\Lambda} \partial_{x} v_{\psi}(x) &=& - \frac{\Lambda^2 (1-x)^{2} v_{\psi}(x)}{w_{\psi}(x)\lambda_{\chi}(x) \varphi(x)}, \label{eq:18}\\
	\frac{1}{\Lambda}  \partial_{x} \lambda_{\chi}(x) &=& -\frac{\varphi(x)}{w_{\psi}(x)}.   \label{eq:19}
	\eqa
	These simplified RG flow equations near the IR boundary result in
	\bqa
	w_{\psi}(x) w_{\psi}'(x)-\lambda_{\chi} (x) \lambda_{\chi}'(x)=0. \label{eq:20}
	\eqa
	Then, we obtain
	\bqa
	w_{\psi}(x)= \sqrt{\lambda_{\chi} (x)^2+w_f^2-\lambda _f^2}. \label{eq:21}
	\eqa

	Substituting Eq. (\ref{eq:16}) and Eq. (\ref{eq:21}) into Eq. (\ref{eq:19}), we find the second order nonlinear differential equation for $\lambda_{\chi} (x)$ as
	\bqa
	\lambda_{\chi} ''(x)+\frac{\lambda_{\chi} (x) \lambda_{\chi} '(x)^2}{\lambda_{\chi} (x)^2+w_f^2-\lambda _f^2}-\frac{k \Lambda }{\lambda_{\chi} (x) \sqrt{\lambda_{\chi} (x)^2+w_f^2-\lambda _f^2}}=0. \label{eq:22}
	\eqa
	To this end, we use the following recursion equation
	\bqa
	\lambda_{\chi}^{(n+1)^{''}} (x)+\frac{\lambda_{\chi}^{(n)} (x) \lambda_{\chi}^{(n+1)^{'}}(x)^2}{\lambda_{\chi}^{(n)} (x)^2+w_f^2-\lambda _f^2}-\frac{k \Lambda }{\lambda_{\chi}^{(n)} (x) \sqrt{\lambda_{\chi}^{(n)} (x)^2+w_f^2-\lambda _f^2}}=0, \label{eq:23}
	\eqa
	starting with
	\bqa
	\lambda_{\chi}^{(0)} (x)=\lambda _f. \label{eq:24}
	\eqa

	For $n=0$, the solution of Eq. (\ref{eq:23}) is given by
	\bqa
	\lambda_{\chi}^{(1)} (x)=\lambda_{ir} (x)=\lambda _f+\frac{w_f^2 \left(\log \left(\cosh \left(\frac{\sqrt{k} \sqrt{\Lambda } \left(x-c_1 \lambda _f w_f^3\right)}{w_f^{3/2}}\right)\right)-\log \left(\cosh \left(\frac{\sqrt{k} \sqrt{\Lambda } \left(1-c_1 \lambda _f w_f^3\right)}{w_f^{3/2}}\right)\right)\right)}{\lambda _f}, \label{eq:25}
	\eqa
	where $c_1$ is an undetermined constant. We point out that Eq. (\ref{eq:19}) gives
	\bqa
	\lambda _f^{'}=-\frac{\Lambda  \varphi _f}{w_f}. \label{eq:26}
	\eqa
	Taking the first derivative in Eq. (\ref{eq:25}), we see that this result should be equivalent to Eq. (\ref{eq:26}). Then, $c_1$ is determined as
	\bqa
	c_1=\frac{\frac{\tanh ^{-1}\left(\frac{\sqrt{\Lambda } \lambda _f \varphi _f}{\sqrt{k} w_f^{3/2}}\right)}{\sqrt{k} \sqrt{\Lambda } w_f^{3/2}}+\frac{1}{w_f^3}}{\lambda _f}. \label{eq:27}
	\eqa

	The IR boundary condition Eq. (\ref{eq:6}) with Eq. (\ref{eq:16}) gives
	\bqa
	k= \frac{\Lambda  \lambda _f \left(\lambda _f+1\right)}{w_f}. \label{eq:28}
	\eqa
	Substituting Eq. (\ref{eq:27}) and Eq. (\ref{eq:28}) into Eq. (\ref{eq:25}), we obtain $\lambda_{\chi}(x)$ for $n=0$ near the IR boundary as
	\bqa
	\lambda_{ir} (x)= \lambda _f+\frac{w_f^2 \left(\frac{1}{2} \log \left(1-\frac{\lambda _f \varphi _f^2}{\left(\lambda _f+1\right) w_f^2}\right)+\log \left(\cosh \left(\tanh ^{-1}\left(\frac{\lambda _f \varphi _f}{\sqrt{\lambda _f \left(\lambda _f+1\right)} w_f}\right)+\frac{\Lambda  (1-x) \sqrt{\lambda _f \left(\lambda _f+1\right)}}{w_f^2}\right)\right)\right)}{\lambda _f}. \label{eq:29}
	\eqa
	From now on, we change the notation as $\lambda_{\chi}^{(1)} (x) \to \lambda_{\chi} (x)$.

\begin{figure}[h]
	\centering
	\includegraphics[width=0.6\linewidth]{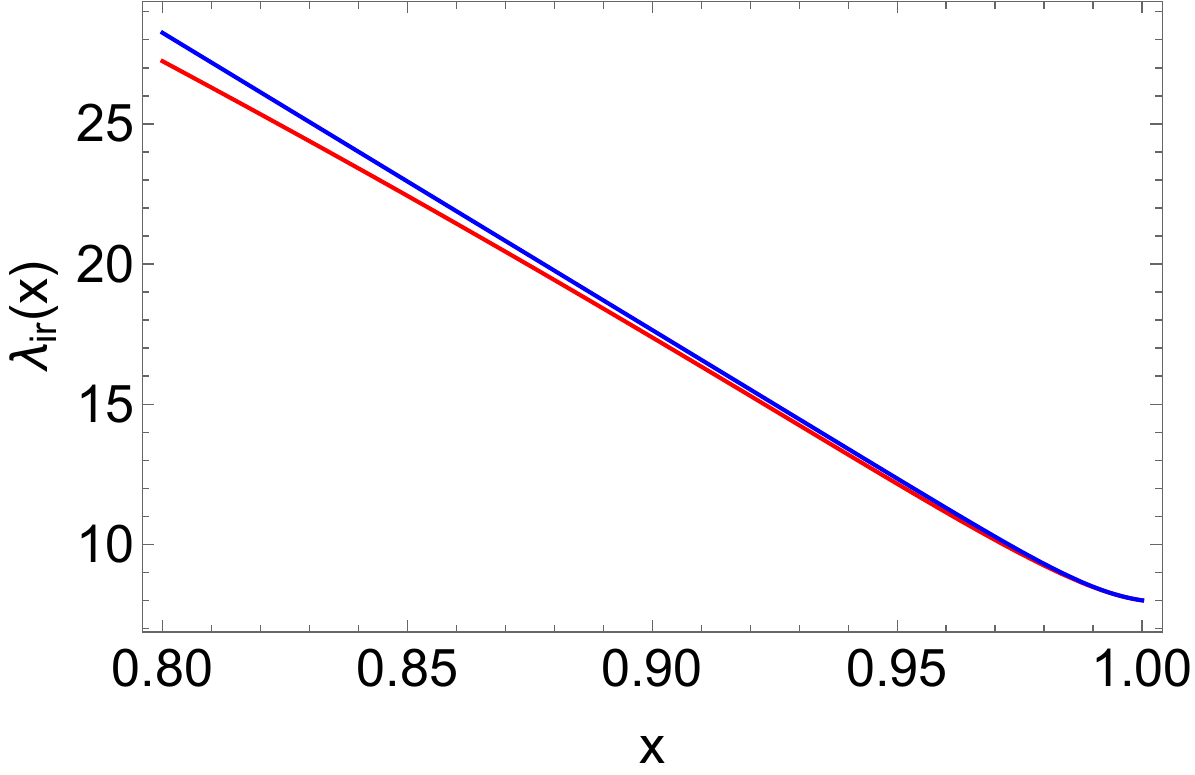}
	\caption{ Comparison between the iteration solution of Eq. (\ref{eq:29}) and the solution of the full differential equation (\ref{eq:22}) near the IR boundary $x \sim 1$ }
	\label{numanaly}
\end{figure}

To justify this solution, we compare it with the solution of the full differential equation (\ref{eq:22}) near the IR boundary $x \sim 1$.
%
%
The red curve in Fig.~\ref{numanaly} is the numerical result of Eq. (\ref{eq:22}). Here, we set $\Lambda =100, ~ \lambda _f=5, ~ \varphi _f=1, ~ w_f=2$. The blue curve is the plot of Eq. (\ref{eq:29}), obtained by the iteration method. As $x \rightarrow 1$, two curves become identical.

	Substituting Eq. (\ref{eq:29}) into Eq. (\ref{eq:21}), we obtain
	\begin{footnotesize}
		\bqa
		w_{ir}(x)= \sqrt{w_f^2-\lambda _f^2+\left(\lambda _f+\frac{w_f^2 \left(\frac{1}{2} \log \left(1-\frac{\lambda _f \varphi _f^2}{\left(\lambda _f+1\right) w_f^2}\right)+\log \left(\cosh \left(\tanh ^{-1}\left(\frac{\lambda _f \varphi _f}{\sqrt{\lambda _f \left(\lambda _f+1\right)} w_f}\right)+\frac{\Lambda  (1-x) \sqrt{\lambda _f \left(\lambda _f+1\right)}}{w_f^2}\right)\right)\right)}{\lambda _f}\right)^2}. \label{eq:30}
		\eqa
	\end{footnotesize}

	Substituting Eq. (\ref{eq:29}) and Eq. (\ref{eq:30}) into Eq. (\ref{eq:19}), we find
	\begin{footnotesize}
		\begin{eqnarray}
			\varphi_{ir}(x)&=&\sqrt{\frac{\lambda _f+1}{\lambda _f}} \sqrt{w_f^2-\lambda _f^2+\left(\lambda _f+\frac{w_f^2 \left(\frac{1}{2} \log \left(1-\frac{\lambda _f \varphi _f^2}{\left(\lambda _f+1\right) w_f^2}\right)+\log \left(\cosh \left(\tanh ^{-1}\left(\frac{\lambda _f \varphi _f}{\sqrt{\lambda _f \left(\lambda _f+1\right)} w_f}\right)+\frac{\Lambda  (1-x) \sqrt{\lambda _f \left(\lambda _f+1\right)}}{w_f^2}\right)\right)\right)}{\lambda _f}\right)^2}\nonumber\\
			&&\times \tanh \left(\tanh ^{-1}\left(\frac{\lambda _f \varphi _f}{\sqrt{\lambda _f \left(\lambda _f+1\right)} w_f}\right)+\frac{\Lambda  (1-x) \sqrt{\lambda _f \left(\lambda _f+1\right)}}{w_f^2}\right)  \label{eq:31}\\
			&=& \sqrt{\frac{1}{\lambda _f}+1} \coth \left(\frac{1}{2} \log \left(\frac{\mathcal{T}+1}{\mathcal{T}-1}\right)+\mathcal{Z}_x\right) \sqrt{-\lambda _f^2+\frac{\left(2 \lambda _f^2+w_f^2 \left(\log (\mathcal{U}-1)+2 \log \left(\sinh \left(\frac{1}{2} \log \left(\frac{\mathcal{T}+1}{\mathcal{T}-1}\right)+\mathcal{Z}_x\right)\right)\right)\right)^2}{4 \lambda _f^2}+w_f^2}\nonumber
		\end{eqnarray}
	\end{footnotesize}
with
\bqa
\mathcal{T}&=&\frac{\lambda _f \varphi _f}{\sqrt{\lambda _f \left(\lambda _f+1\right)} w_f}, 	\nonumber\\
\mathcal{U}&=&\frac{\lambda _f \varphi _f^2}{\left(\lambda _f+1\right) w_f^2}, \nonumber\\
\mathcal{Z}_x &=&\frac{\Lambda  (1-x) \sqrt{\lambda _f \left(\lambda _f+1\right)}}{w_f^2}.   \nonumber
\eqa

To find $v_{ir}(x)$, we consider Eq. (\ref{eq:29})--Eq. (\ref{eq:31}) to expand $-\frac{\Lambda ^3}{w_{\psi}(x) \lambda_{\chi}(x)\varphi(x)}$ at $x=1$. Then, we obtain
	\bqa
	-\frac{\Lambda ^3}{w_{\psi}(x) \lambda_{\chi}(x)\varphi(x)}=-\frac{\Lambda ^3 \sum _{n=0}^{\infty } (-1)^n \left(\frac{(1-x) \left(\Lambda  \left(\lambda _f+1\right)\right)}{w_f \varphi _f}\right)^n}{\lambda _f w_f \varphi _f}=\frac{\Lambda ^3}{\lambda _f \left(\Lambda  (x-1) \left(\lambda _f+1\right)-w_f \varphi _f\right)}. \label{eq:32}
	\eqa
	Substituting Eq. (\ref{eq:32}) into Eq. (\ref{eq:21}), we obtain
	\bqa
	v_{ir}(x)=v_f \left(\frac{\Lambda  (1-x) \left(\lambda _f+1\right)}{w_f \varphi _f}+1\right)^{\frac{w_f^2 \varphi _f^2}{\lambda _f \left(\lambda _f+1\right)^3}} \exp \left(\frac{2 \Lambda  (x-1) \left(\lambda _f+1\right) w_f \varphi _f+\Lambda ^2 (x-1)^2 \left(\lambda _f+1\right)^2}{2 \lambda _f \left(\lambda _f+1\right)^3}\right). \label{eq:33}
	\eqa

%
%

All these procedures are rather tedious but straightforward. However, we show them explicitly to demonstrate how all these analytic solutions can be found.

	\subsubsection{$ w_{\psi}(x), v_{\psi}(x)$,  $\lambda_{\chi}(x)  $ and $\varphi(x) $ near UV boundary}

	For sufficiently large $\Lambda$, the RG flow equations of Eq. (\ref{eq:1})--Eq. (\ref{eq:3}) and the UV boundary condition Eq. (\ref{eq:7}) become simplified as
	\bqa  w_{\psi}^{2}(x) \partial_{x} w_{\psi}(x) &=& - \Lambda^2 (1 - x)v_{\psi}(x), 	\label{eq:35}\\
	\partial_{x} v_{\psi}(x) &=& - \frac{\Lambda^2  (1 - x)  }{w_{\psi}(x)} , \label{eq:36}\\
	\frac{ \partial_{x} \lambda_{\chi}(x)}{\lambda_{\chi}(x) }   &=& -\frac{   \varphi^{2}(x)}{   (1 - x)  v_{\psi}(x)w_{\psi}(x)}, \label{eq:37}
	\eqa
	and
	\bqa   \partial_{x} \varphi(x) \Big|_{x = 0}=\Lambda \varphi_0.\label{eq:38}
	\eqa

	Substituting Eq. (\ref{eq:1})--Eq. (\ref{eq:4}) into Eq. (\ref{eq:5}), we obtain
	\begin{eqnarray}
		&&\varphi ''(x)+\varphi '(x) \left(\frac{\lambda_{\chi} '(x)}{\lambda_{\chi} (x)}+\frac{\lambda_{\chi}(x)^6 \varphi(x)^4}{\Lambda ^4 (x-1)^5 v_{\psi}(x)^5 w_{\psi} (x)}\right)\nonumber\\
		&&-\frac{\lambda_{\chi}(x)^3 \varphi (x)^3 \lambda_{\chi}'(x)}{\Lambda ^2 (1-x)^3 v_{\psi}(x)^3 w_{\psi}(x)}-\frac{\lambda_{\chi}(x) \varphi (x) \left(\Lambda -3\lambda_{\chi}'(x)\right)}{(1-x)v_{\psi}(x) w_{\psi}(x)}-\frac{\lambda_{\chi}(x)^4 \varphi (x) \left(\Lambda ^2 (1-x) v_{\psi}(x)+\varphi (x)^2 w_{\psi}'(x)\right)}{\Lambda ^2 (1-x)^3 v_{\psi}(x)^3 w_{\psi}(x)^2}\nonumber\\
		&&+\frac{\lambda_{\chi}(x)^2 \varphi (x) \left((x-1) w_{\psi}(x) v_{\psi}'(x)+v_{\psi}(x) \left(w_{\psi}(x)-(1-x) w_{\psi} '(x)\right)\right)}{(1-x)^2 v_{\psi}(x)^2 \omega (x)^2}=0.\label{eq:39}
	\end{eqnarray}
	Here, we point out $\varphi_0 < 0$ although this sign does not change the spectrum. We further simplify this equation, neglecting higher-power terms in $\varphi^n(x)$.
%
%
As a result, Eq. (\ref{eq:39}) is more simplified as
	\begin{eqnarray}
		&&\varphi ''(x) +\left( \frac{\lambda_{\chi} (x)^2}{(1-x)^2 v_{\psi}(x) w_{\psi} (x)}-\frac{\Lambda  \lambda_{\chi}(x)}{(1-x) v_{\psi}(x) w_{\psi}(x)}\right.\nonumber\\
		&&-\left.\frac{\lambda_{\chi}(x)^2 \left(w_{\psi} (x) v_{\psi}'(x)+v_{\psi}(x) w_{\psi}'(x)\right)}{(1-x) v_{\psi}(x)^2w_{\psi}(x)^2}-\frac{\lambda_{\chi} (x)^4}{(1-x)^2 v_{\psi}(x)^2 w_{\psi}(x)^2} \right)\varphi(x)=0.\label{eq:40}
	\end{eqnarray}

	Considering the $x\rightarrow 0$ limit (UV boundary), we find $w_{\psi}'(x), ~ v_{\psi} '(x) \rightarrow 0$. Then, this equation becomes further simplified as
	\begin{equation}
		\varphi ''(x) +\left( \frac{\lambda_{\chi} (x)^2}{(1-x)^2 v_{\psi}(x) w_{\psi} (x)}-\frac{\Lambda \; \lambda_{\chi}(x)}{(1-x) v_{\psi}(x) w_{\psi}(x)}\right. -\left.\frac{\lambda_{\chi}(x)^4}{(1-x)^2 v_{\psi} (x)^2 w_{\psi}(x)^2} \right)\varphi(x)=0 \label{eq:41}
	\end{equation}
near $x=0$. To this end, we consider the following recursion equation
	\begin{equation}
		\varphi^{(n+1)^{''}}(x) +\left( \frac{\lambda_{\chi}^{(n)} (x)^2}{(1-x)^2 v_{\psi}^{(n)}(x) w_{\psi}^{(n)} (x)}-\frac{\Lambda \; \lambda_{\chi}^{(n)}(x)}{(1-x) v_{\psi}^{(n)}(x) w_{\psi}^{(n)}(x)}\right. -\left.\frac{\lambda_{\chi}^{(n)}(x)^4}{(1-x)^2 v_{\psi}^{(n)} (x)^2 w_{\psi}^{(n)}(x)^2} \right)\varphi^{(n+1)}(x)=0,\label{eq:42}
	\end{equation}
	starting with
	\bqa
	\lambda_{\chi}^{(0)} (x)&=&\lambda _0, 	\nonumber\\
	v_{\psi}^{(0)}(x) &=& v_0, \nonumber\\
	w_{\psi}^{(0)}(x) &=& w_0.   \nonumber
	\eqa
	As a result, we find
%
%
	\begin{eqnarray}
		\varphi^{(1)}(x)&=&\varphi_{uv}(x) =\frac{ \varphi_0\sqrt{1-x}}{\left(v_0 w_0\right)^{5/2}}\left( -\frac{\pi  \lambda _0^{3/2} \sqrt{\Lambda } v_0 w_0 p^{\frac{j}{2}-1} \csc (\pi  j) I_{1-j}\left(2 \sqrt{p(1-x)}\right) \left(\lambda _0 (j+2 \Lambda ) \, _0F_1(;j;p)+\Lambda  \, _0F_1(;j+1;p)\right)}{\Gamma (j+1)}\right. \nonumber\\
		&&-\left.\frac{\left(v_0 w_0\right)^{5/2} I_{j-1}\left(2 \sqrt{p(1-x)}\right) \left(\, _0F_1(;2-j;p) \left(\lambda _0 (j+2 \Lambda ) \, _0F_1(;j;p)+\Lambda  \, _0F_1(;j+1;p)\right)-2 j \lambda _0+2 \lambda _0\right)}{2 (j-1) \lambda _0 I_{j-1}\left(2 \sqrt{p}\right)}\right)
		\label{eq:43}
	\end{eqnarray}
	for $n = 0$, where $j=\frac{2 \lambda _0^2}{v_0 w_0}$ and $p=\frac{\lambda _0 \Lambda }{v_0 w_0}$. The $F-$symbol ($I$) represents a hypergeometric (modified Bessel) function, given by the Mathematica.
	
	Similarly, Eq. (\ref{eq:35})--Eq. (\ref{eq:37}) become
	\bqa
	w_{\psi}^{(n+1)^{2}}(x)  w_{\psi}^{(n+1)^{'}}(x) &=& - \Lambda^2 (1-x)v_{\psi}^{(n)}(x), 	\label{eq:44}\\
	v_{\psi}^{(n+1)^{'}}(x) &=& - \frac{\Lambda^2  (1 - x)  }{w_{\psi}^{(n)}(x)} , \label{eq:45}\\
	\frac{   \lambda_{\chi}^{(n+1)^{'}}(x)}{\lambda_{\chi}^{(n+1)}(x) }   &=& -\frac{   \varphi^{(n)^{2}}(x)}{   (1 - x)  v_{\psi}^{(n)}(x)w_{\psi}^{(n)}(x)}, \label{eq:46}
	\eqa
	starting with
	\bqa
	w_{\psi}^{(0)}(x)&=& w_0-\frac{\Lambda ^2 v_0 x}{w_0^2}, 	\label{eq:47}\\
	v_{\psi}^{(0)}(x) &=& v_0-\frac{\Lambda ^2 x}{w_0}, \label{eq:48}\\
	\varphi^{(0)}(x) &=& \phi _0 +  \phi _0 \Lambda  x .   \label{eq:49}
	\eqa
%
%
As a result, we find
%
%
	\bqa
	v_{\psi}^{(1)}(x)=v_{uv}(x)&=& v_0+\frac{w_0^5 \log \left(\frac{w_0^3}{w_0^3-\Lambda ^2 v_0 x}\right)}{\Lambda ^2 v_0^2}-\frac{w_0^2 \left(\log \left(\frac{w_0^3}{w_0^3-\Lambda ^2 v_0 x}\right)+x\right)}{v_0}, 	\label{eq:50}\\
	w_{\psi}^{(1)}(x)=w_{uv}(x) &=& \sqrt[3]{\frac{3 \Lambda ^2 v_0 w_0 (x-2) x+2 w_0^4+\Lambda ^4 x^2 (3-2 x)}{2 w_0}} \approx w_0-\frac{\Lambda ^2 v_0 x}{w_0^2}\;\; \mbox{for} \;\mbox{near}\; x=0, \label{eq:51}\\
	\lambda_{\chi}^{(1)}(x)=\lambda_{uv}(x) &=& \lambda _0 (1-x)^{\frac{\varphi_0^2}{v_0 w_0}}  \label{eq:52}
	\eqa
for $n = 0$.

%
%

\begin{figure}[!htb]
	\centering
	\subfigure[]
	{ \includegraphics[width=0.49\linewidth]{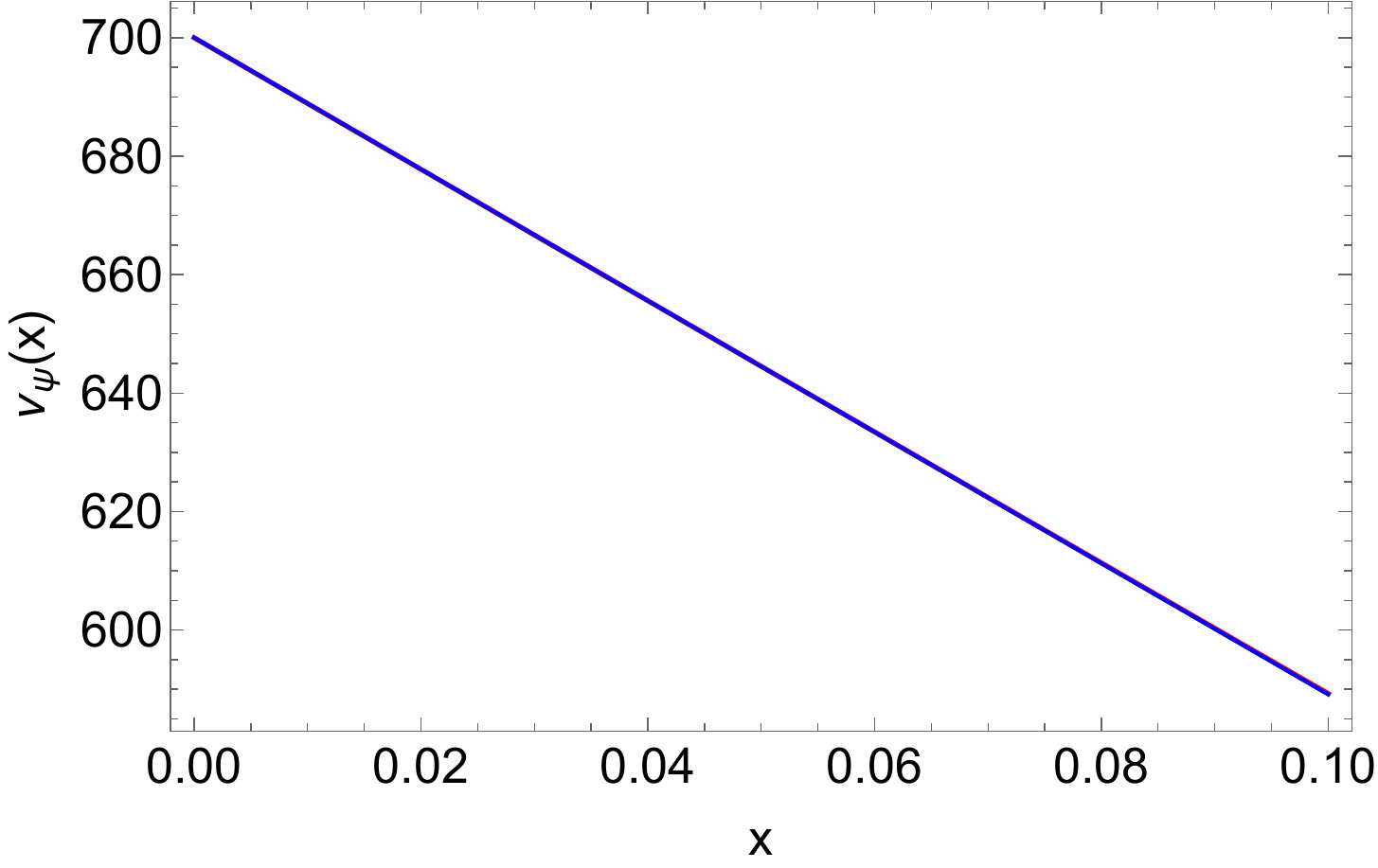}}
	\subfigure[]
	{ \includegraphics[width=0.49\linewidth]{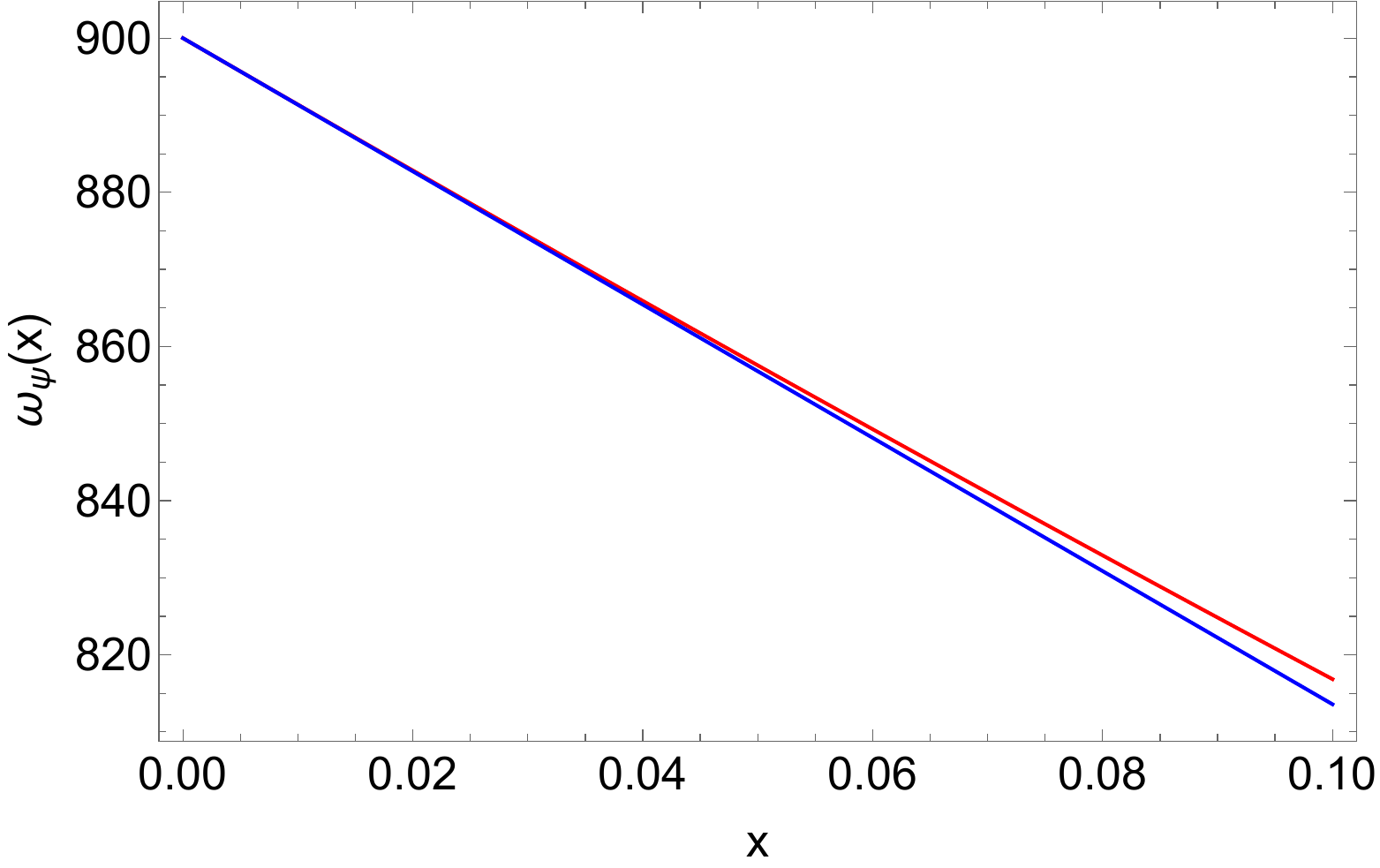}}
	
	\caption{ Verification of the iteration solution near the UV boundary $x \sim 0$. (a) $v_{\psi}(x)$ vs. $x$. (b) $w_{\psi}(x)$ vs. $x$. }
	\label{vwnum}
\end{figure}

In the same way as discussed before, all these $n = 0$ solutions are verified in the UV limit. The red curve in Fig.~\ref{vwnum} (a) is the result of the full differential equation for $v_{\psi}(x)$. Here, we set $\Lambda =1000, ~ w_0=0.9 \Lambda, ~ v_0=0.7 \Lambda$. The blue curve is the plot of Eq. (\ref{eq:50}), obtained by the iteration method. As $x \rightarrow 0$, two curves are identical to each other. Similarly, the red (blue) curve in Fig.~\ref{vwnum} (b) is the plot of the full differential equation (Eq. (\ref{eq:51})) for $w_{\psi}(x)$.

\subsubsection{The Matching solution for   $\lambda_{\chi}(x)  $ } 	

Substituting Eq. (\ref{eq:29}) and Eq. (\ref{eq:52}) into $\lambda_{uv}(x_1)=\lambda_{ir}(x_1)$ and $ \partial_{x}\lambda_{uv}(x) \Big|_{x = x_1} =\partial_{x}\lambda_{ir}(x) \Big|_{x = x_1}$, we obtain two matching equations as follows
\begin{enumerate}
	\item
\begin{small}
\bqa
\lambda _0 \left(1-x_1\right)^{\frac{\varphi _0^2}{v_0 w_0}}=\lambda _f+\frac{w_f^2 \left(\frac{1}{2} \log \left(1-\frac{\lambda _f \varphi _f^2}{\left(\lambda _f+1\right) w_f^2}\right)+\log \left(\cosh \left(\tanh ^{-1}\left(\frac{\lambda _f \varphi _f}{\sqrt{\lambda _f \left(\lambda _f+1\right)} w_f}\right)+\frac{\Lambda  \left(1-x_1\right) \sqrt{\lambda _f \left(\lambda _f+1\right)}}{w_f^2}\right)\right)\right)}{\lambda _f},   \hspace{1cm}\label{eq:53}
\eqa
\end{small}
	\item
\begin{small}
\bqa
\frac{\lambda _0 \varphi _0^2 \left(1-x_1\right)^{\frac{\varphi _0^2}{v_0 w_0}-1}}{v_0 w_0}=\Lambda  \sqrt{\frac{\lambda _f+1}{\lambda _f}} \tanh \left(\tanh ^{-1}\left(\frac{\lambda _f \varphi _f}{\sqrt{\lambda _f \left(\lambda _f+1\right)} w_f}\right)+\frac{\Lambda  \left(1-x_1\right) \sqrt{\lambda _f \left(\lambda _f+1\right)}}{w_f^2}\right)-\lambda _f.   \label{eq:54}
\eqa
\end{small}
\end{enumerate}  	

Our numerical results show that the condition of $\lambda _0 \ll v_0, ~ w_0$ should be satisfied. Otherwise, these two functions $\lambda_{uv}(x)$ and  $\lambda_{ir}(x)$ do not meet at any point in $0 \leq x \leq 1$. In this respect we consider $\lambda _0 \sim 1$ and $v_0, ~ w_0 \sim \Lambda$ as UV boundary values.
%
%
We discuss $v_0, ~ w_0 \sim \Lambda$ below in more details.

	\begin{figure}[!htb]
	\centering
	\subfigure[]
	{ \includegraphics[width=0.49\linewidth]{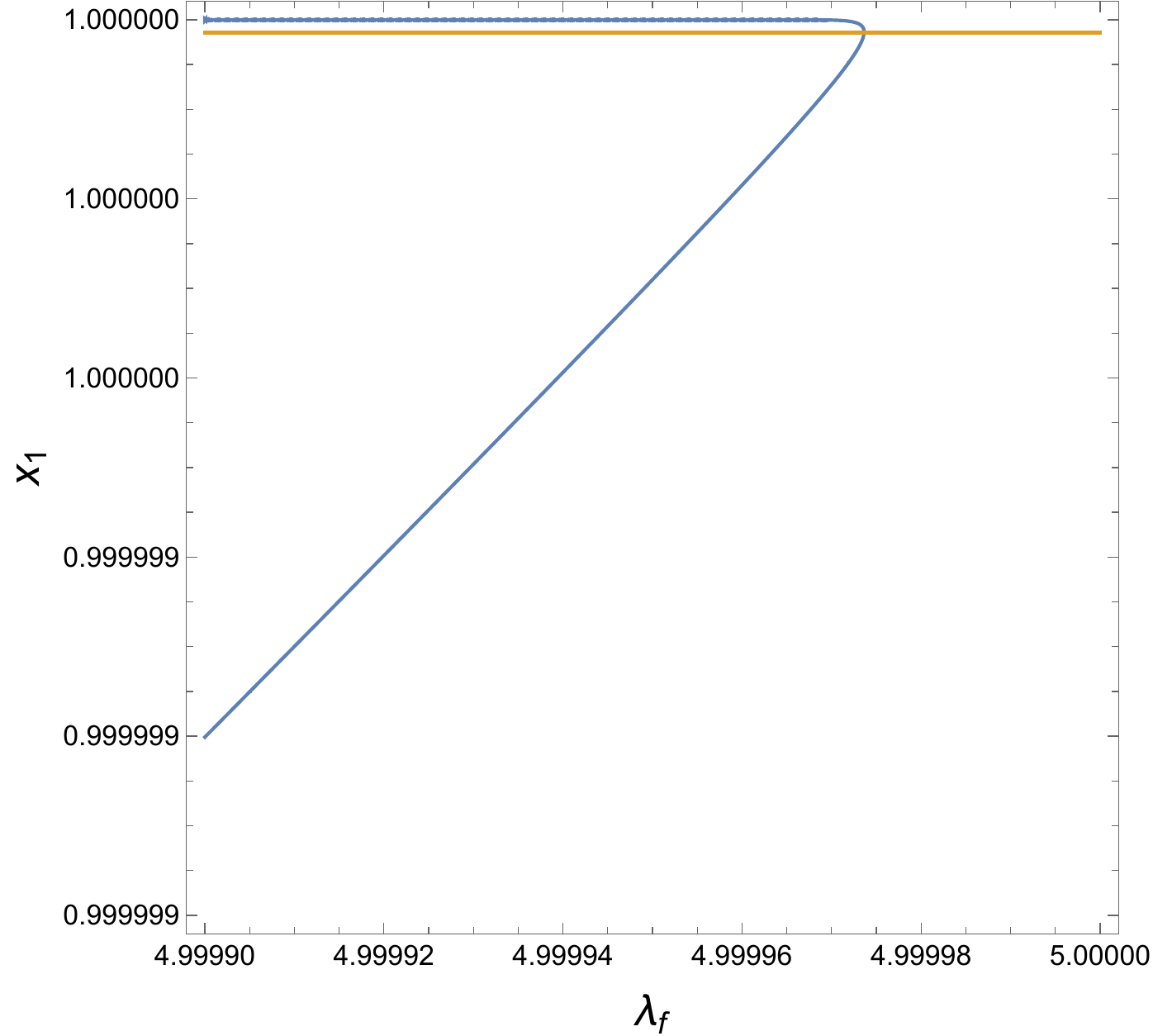}}
	\subfigure[]
	{ \includegraphics[width=0.49\linewidth]{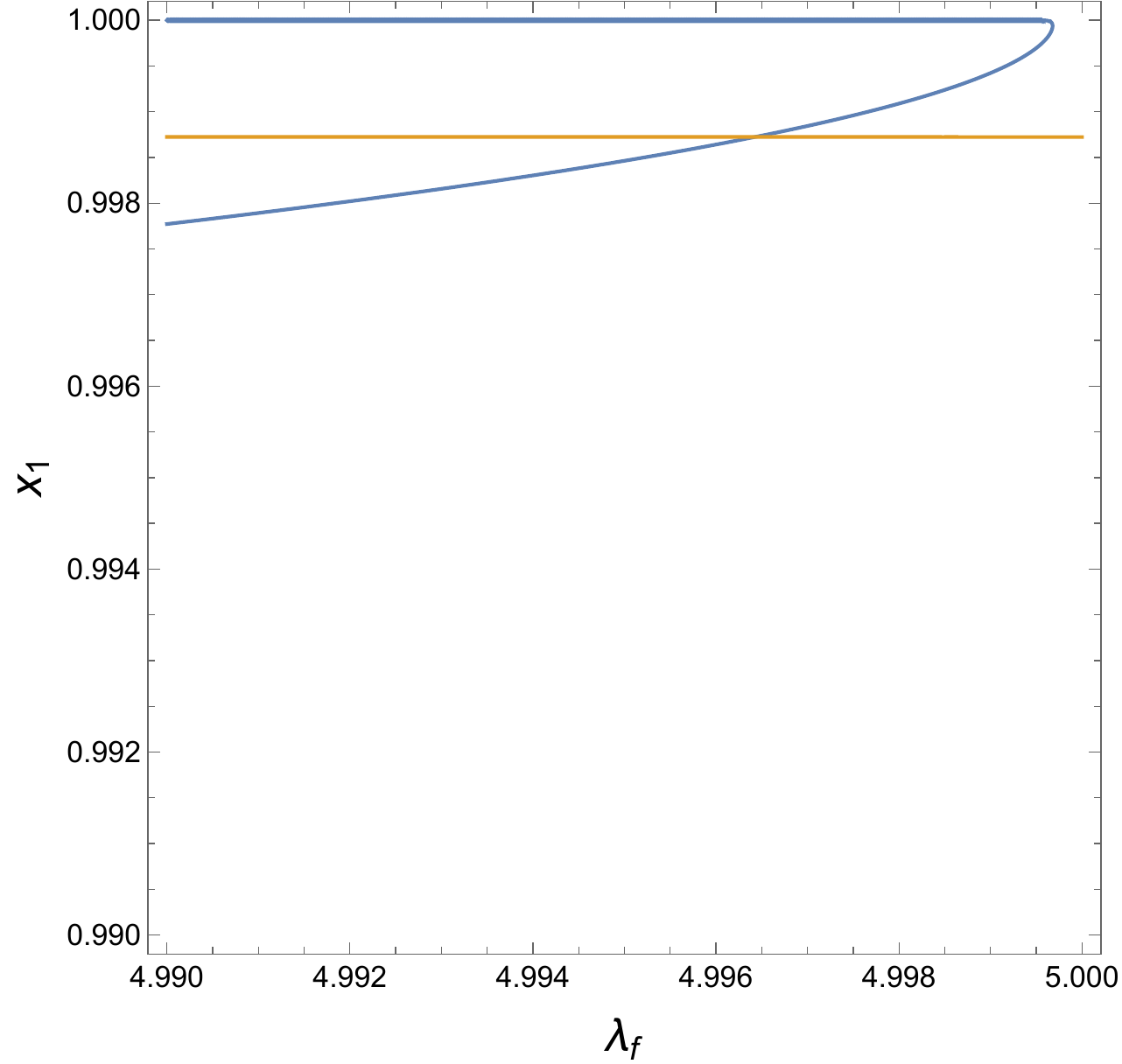}}
	
	\caption{ Self-consistent solution for the matching point $x_{1}$ and the IR boundary value $\lambda_{f}$. (a) We set $\Lambda =50, ~ \lambda _0=5, ~ w_0=0.9\Lambda, ~ v_0=0.7209\Lambda, ~ w_f=0.5, ~ \varphi _f =1.007, ~ \varphi _0=0.0211605$. (b) We take $\Lambda =50, ~ \lambda _0=5, ~ w_0=0.9\Lambda, ~ v_0=0.7209\Lambda, ~ w_f=2, ~ \varphi_f =0.01, ~ \varphi _0 =0.1$. }
	\label{lambda}
\end{figure} 	

Blue (yellow) curves in Fig.~\ref{lambda} (a) and Fig.~\ref{lambda} (b) represent contour plots of Eq. (\ref{eq:53}) (Eq. (\ref{eq:54})). We notice that the intersection of these two curves occurs near $x_1=1$ and $\lambda _0=\lambda _f$. Then, Eq. (\ref{eq:53}) and Eq. (\ref{eq:54}) near $x_1=1$ become
	\begin{enumerate}
		\item
			\bqa
			\lambda _0 \left(1-x_1\right)^{\frac{\varphi _0^2}{v_0 w_0}}=\lambda _f-\frac{\Lambda  \left(x_1-1\right) \varphi _f}{w_f},   \hspace{1cm}\label{eq:55}
			\eqa
		\item
			\bqa
			\frac{\lambda _0 \phi _0^2 \left(1-x_1\right)^{\frac{\varphi _0^2}{v_0 w_0}-1}}{v_0 w_0}=\frac{\Lambda  \varphi _f}{w_f}.   \label{eq:56}
			\eqa
	\end{enumerate}

	Based on these simplified equations, we find an analytic expression for $x_1$ and $\lambda _f$ as follows
	\bqa
	x_1 &=& 1-\frac{\varphi _0^2 \lambda _f w_f}{\Lambda  \varphi _f \left(v_0 w_0-\varphi _0^2\right)}, \label{eq:57}\\
	\lambda _f &=& \frac{\lambda _0 \left(v_0 w_0-\varphi _0^2\right) \left(\frac{\lambda _0 \varphi _0^2 w_f}{\Lambda  \varphi _f \left(v_0 w_0-\varphi _0^2\right)}\right)^{\frac{\varphi _0^2}{v_0 w_0}}}{v_0 w_0}.   \label{eq:58}
	\eqa
	For small $\varphi _0$, Eq. (\ref{eq:58}) becomes
	\bqa
	\lambda _f=\lambda _0 \left(1+\frac{\varphi _0^2 }{v_0 w_0} \log \left(\frac{\varphi _0^2 w_f}{\Lambda  v_0 w_0 \varphi _f}\right)\right)\approx \lambda _0.   \label{eq:59}
	\eqa

	\begin{figure}[!htb]
		\centering
		\subfigure[]
		{ \includegraphics[width=0.49\linewidth]{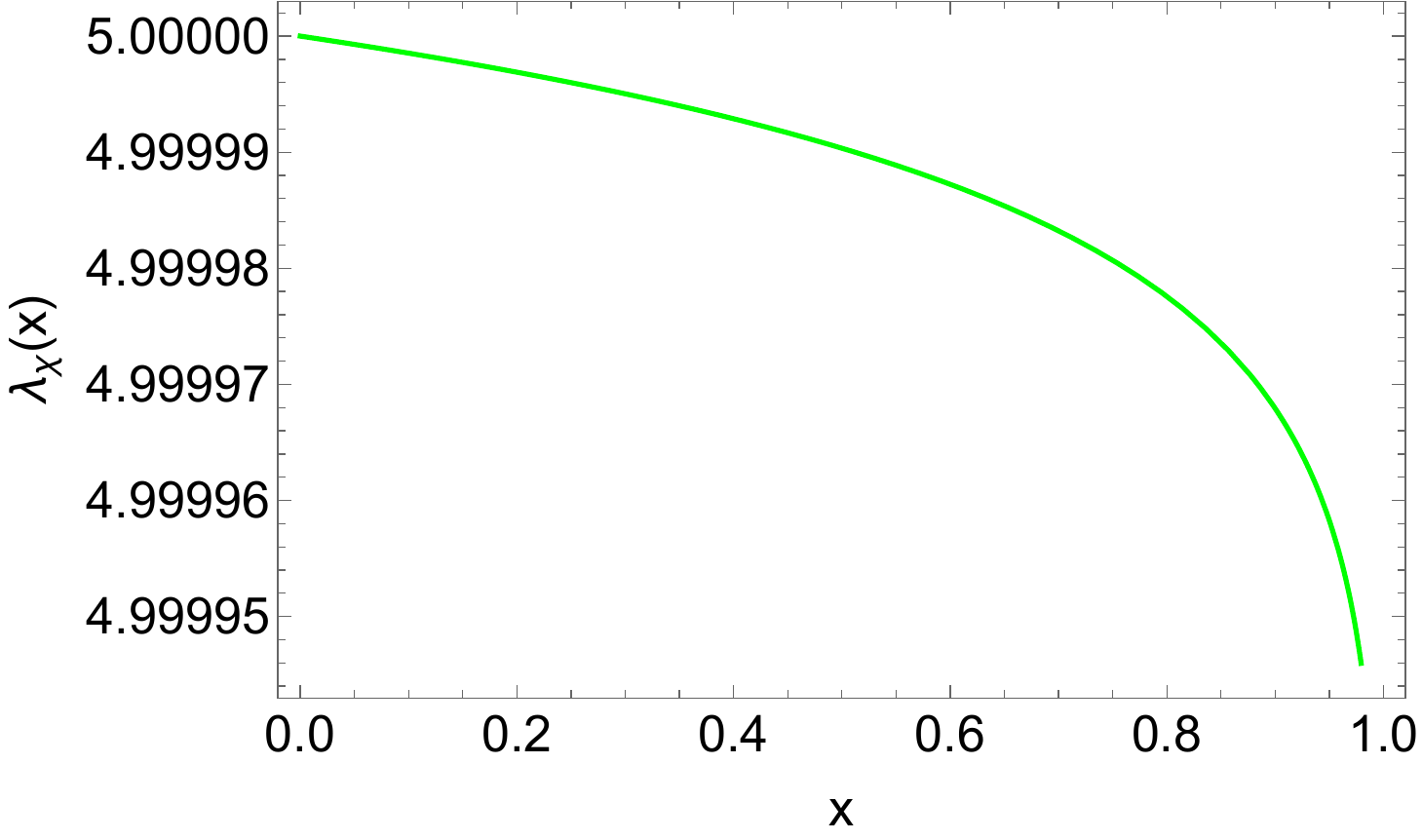}}
		\subfigure[]
		{ \includegraphics[width=0.49\linewidth]{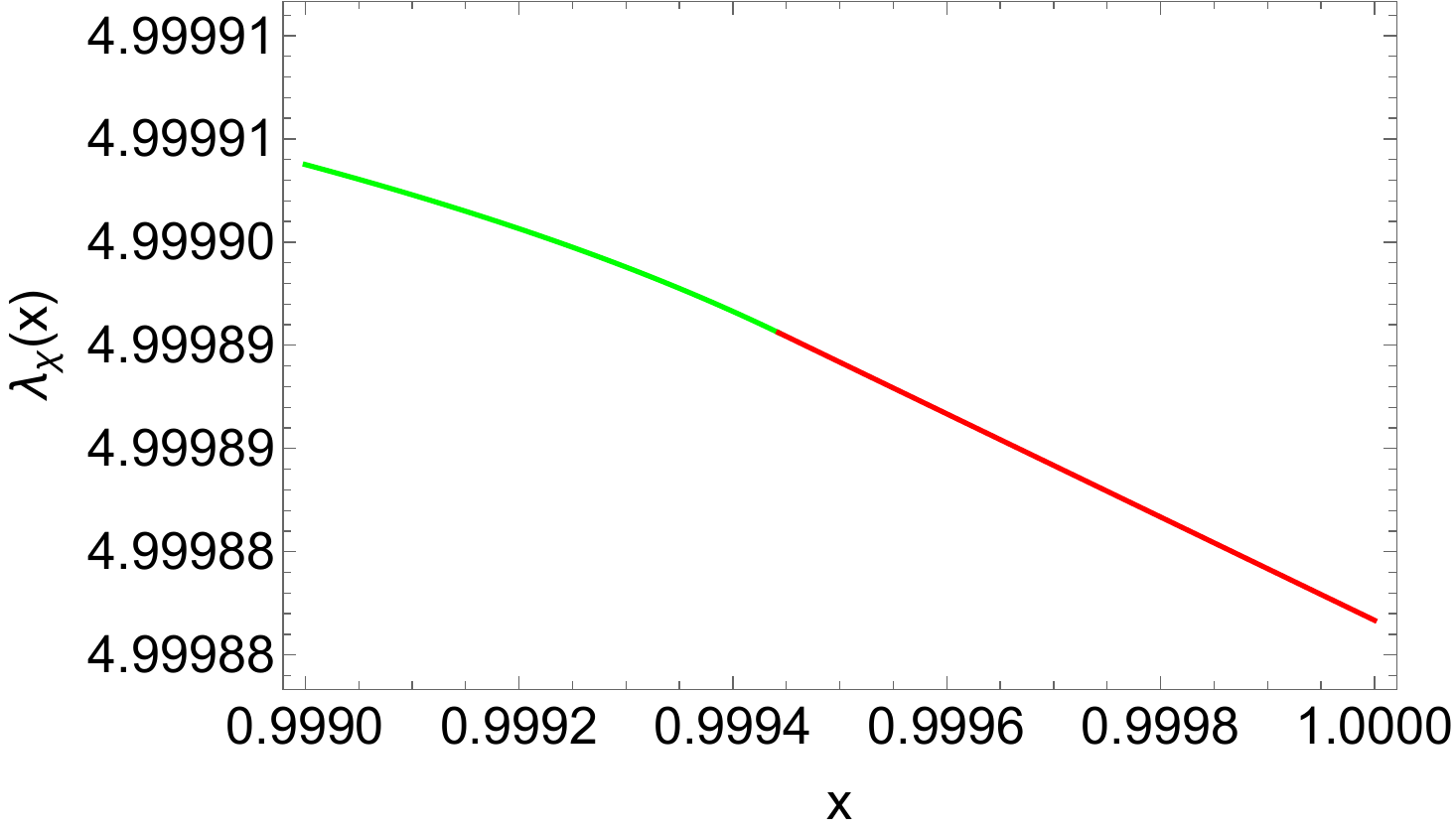}}
		
		\caption{ RG flow of the interaction parameter $\lambda_{\chi}(x)$. (a) $\lambda_{\chi}(x)$ vs. $x$ overall. (b) $\lambda_{\chi}(x)$ vs. $x$ zoomed.	}
		\label{lambda1}
	\end{figure} 	

 Figure \ref{lambda1} shows an RG flow of the interaction parameter $\lambda_{\chi}(x)$ in the parameter range of $\Lambda =50, ~ \lambda _0=5, ~ w_0=1.1 \Lambda , ~ v_0=1.305 \Lambda , ~ w_f=2, ~ \varphi _f=0.001, ~ \varphi _0=0.1$. The green curve in Fig.~\ref{lambda1} (b) is a plot of Eq. (\ref{eq:52}) while the red one is a plot of Eq. (\ref{eq:29}). The matching point is given by Eq. (\ref{eq:57}) and Eq. (\ref{eq:58}).

\subsubsection{The Matching solution for   $	w_{\psi}(x)  $ } 	

Substituting Eq. (\ref{eq:30}) and Eq. (\ref{eq:51}) into $w_{uv}(x_2)=w_{ir}(x_2)$ and $ \partial_{x}w_{uv}(x) \Big|_{x = x_2} =\partial_{x}w_{ir}(x) \Big|_{x = x_2}$, we obtain two matching equations as follows
\begin{enumerate}
	\item
	\begin{scriptsize}
		\bqa
		w_0-\frac{\Lambda ^2 v_0 x_2}{w_0^2}=\sqrt{w_f^2-\lambda _f^2+\left(\lambda _f+\frac{w_f^2 \left(\frac{1}{2} \log \left(1-\frac{\lambda _f \varphi _f^2}{\left(\lambda _f+1\right) w_f^2}\right)+\log \left(\cosh \left(\tanh ^{-1}\left(\frac{\lambda _f \varphi _f}{\sqrt{\lambda _f \left(\lambda _f+1\right)} w_f}\right)+\frac{\Lambda  \left(1-x_2\right) \sqrt{\lambda _f \left(\lambda _f+1\right)}}{w_f^2}\right)\right)\right)}{\lambda _f}\right)^2},   \hspace{1cm}\label{eq:60}
		\eqa
	\end{scriptsize}
	\item
	\begin{small}
		\bqa
		\frac{\Lambda  v_0}{w_0^2}=\frac{\sqrt{\frac{\lambda _f+1}{\lambda _f}} \sqrt{\lambda _f^2-w_f^2+\left(w_0-\frac{\Lambda ^2 v_0 x_2}{w_0^2}\right)^2} \tanh \left(\tanh ^{-1}\left(\frac{\lambda _f \varphi _f}{\sqrt{\lambda _f \left(\lambda _f+1\right)} w_f}\right)+\frac{\Lambda  \left(1-x_2\right) \sqrt{\lambda _f \left(\lambda _f+1\right)}}{w_f^2}\right)}{w_0-\frac{\Lambda ^2 v_0 x_2}{w_0^2}}.   \label{eq:61}
		\eqa
	\end{small}
\end{enumerate}

	\begin{figure}[!htb]
	\centering
	\subfigure[]
	{ \includegraphics[width=0.3\linewidth]{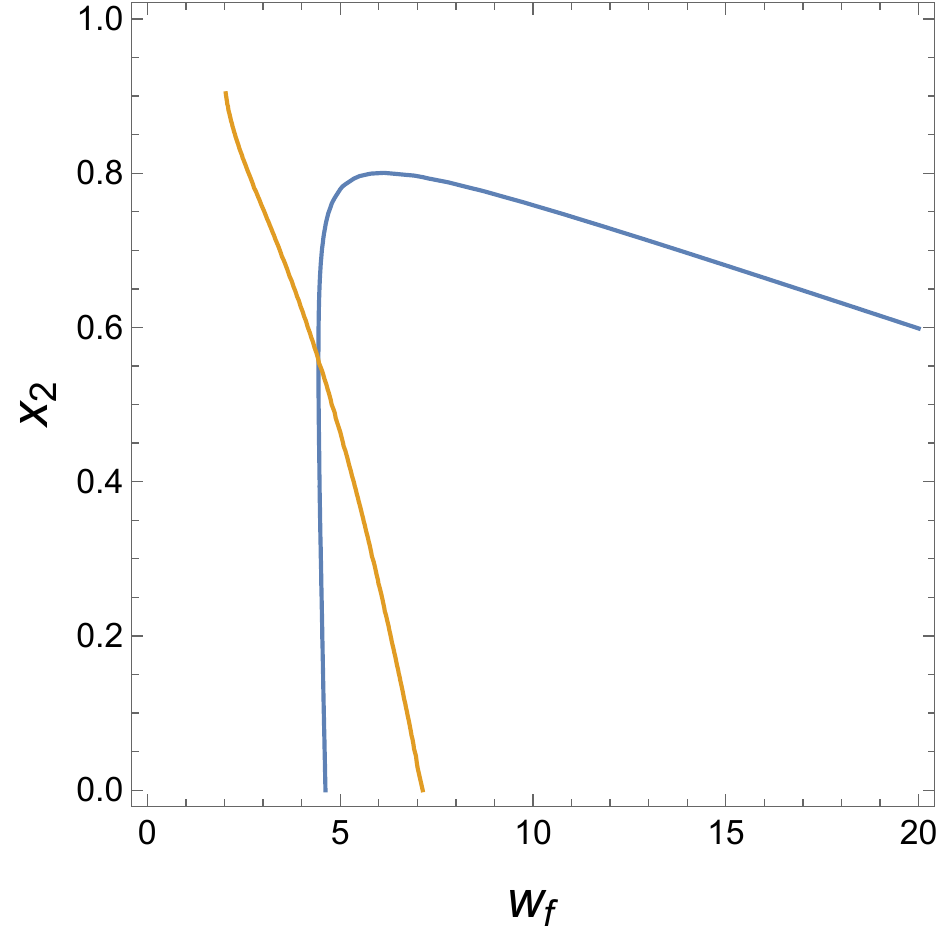}}
	\subfigure[]
	{ \includegraphics[width=0.3\linewidth]{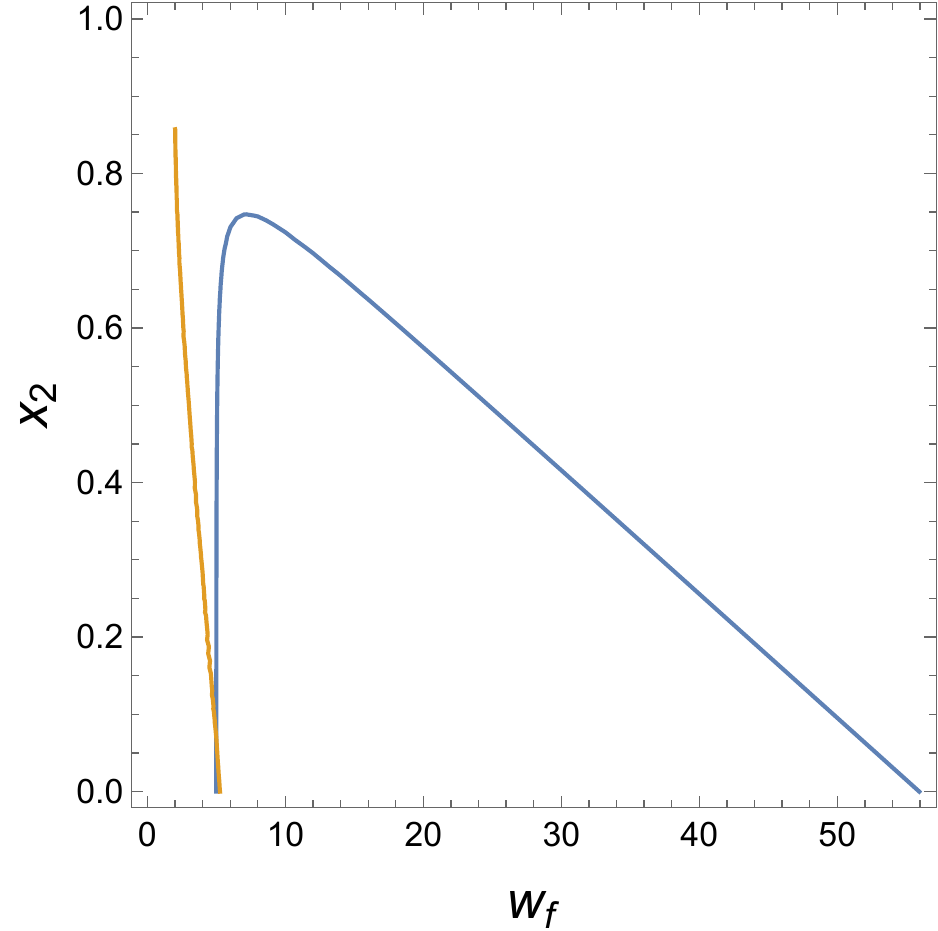}}
	\subfigure[]
	{ \includegraphics[width=0.3\linewidth]{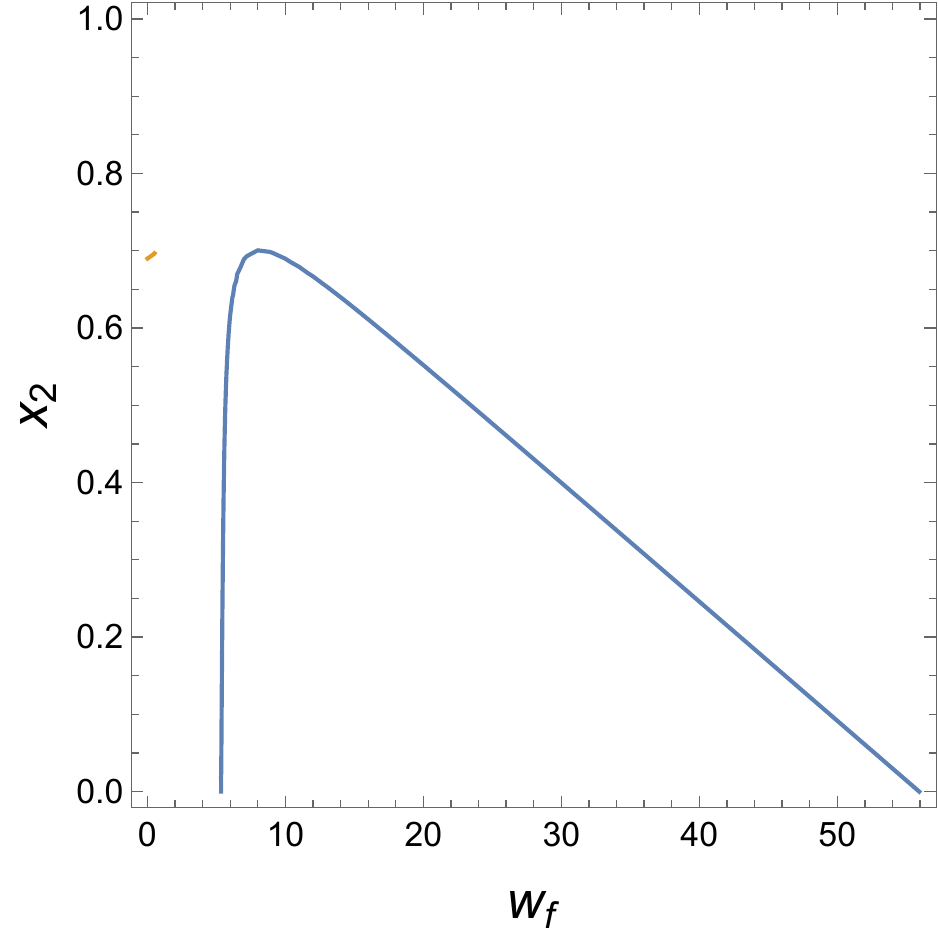}}
	
	\caption{ Self-consistent solution for the matching point $x_{2}$ and the IR boundary value $w_{f}$, given $v_0$ and $w_0$. (a) $x_2$ vs. $w_f$ at $\Lambda =50$, (b) $x_2$ vs. $w_f$ at $\Lambda =51$, and (c) $x_2$ vs. $w_f$ at $\Lambda =52$.	}
	\label{omega1}
\end{figure} 	

Figure \ref{omega1} shows a self-consistent solution for the matching point $x_{2}$ and the IR boundary value $w_{f}$. Blue (yellow) curves in Fig.~\ref{omega1} (a), Fig.~\ref{omega1} (b), and Fig.~\ref{omega1} (c) are contour plots of Eq. (\ref{eq:60}) (Eq. (\ref{eq:61})) with different values of $\Lambda$. Here, we set $ \lambda _0=2, ~ \lambda _f=2, ~ v_0=75, ~ w_0=56, ~ \varphi _f=0.0001$. We notice that the intersection point disappears as $\Lambda$ increases, given $v_0$ and $w_0$.
%
%

	\begin{figure}[!htb]
		\centering
		\subfigure[]
		{ \includegraphics[width=0.3\linewidth]{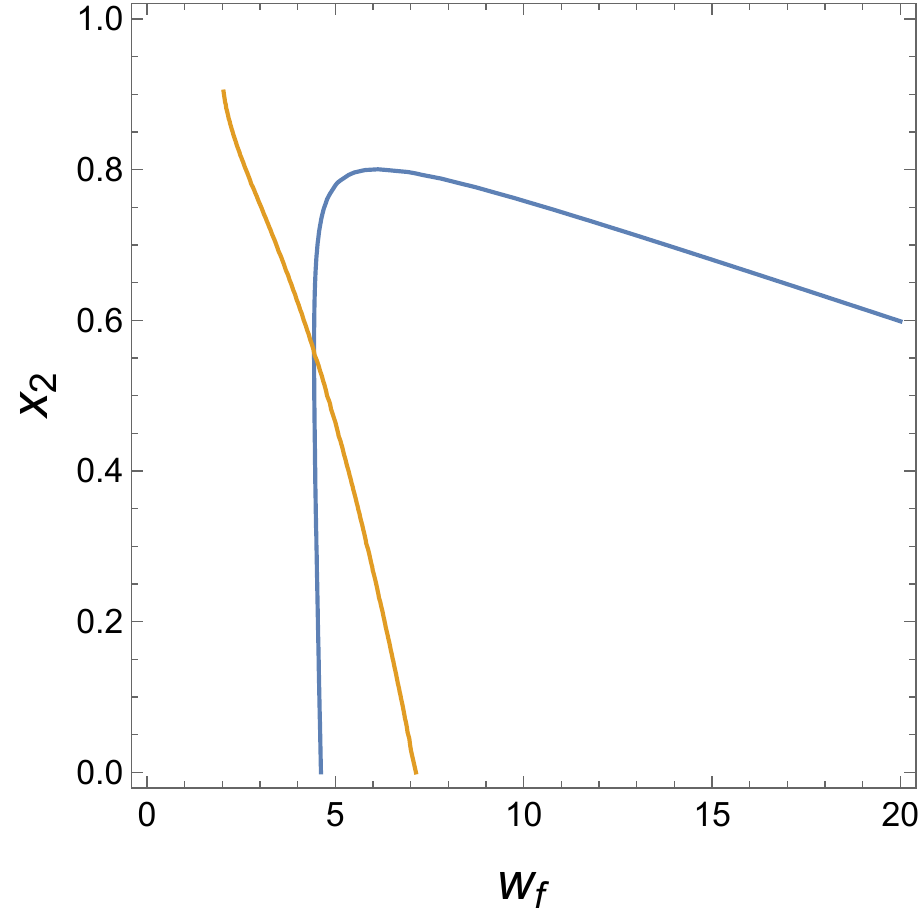}}
		\subfigure[]
		{ \includegraphics[width=0.3\linewidth]{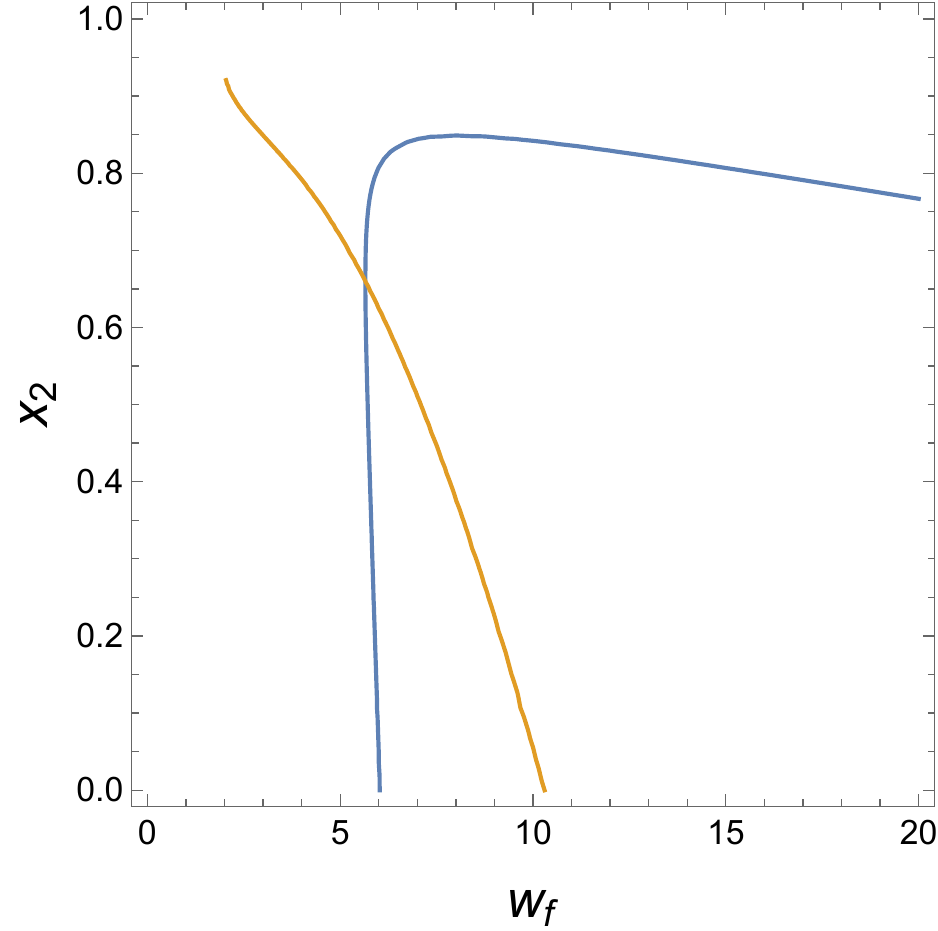}}
		\subfigure[]
		{ \includegraphics[width=0.3\linewidth]{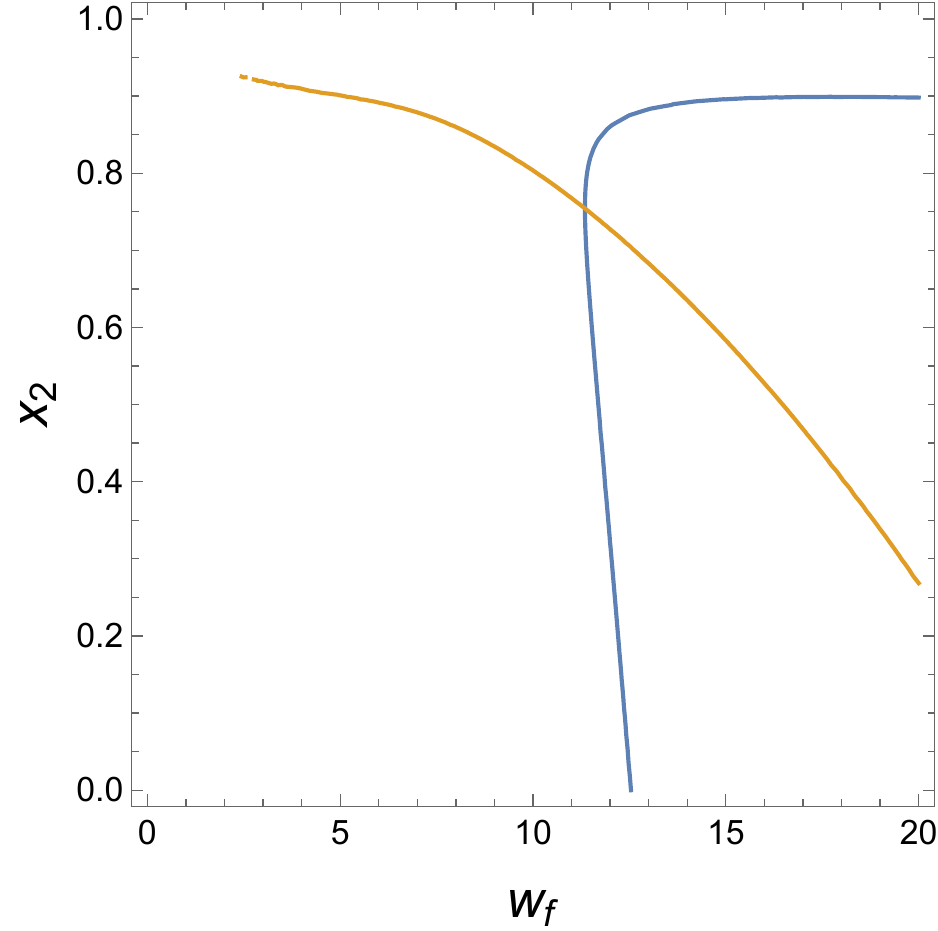}}\subfigure[]
		{ \includegraphics[width=0.3\linewidth]{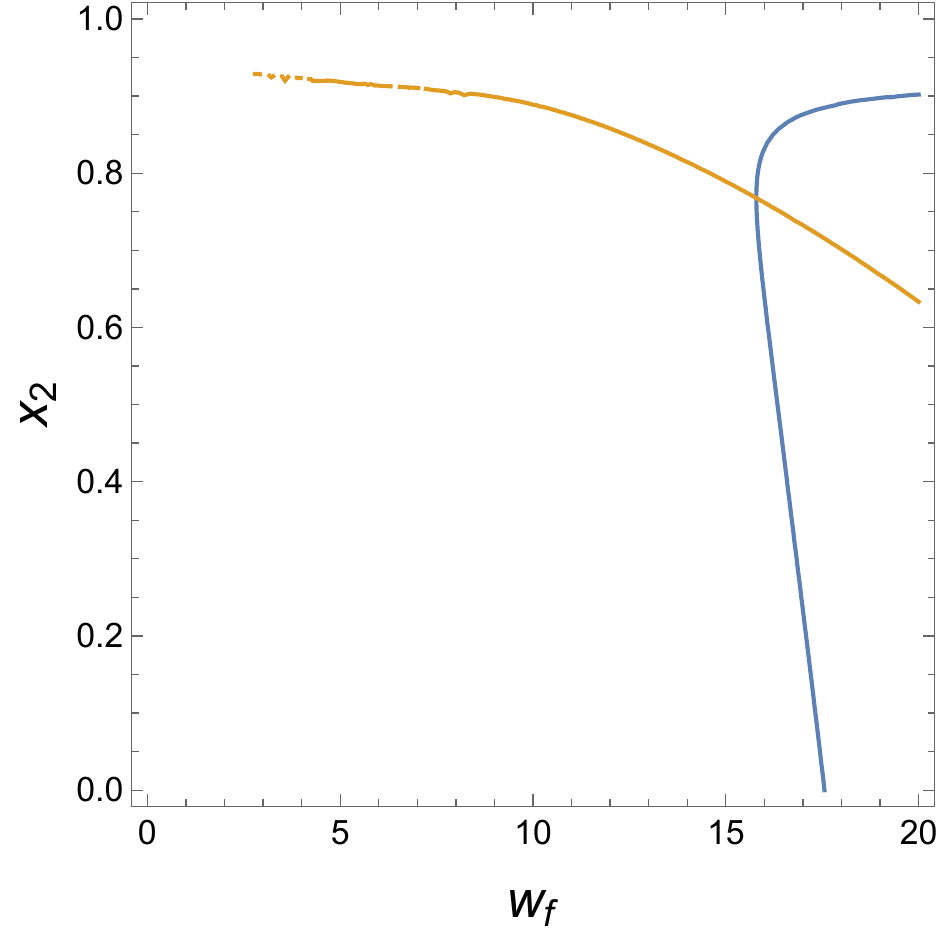}}
		
		\caption{ Self-consistent solution for the matching point $x_{2}$ and the IR boundary value $w_{f}$ as a function of the UV cutoff $\Lambda$. (a) $x_2$ vs. $w_f$ at $\Lambda =50$, (b) $x_2$ vs. $w_f$ at $\Lambda =100$, (c) $x_2$ vs. $w_f$ at $\Lambda =500$, and (d) $x_2$ vs. $w_f$ at $\Lambda =1000$. }
		\label{omega2}
	\end{figure} 	

\begin{figure}[!htb]
	\centering
	\subfigure[]
	{ \includegraphics[width=0.49\linewidth]{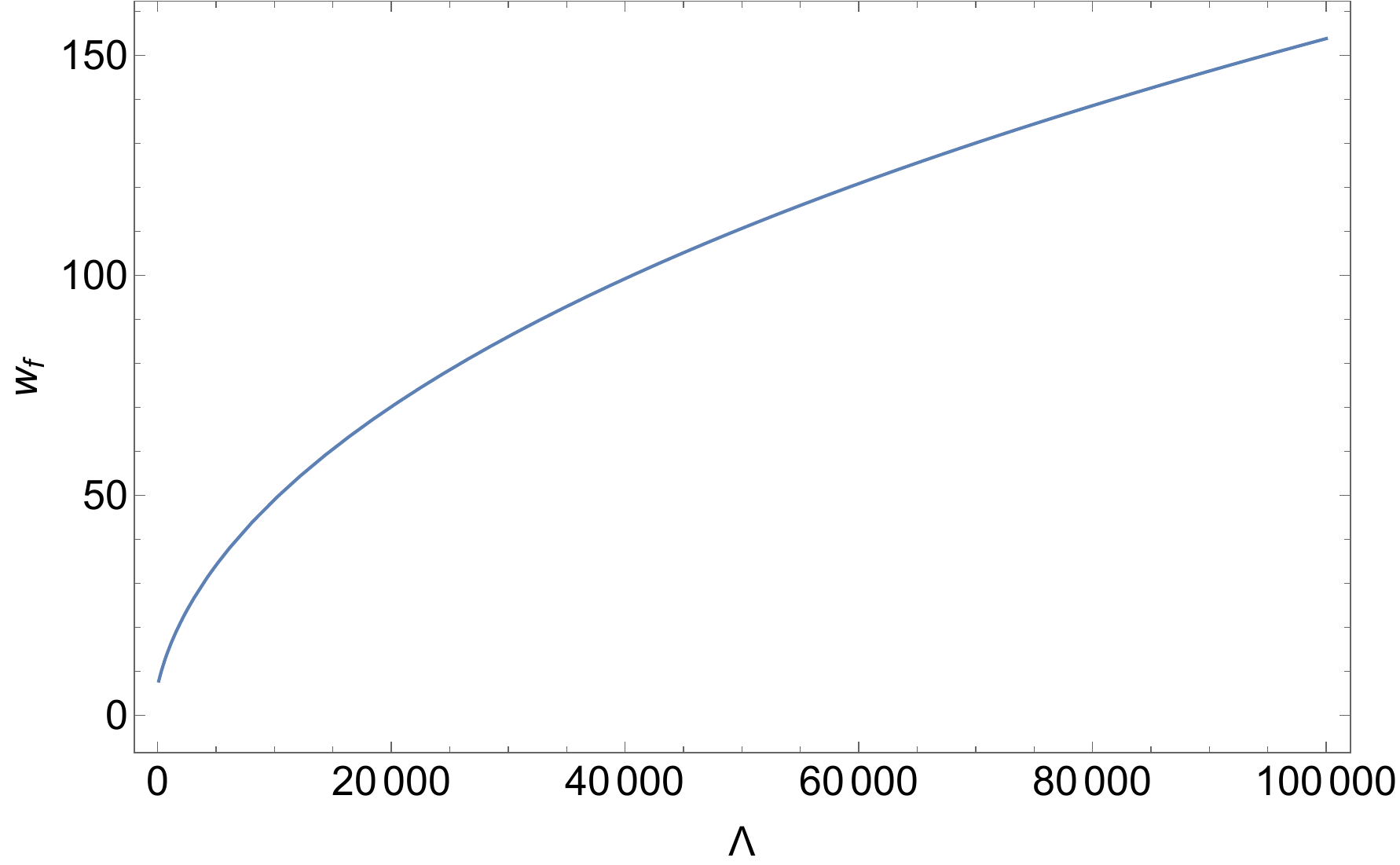}}
	\subfigure[]
	{ \includegraphics[width=0.49\linewidth]{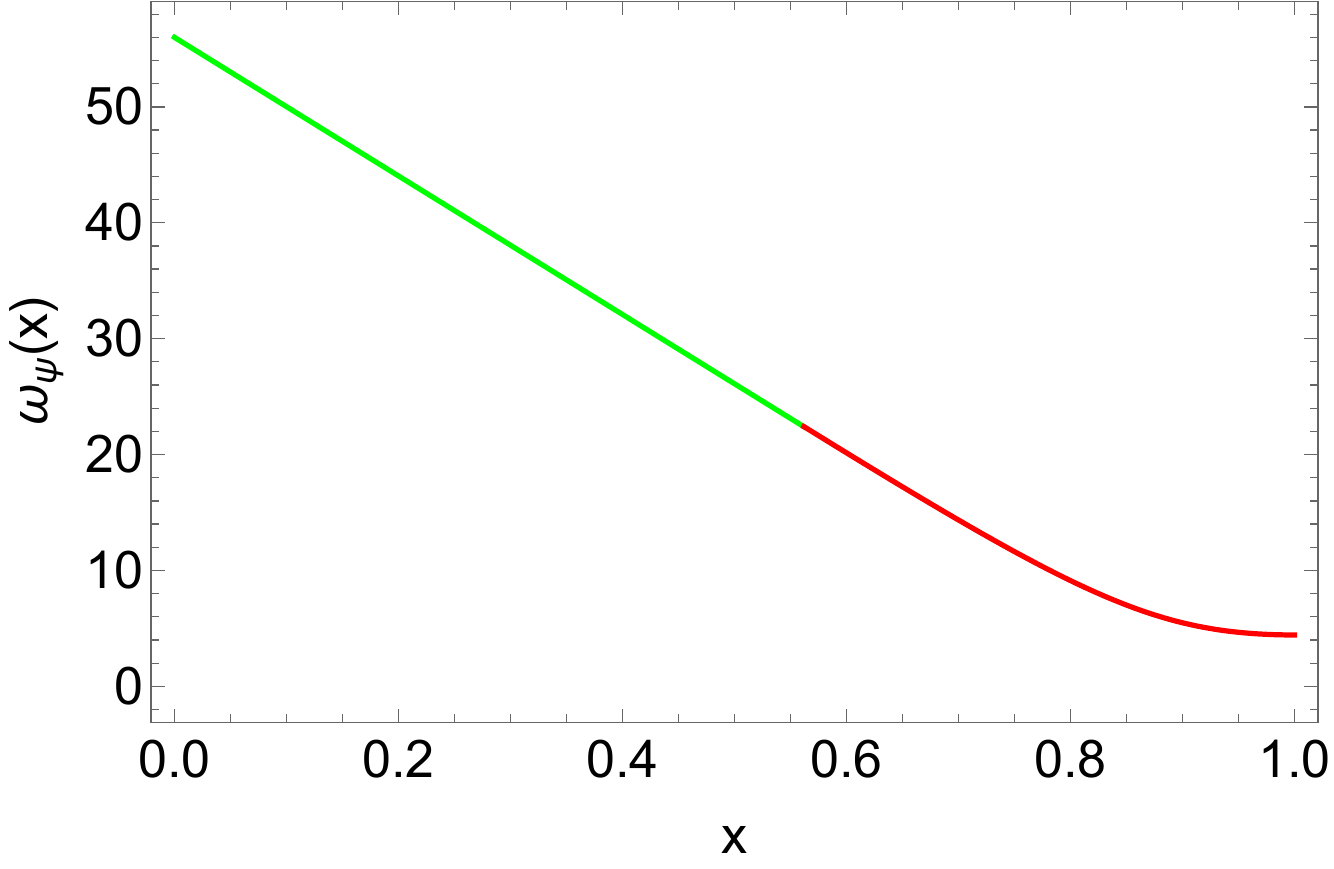}}
	
	\caption{ RG flow of the wave-function renormalization constant $w_{\psi}(x)$. (a) $w_f$ vs. $\Lambda$ overall with $\lambda _0=2, ~ \lambda _f=2, ~ v_0=1.5 \Lambda , ~ w_0=1.12 \Lambda , ~ \varphi _f=0.0001$. (b) $w_{\psi}(x)$ vs. $x$ overall: The matching point $x_2$ and the IR renormalized value $w_f$ are obtained from Fig.~\ref{omega2} (a).	}
	\label{wf}
\end{figure}

Now, considering $\lambda _0=2, ~ \lambda _f=2, ~ v_0=1.5 \Lambda , ~ w_0=1.12 \Lambda , ~ \varphi_f=0.0001$, we investigate the UV cutoff $\Lambda$ dependence of $w_{f}$, shown in Fig.~\ref{omega2}. It turns out that $w_f$ increases quite slowly as a function of $\Lambda$, confirmed in Fig.~\ref{wf} (a).
%
%
However, we point out $w_f \ll w_0$, not easy to obtain in the conventional quantum-field-theory calculation.

Finally, we discuss an RG flow of the wave-function renormalization constant $w_{\psi}(x)$. The green curve in Fig.~\ref{lambda1} (b) is a plot of  Eq. (\ref{eq:51}) while the red one is a plot of Eq. (\ref{eq:30}). The matching point is obtained by Eq. (\ref{eq:60}) and Eq. (\ref{eq:61}).
	
For sufficiently large $\Lambda$, we find an analytic expression for the matching point $x_2$ and the IR renormalized wave-function constant $w_f$ from Eq. (\ref{eq:60}) and Eq. (\ref{eq:61}) as follows
\bqa
x_2 &=& 1-\frac{w_f^2 \tanh ^{-1}\left(\frac{v_r}{w_r^2}\sqrt{\frac{\lambda _f}{\lambda _f+1}}\right)}{\Lambda  \sqrt{\lambda _f \left(\lambda _f+1\right)}}, \label{eq:62}\\
w_f &=& 2 \sqrt{2} \sqrt{-\frac{\Lambda ^2 \lambda _f^2 \left(\lambda _f+1\right) \left(v_r-w_r^3\right)^2}{\sqrt{-\lambda _f^2 \left(16 \Lambda ^2 \left(\lambda _f+1\right) \left(v_r-w_r^3\right)^2 \left(4 \lambda _f \tanh ^{-1}(M)^2 v_r^2-\left(\lambda _f+1\right) \log ^2(1-J) w_r^4\right)-K^2\right)}+K \lambda _f}}  \label{eq:63}\\
&\approx & \sqrt{\Lambda } \sqrt{\frac{2 \lambda _f \left(\lambda _f+1\right) \left(v_r-w_r^3\right)}{\left(\lambda _f+1\right) w_r^2 \log \left(1-\frac{\lambda _f v_r^2}{\left(\lambda _f+1\right) w_r^4}\right)+2 \sqrt{\lambda _f \left(\lambda _f+1\right)} v_r \tanh ^{-1}\left(\frac{\sqrt{\frac{\lambda _f}{\lambda _f+1}} v_r}{w_r^2}\right)}} \label{eq:64}
\eqa
with
\bqa
v_0 &=& v_r \Lambda, \nonumber\\
w_0 &=& w_r \Lambda,   \nonumber\\
J &=& \frac{\lambda _f v_r^2}{\left(\lambda _f+1\right) w_r^4}, \nonumber\\
K &=& 4 \lambda _f \left(\lambda _f+1\right) (\log (1-J)-1) w_r^4+8 \Lambda  \sqrt{\lambda _f \left(\lambda _f+1\right)} v_r \left(w_r^3-v_r\right) \tanh ^{-1}\left(\frac{\sqrt{\frac{\lambda _f}{\lambda _f+1}} v_r}{w_r^2}\right).   \nonumber
\eqa
We point out that Eq. (\ref{eq:64}) confirms $w_f \sim \sqrt{\Lambda}$.

\begin{figure}[!htb]
	\centering
	\subfigure[]
	{ \includegraphics[width=0.3\linewidth]{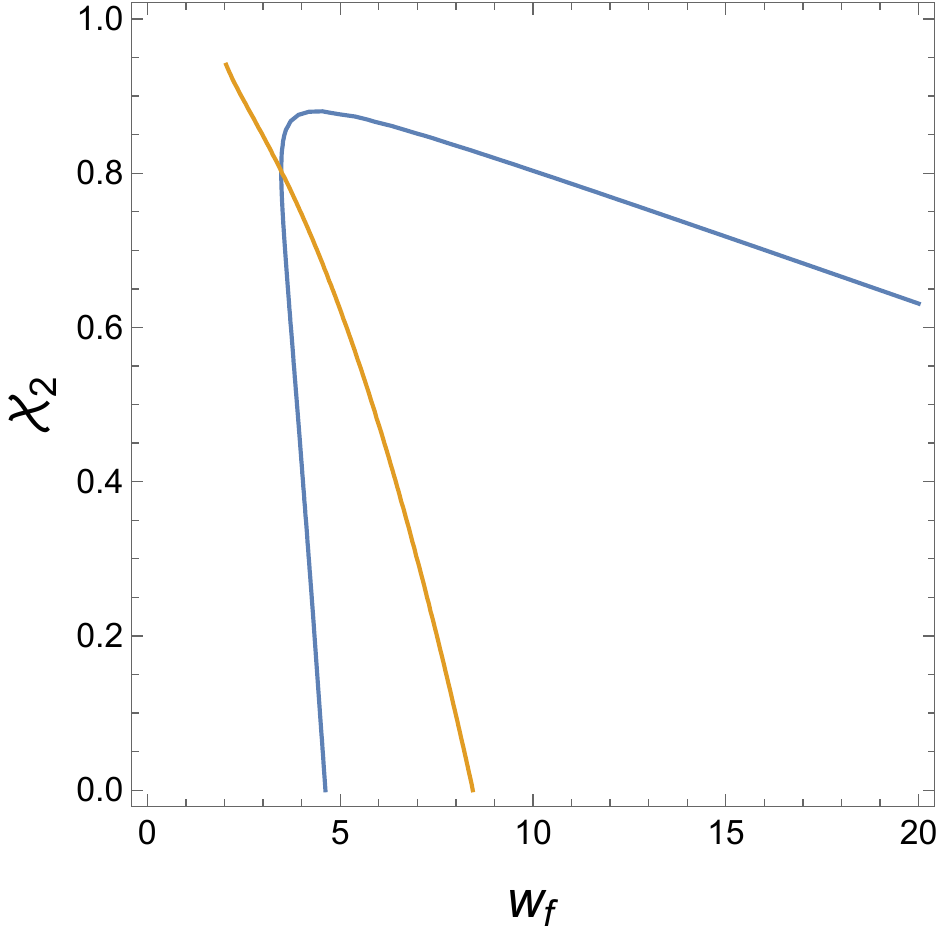}}
	\subfigure[]
	{ \includegraphics[width=0.3\linewidth]{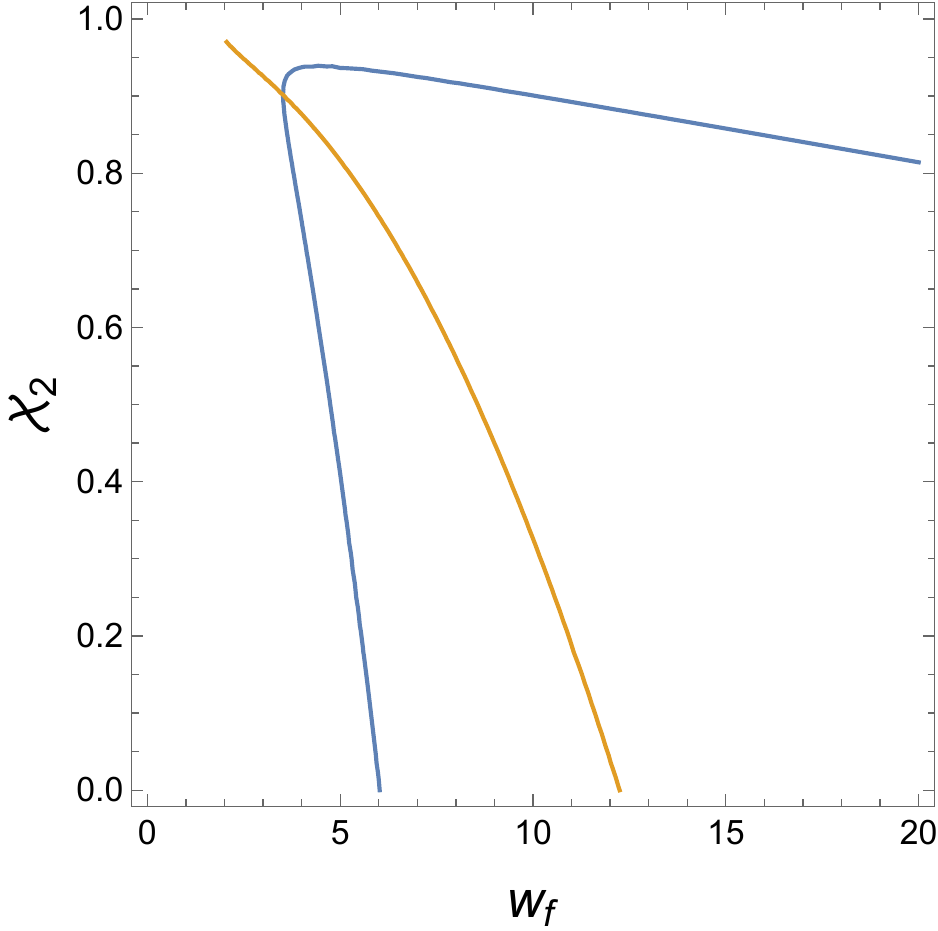}}
	\subfigure[]
	{ \includegraphics[width=0.3\linewidth]{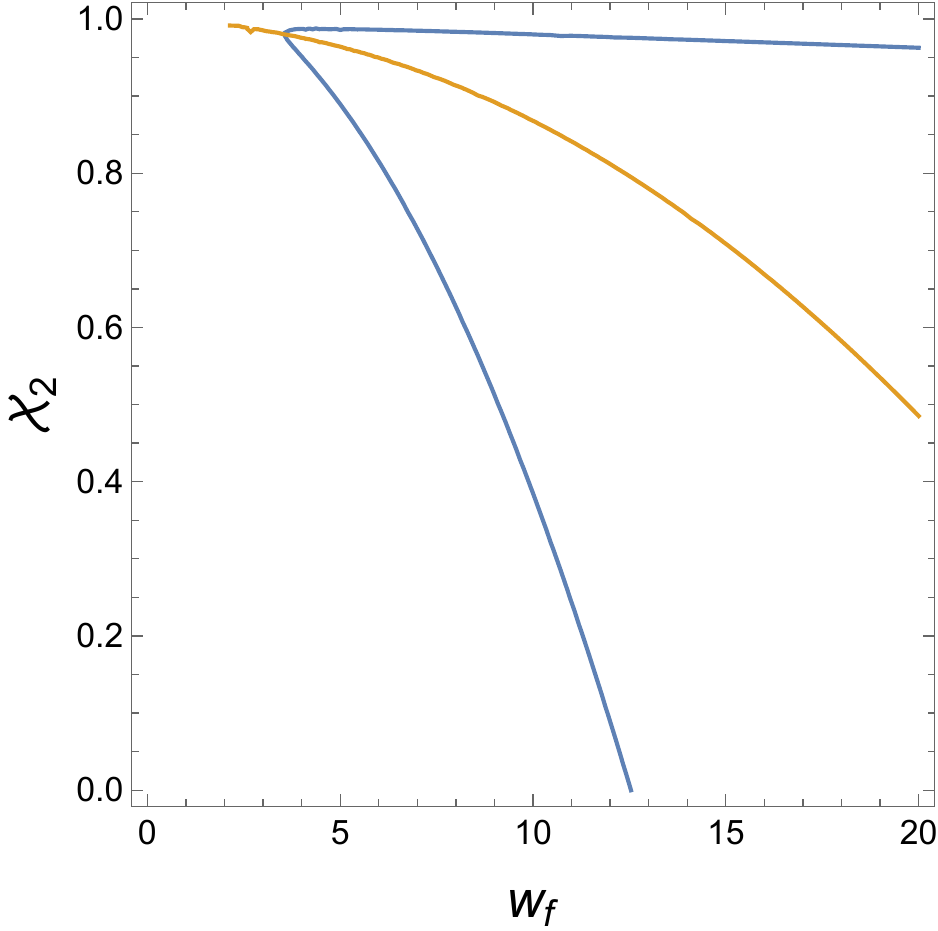}}\subfigure[]
	{ \includegraphics[width=0.3\linewidth]{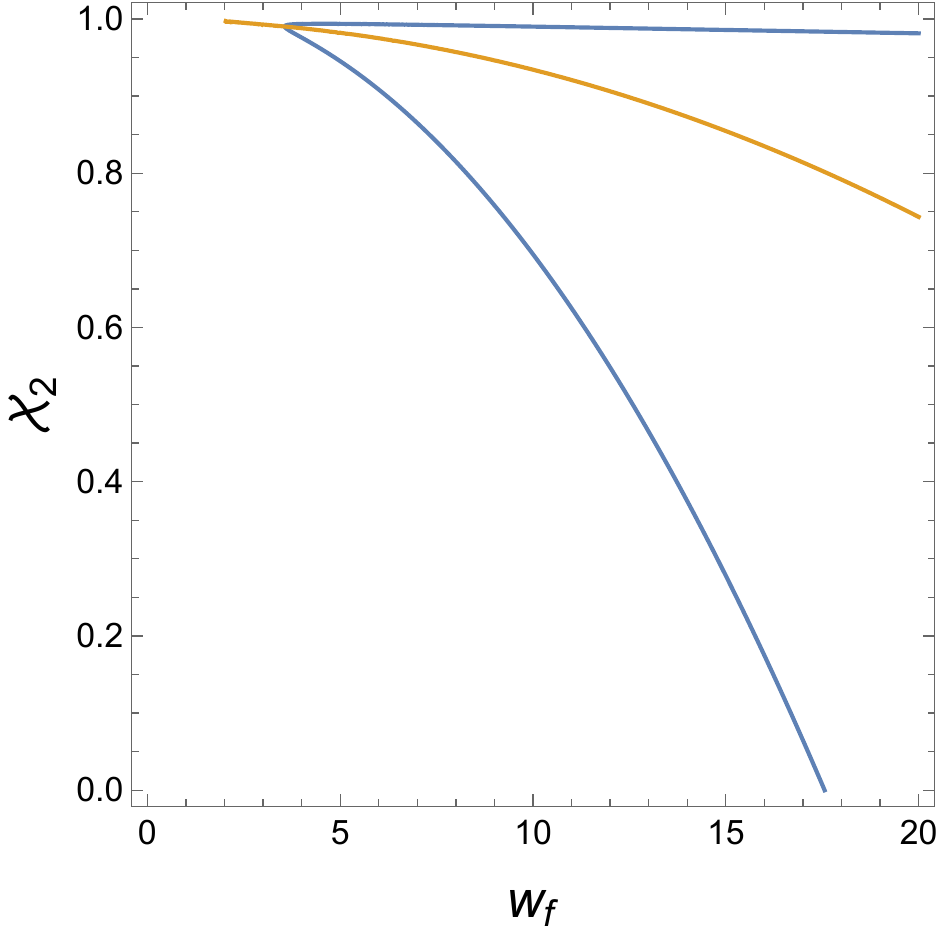}}
	
	\caption{The case when the IR renormalized wave-function constant $w_{f}$ does not depend on the UV cutoff $\Lambda$. Here, we take $v_r=\frac{v_m}{\Lambda }+w_r^3$ with $\lambda _0=2, ~ \lambda_f=2, ~ v_m=1, ~ w_r=1.12, ~ \varphi _f=0.0001$: (a) $x_2$ vs. $w_f$  at $\Lambda =50$,	 (b) $x_2$ vs. $w_f$ at $\Lambda =100$, (c) $x_2$ vs. $w_f$ at $\Lambda =500$, and (d) $x_2$ vs. $w_f$  at $\Lambda =1000$. }
	\label{wfremix}
\end{figure}

We have to mention that there is a case in which $w_f$ does not depend on $\Lambda$. If we consider $v_r=\frac{v_m}{\Lambda }+w_r^3$ as a UV boundary value in Eq. (\ref{eq:63}), we obtain
\begin{footnotesize}
\bqa
w_f &=&  2 \sqrt{2} \sqrt{-\frac{\lambda _f^2 \left(\lambda _f+1\right) v_m^2}{\sqrt{-\lambda _f^2 \left(16 \left(\lambda _f+1\right) v_m^2 \left(4 \lambda _f \tanh ^{-1}(U)^2 \left(\frac{v_m}{\Lambda }+w_r^3\right)^2-\left(\lambda _f+1\right) w_r^4 \log ^2(1-T)\right)-\frac{16 P^2}{\Lambda ^2}\right)}+Q}}, \label{eq:65}\\
&\approx &\sqrt{2} \sqrt{-\frac{\lambda _f \left(\lambda _f+1\right) v_m^2}{w_r^2 \left(-2 \mathcal{F}_1 \sqrt{\lambda _f \left(\lambda _f+1\right)} v_m w_r+\sqrt{\lambda _f+1} \sqrt{v_m^2 \left(\mathcal{F}_0^2 \left(\lambda _f+1\right)-4 \mathcal{F}_1^2 \lambda _f w_r^2\right)+\left(\mathcal{F}_0-1\right)^2 \lambda _f^2 \left(\lambda _f+1\right) w_r^4}+\left(\mathcal{F}_0-1\right) \lambda _f \left(\lambda _f+1\right) w_r^2\right)}} , \hspace{0.5cm}  \label{eq:66}
\eqa
\end{footnotesize}	
where
\bqa
T &=& \frac{\lambda _f \left(\frac{v_m}{\Lambda }+w_r^3\right)^2}{\left(\lambda _f+1\right) w_r^4}, \nonumber\\
U &=& \frac{\frac{v_m}{\Lambda }+w_r^3}{\sqrt{\frac{1}{\lambda _f}+1} w_r^2},   \nonumber\\
P &=& \Lambda  w_r^3 \left(2 \sqrt{\lambda _f \left(\lambda _f+1\right)} v_m \tanh ^{-1}(U)-\lambda _f \left(\lambda _f+1\right) w_r (\log (1-T)-1)\right)+2 \sqrt{\lambda _f \left(\lambda _f+1\right)} v_m^2 \tanh ^{-1}(U), \nonumber\\
Q &=&4 \lambda _f^2 \left(\lambda _f+1\right) w_r^4 (\log (1-T)-1)-8 \lambda _f \sqrt{\lambda _f \left(\lambda _f+1\right)} v_m \tanh ^{-1}(U) \left(\frac{v_m}{\Lambda }+w_r^3\right),   \nonumber\\
\mathcal{F}_0 &=& \log \left(1-\frac{\lambda _f w_r^2}{\lambda _f+1}\right), \nonumber\\
\mathcal{F}_1 &=& \tanh ^{-1}\left(\sqrt{\frac{\lambda _f}{\lambda _f+1}} w_r\right). \nonumber
\eqa   	
We see that Eq. (\ref{eq:66}) is independent of $\Lambda$. Figure \ref{wfremix} confirms that $w_f$ is independent of $\Lambda$ if we set $v_r=\frac{v_m}{\Lambda }+w_r^3$. Roughly, we have $w_f \approx 3.607$. But, the matching point depends on $\Lambda$, given by $x_2 \rightarrow 1$ as $\Lambda \rightarrow \infty$.

\subsubsection{The Matching solution for $v_{\psi}(x)$} 	
	
Substituting Eq. (\ref{eq:33}) and Eq. (\ref{eq:50}) into $v_{uv}(x_3)=v_{ir}(x_3)$ and $\partial_{x}v_{uv}(x) \Big|_{x = x_3} =\partial_{x}v_{ir}(x) \Big|_{x = x_3}$, we obtain two matching equations as follows
	\begin{enumerate}
		\item
		\begin{small}
			\bqa
			&&\Lambda  v_r+\frac{\Lambda  w_r^5 \log \left(\frac{w_r^3}{w_r^3-x_3 v_r}\right)}{v_r^2}-\frac{\Lambda  w_r^2 \left(\log \left(\frac{w_r^3}{w_r^3-x_3 v_r}\right)+x_3\right)}{v_r}\nonumber\\
			&&=v_f \left(\frac{\Lambda  \left(1-x_3\right) \left(\lambda _f+1\right)}{\varphi _f w_f}+1\right)^{\frac{\varphi _f^2 w_f^2}{\lambda _f \left(\lambda _f+1\right)^3}} \exp \left(\frac{2 \Lambda  \left(x_3-1\right) \left(\lambda _f+1\right) \varphi _f w_f+\Lambda ^2 \left(1-x_3\right)^2 \left(\lambda _f+1\right)^2}{2 \lambda _f \left(\lambda _f+1\right)^3}\right),   \hspace{1cm}\label{eq:67}
			\eqa
		\end{small}
		\item
		\begin{small}
			\bqa
			\frac{\Lambda ^2 \left(x_3-1\right) v_f \left(1-\frac{\Lambda  \left(x_3-1\right) \left(\lambda _f+1\right)}{w_f \varphi _f}\right)^{\frac{w_f^2 \varphi _f^2}{\lambda _f \left(\lambda _f+1\right)^3}} \exp \left(\frac{\Lambda  \left(x_3-1\right) \left(2 w_f \varphi _f+\Lambda  \left(x_3-1\right) \left(\lambda _f+1\right)\right)}{2 \lambda _f \left(\lambda _f+1\right)^2}\right)}{\lambda _f \left(\Lambda  \left(x_3-1\right) \left(\lambda _f+1\right)-w_f \varphi _f\right)}+\frac{w_r^2}{x_3 v_r-w_r^3}=0.   \label{eq:68}
			\eqa
		\end{small}
	\end{enumerate}

With the choice of the UV boundary value $v_r=\frac{v_m}{\Lambda }+w_r^3$, we find an analytic expression for the matching point $x_3$ and the IR renormalized velocity $v_f$ from Eq. (\ref{eq:67}) and Eq. (\ref{eq:68}) at sufficiently large $\Lambda$ as follows 	
\begin{small}
	\bqa
	x_3 &=& 1-\frac{v_m W\left(\frac{\Lambda  w_r^3 e^{\frac{w_r^3 \left(\Lambda -w_r \left(v_m+\Lambda  w_r^3\right)\right)}{v_m}}}{v_m}\right)}{\Lambda  w_r^3},  \label{eq:69}\\
v_f	&= & \frac{\lambda _f \left(1-\frac{\Lambda  \left(x_3-1\right) \left(\lambda _f+1\right)}{\varphi _f w_f}\right)^{-\frac{\varphi _f^2 w_f^2}{\lambda _f \left(\lambda _f+1\right)^3}} \left(\varphi _f w_f-\Lambda  \left(x_3-1\right) \left(\lambda _f+1\right)\right) \exp \left(-\frac{\Lambda  \left(x_3-1\right) \left(2 \varphi _f w_f+\Lambda  \left(x_3-1\right) \left(\lambda _f+1\right)\right)}{2 \lambda _f \left(\lambda _f+1\right)^2}\right)}{\Lambda ^2 \left(x_3-1\right)^2 w_r},   \label{eq:70}\\
&\approx & \frac{\exp \left(-\frac{v_m^2 W\left(\frac{ \Lambda  w_r^3  e^{\frac{w_r^3 \left(\Lambda -w_r \left(v_m+\Lambda  w_r^3\right)\right)}{v_m}}}{v_m}\right)^2}{2 w_r^6 \lambda _f \left(\lambda _f+1\right)}\right) \left(\lambda _f \left(\lambda _f+1\right) w_r^2 \left(\frac{\left(\lambda _f+1\right) v_m W\left(\frac{\left(\Lambda  w_r^3\right) e^{\frac{w_r^3 \left(\Lambda -w_r \left(v_m+\Lambda  w_r^3\right)\right)}{v_m}}}{v_m}\right)}{\varphi _f w_f w_r^3}+1\right)^{-\frac{\varphi _f^2 w_f^2}{\lambda _f \left(\lambda _f+1\right)^3}}\right)}{v_m W\left(\frac{\left(\Lambda  w_r^3\right) e^{\frac{w_r^3 \left(\Lambda -\omega _0 \left(v_m+\Lambda  w_r^3\right)\right)}{v_m}}}{v_m}\right)}.\hspace{0.5cm} \label{eq:71}
	\eqa
\end{small}	
Here, $W(z)$ is the Lambert W function. We observe
	\bqa
v_f &\sim &\exp \left(-\frac{v_m^2 W\left(\frac{ \Lambda  w_r^3  e^{\frac{w_r^3 \left(\Lambda -w_r \left(v_m+\Lambda  w_r^3\right)\right)}{v_m}}}{v_m}\right)^2}{2 w_r^6 \lambda _f \left(\lambda _f+1\right)}\right) \rightarrow 0
%
%
\label{eq:72}
\eqa
at sufficiently large $\Lambda$. This indicates appearance of a strong coupling IR fixed point, where the kinetic energy vanishes.

\begin{figure}[!htb]
	\centering
	\subfigure[]
	{ \includegraphics[width=0.45\linewidth]{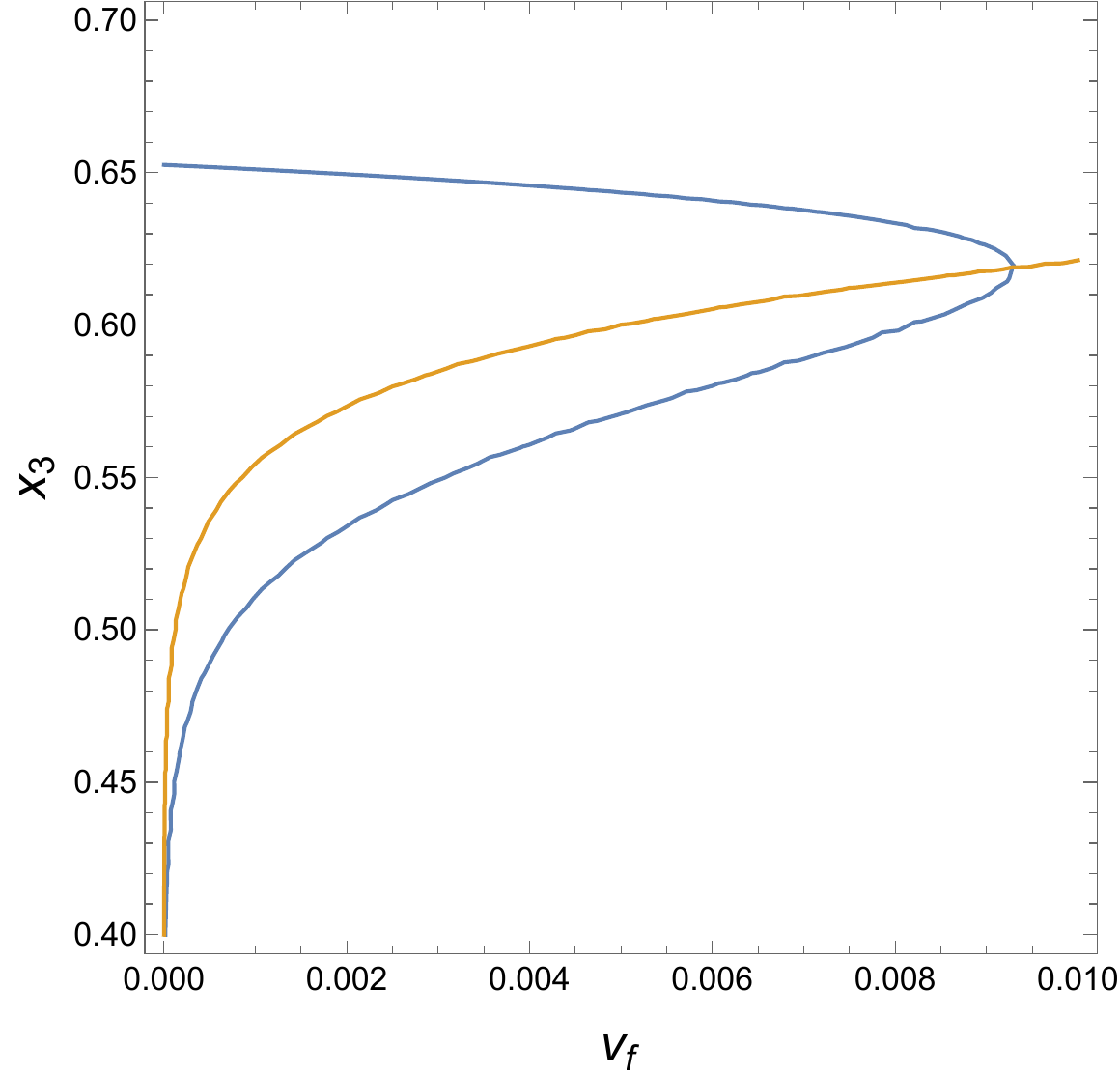}}
	\subfigure[]
	{ \includegraphics[width=0.45\linewidth]{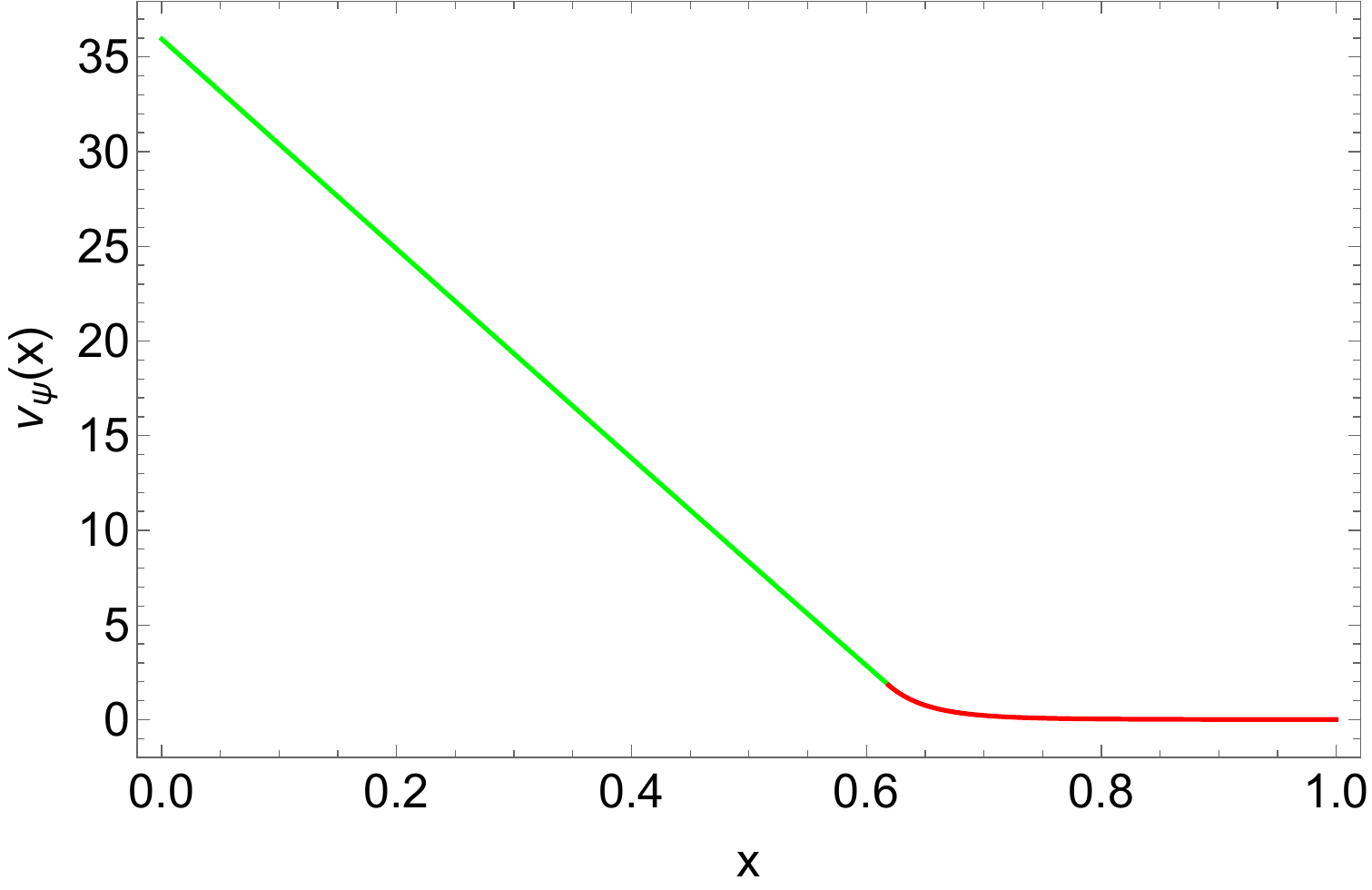}}
	\caption{(a) Self-consistent solution for the matching point $x_{3}$ and the IR boundary value $v_{f}$ and (b) RG flow of the renormalized velocity $v_{\psi}(x)$. Here, we take $v_r=\frac{v_m}{\Lambda }+w_r^3$ as a UV boundary value with $\Lambda =50, ~ \lambda _0=5, ~ \lambda_f=5, ~ w_r=0.9, ~ w_f=0.5, ~ \varphi_f=1.007$
%
%
}
	\label{vf1}
\end{figure}

The blue (yellow) curve in Fig.~\ref{vf1} (a) is a contour plot of Eq. (\ref{eq:67}) (Eq. (\ref{eq:68})). We notice that the intersection of these two curves occurs near $v_f=0.0065$. The green curve in Fig.~\ref{vf1} (b) is a plot of Eq. (\ref{eq:50}) while the red one is a plot of Eq. (\ref{eq:33}). The matching point is obtained by Eq. (\ref{eq:67}) and Eq. (\ref{eq:68}).	

\subsubsection{The Matching solution for $\varphi(x)$} 		

Substituting Eq. (\ref{eq:31}) and Eq. (\ref{eq:43}) into (1) $\varphi_{uv}(x_4)=\varphi_{ir}(x_4)$, (2) $\partial_{x}\varphi_{uv}(x) \Big|_{x = x_4} =\partial_{x}\varphi_{ir}(x) \Big|_{x = x_4}$, and (3) $ \partial_{x}^2\varphi_{uv}(x) \Big|_{x = x_4} =\partial_{x}^2\varphi_{ir}(x) \Big|_{x = x_4}$, we obtain three matching equations as follows
	
\begin{enumerate}
	\item
	\begin{small}
		\bqa
		&& \sqrt{\frac{1}{\lambda _f}+1} \coth \left(\frac{1}{2} \log \left(\frac{\mathcal{T}+1}{\mathcal{T}-1}\right)+\mathcal{Z}_x\right) \sqrt{-\lambda _f^2+\frac{\left(2 \lambda _f^2+w_f^2 \left(\log (\mathcal{U}-1)+2 \log \left(\sinh \left(\frac{1}{2} \log \left(\frac{\mathcal{T}+1}{\mathcal{T}-1}\right)+\mathcal{Z}_x\right)\right)\right)\right)^2}{4 \lambda _f^2}+w_f^2}\Bigg|_{x = x_4}\nonumber\\
		&&=\frac{  \varphi _0\sqrt{1-x}  \left(\mathfrak{Q}_{\text{in}} I_{1-2 \mathcal{B}_1}\left(\mathcal{Y}_x\right)+\mathfrak{Q}_{\text{re}} I_{2 \mathcal{B}_1-1}\left(\mathcal{Y}_x\right)\right)}{\left(v_0 w_0\right)^{5/2}}\Bigg|_{x = x_4}\label{eq:73}
		\eqa
	\end{small}
	\item
	\begin{small}
		\bqa
		  && \sqrt{\frac{1}{\lambda _f}+1}\frac{\partial}{\partial x}\left( \coth \left(\frac{1}{2} \log \left(\frac{\mathcal{T}+1}{\mathcal{T}-1}\right)+\mathcal{Z}_x\right) \sqrt{-\lambda _f^2+\frac{\left(2 \lambda _f^2+w_f^2 \left(\log (\mathcal{U}-1)+2 \log \left(\sinh \left(\frac{1}{2} \log \left(\frac{\mathcal{T}+1}{\mathcal{T}-1}\right)+\mathcal{Z}_x\right)\right)\right)\right)^2}{4 \lambda _f^2}+w_f^2}\right)\Bigg|_{x = x_4} \nonumber\\
		  &&=\frac{\varphi _0}{\left(v_0 w_0\right)^{5/2}}\frac{\partial}{\partial x}\left(  \sqrt{1-x}  \left(\mathfrak{Q}_{\text{in}} I_{1-2 \mathcal{B}_1}\left(\mathcal{Y}_x\right)+\mathfrak{Q}_{\text{re}} I_{2 \mathcal{B}_1-1}\left(\mathcal{Y}_x\right)\right)  \right)\Bigg|_{x = x_4}
	  \label{eq:74}
		\eqa
	\end{small}
\item
\begin{footnotesize}
	\bqa
&& \sqrt{\frac{1}{\lambda _f}+1}\frac{\partial^2}{\partial x^2}\left( \coth \left(\frac{1}{2} \log \left(\frac{\mathcal{T}+1}{\mathcal{T}-1}\right)+\mathcal{Z}_x\right) \sqrt{-\lambda _f^2+\frac{\left(2 \lambda _f^2+w_f^2 \left(\log (\mathcal{U}-1)+2 \log \left(\sinh \left(\frac{1}{2} \log \left(\frac{\mathcal{T}+1}{\mathcal{T}-1}\right)+\mathcal{Z}_x\right)\right)\right)\right)^2}{4 \lambda _f^2}+w_f^2}\right)\Bigg|_{x = x_4} \nonumber\\
&&=\frac{\varphi _0}{\left(v_0 w_0\right)^{5/2}}\frac{\partial^2}{\partial x^2}\left(  \sqrt{1-x}  \left(\mathfrak{Q}_{\text{in}} I_{1-2 \mathcal{B}_1}\left(\mathcal{Y}_x\right)+\mathfrak{Q}_{\text{re}} I_{2 \mathcal{B}_1-1}\left(\mathcal{Y}_x\right)\right)  \right)\Bigg|_{x = x_4},	 \label{eq:75}
	\eqa
\end{footnotesize}
\end{enumerate}  	
where
\bqa
\mathcal{T} &=&\frac{\lambda _f \varphi _f}{\sqrt{\lambda _f \left(\lambda _f+1\right)} w_f}, \nonumber\\
\mathcal{U}&=&\frac{\lambda _f \varphi _f^2}{\left(\lambda _f+1\right) w_f^2},   \nonumber\\
\mathcal{Z}_x&=&\frac{\Lambda  (1-x) \sqrt{\lambda _f \left(\lambda _f+1\right)}}{w_f^2}, \nonumber\\
\mathcal{B}_0&=&\frac{\lambda _0 \Lambda }{v_0 w_0},   \nonumber\\
\mathcal{B}_1&=&\frac{\lambda _0^2}{v_0 w_0}, \nonumber\\
\mathcal{Y}_x&=&2 \sqrt{\mathcal{B}_0 (1-x)}, \nonumber\\
\mathfrak{Q}_{\text{in}}&=&-\frac{\pi  \lambda _0^{3/2} \Lambda ^{3/2} v_0 w_0 \mathcal{B}_0^{\mathcal{B}_1-2} \csc \left(2 \pi  \mathcal{B}_1\right) \left(2 \mathcal{B}_1 \, _0F_1\left(;2 \mathcal{B}_1;\mathcal{B}_0\right) \left(\Lambda +\mathcal{B}_1\right)+\mathcal{B}_0 \, _0F_1\left(;2 \mathcal{B}_1+1;\mathcal{B}_0\right)\right)}{\Gamma \left(2 \mathcal{B}_1+1\right)}, \nonumber\\
\mathfrak{Q}_{\text{re}}&=&-\frac{\left(v_0 w_0\right)^{5/2} \left(\, _0F_1\left(;2-2 \mathcal{B}_1;\mathcal{B}_0\right) \left(2 \lambda _0 \, _0F_1\left(;2 \mathcal{B}_1;\mathcal{B}_0\right) \left(\Lambda +\mathcal{B}_1\right)+\Lambda  \, _0F_1\left(;2 \mathcal{B}_1+1;\mathcal{B}_0\right)\right)+2 \lambda _0 \left(1-2 \mathcal{B}_1\right)\right)}{\left(2 \lambda _0 \left(2 \mathcal{B}_1-1\right)\right) I_{2 \mathcal{B}_1-1}\left(2 \sqrt{\mathcal{B}_0}\right)}. \nonumber
\eqa   		

\begin{figure}[!htb]
	\centering
	\subfigure[]
	{ \includegraphics[width=0.4\linewidth]{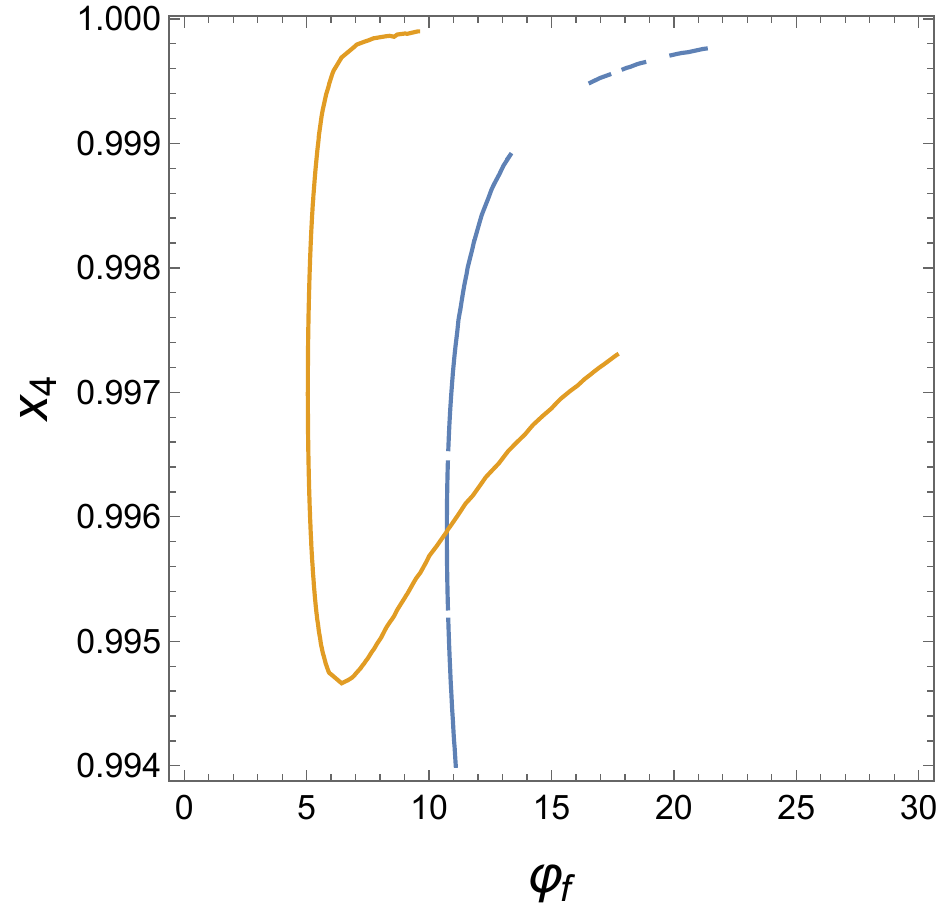}}
	\subfigure[]
	{ \includegraphics[width=0.4\linewidth]{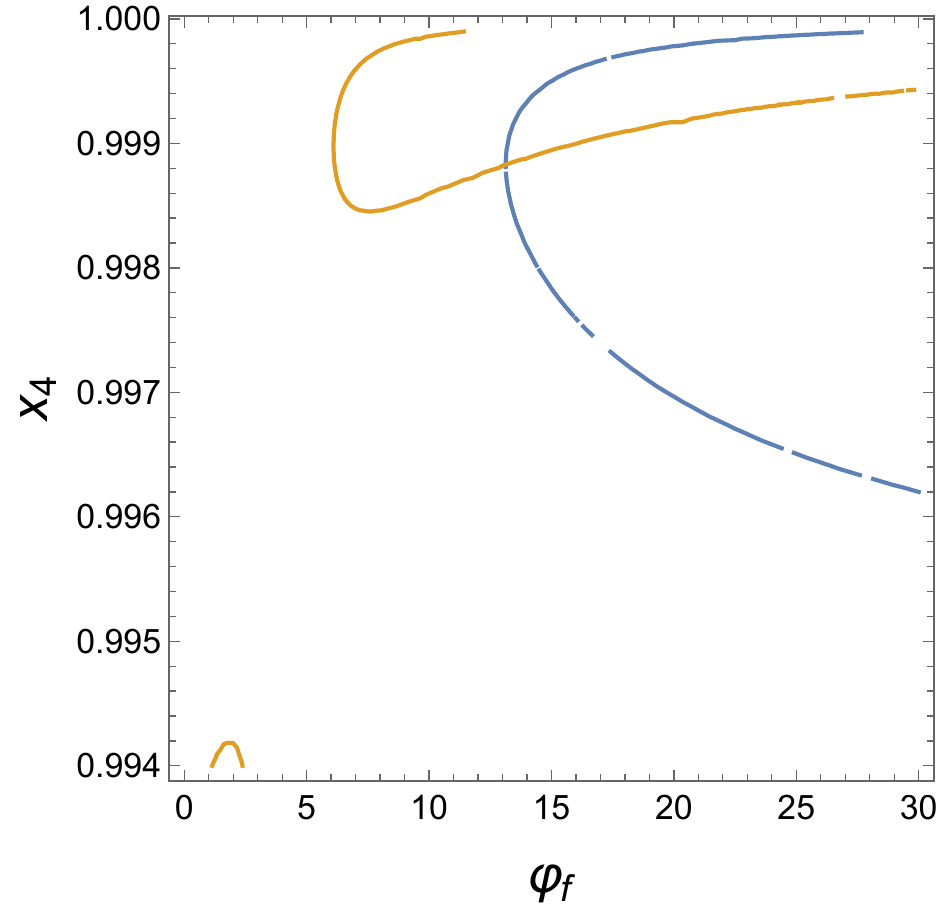}}
	\subfigure[]
	{ \includegraphics[width=0.4\linewidth]{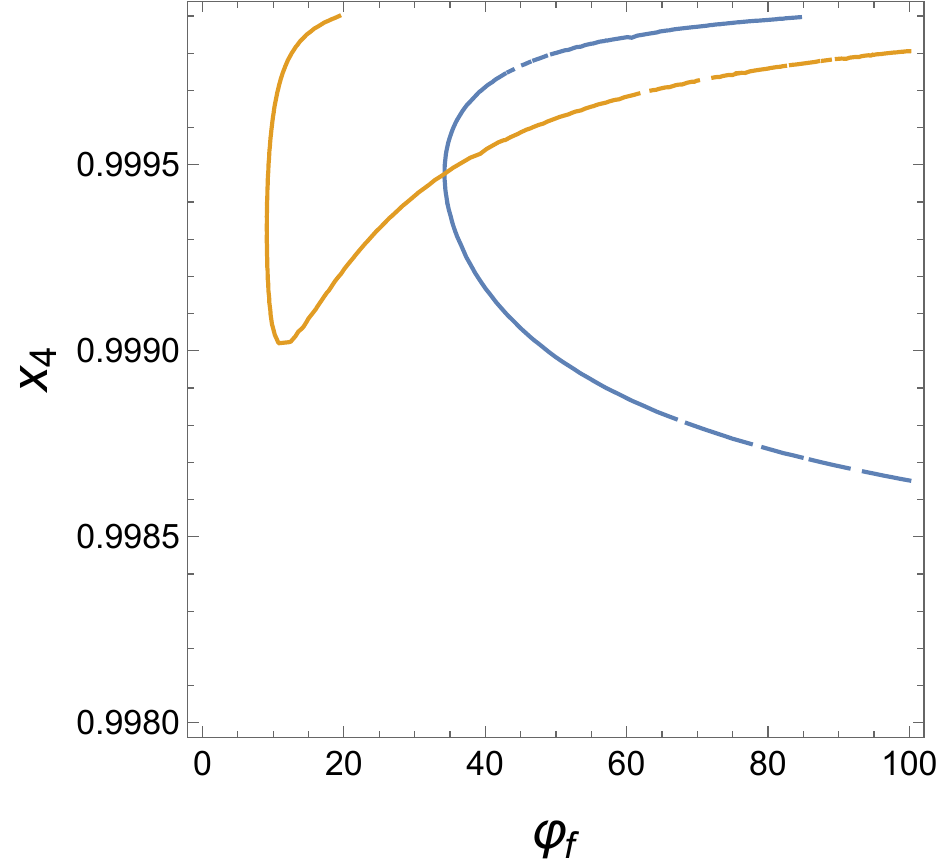}}
	\subfigure[]
	{ \includegraphics[width=0.4\linewidth]{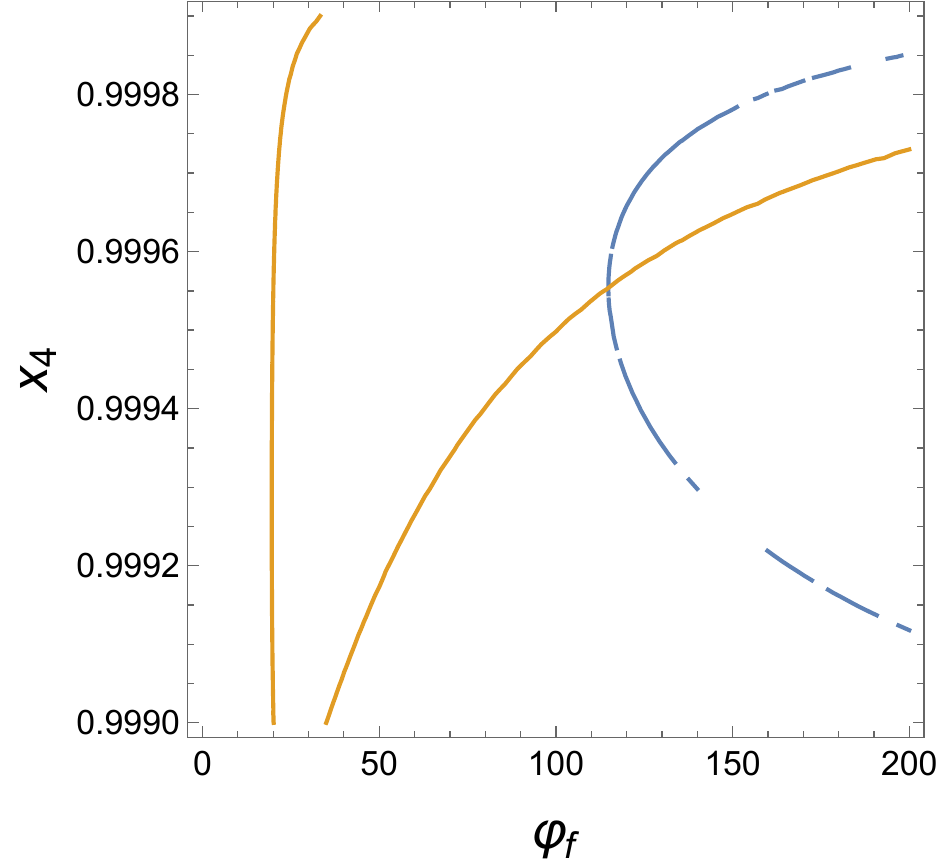}}
	
	\caption{Self-consistent solution for the matching point $x_{4}$ and the IR renormalized value $\varphi_{f}$ of the chiral condensate as a function of the UV cutoff $\Lambda$. Here, we take $w_f \approx 3.11568$, given by Eq. (\ref{eq:63}). We use $v_r=\frac{v_m}{\Lambda }+w_r^3$ with $\lambda_0=2.1, ~ \lambda _f=2.1, ~ w_r=1.2, ~ v_m=1$: (a) $x_4$ vs. $\varphi_f$ at $\Lambda =50$,	(b) $x_4$ vs. $\varphi_f$ at $\Lambda =100$, (c) $x_4$ vs. $\varphi_f$ at $\Lambda =500$, and (d)  $x_4$ vs. $\varphi_f$ at $\Lambda =1000$.}
	\label{phicont}
\end{figure}

Blue (yellow) curves in Fig.~\ref{phicont} (a)--(d) are contour plots of Eq. (\ref{eq:74}) (Eq. (\ref{eq:75})) with different values of $\Lambda$. $\varphi_0$ is obtained from Eq. (\ref{eq:73}). We notice that the intersection of these two curves occurs near $\varphi_f\approx 13.11$ and $x_4\approx 1 $. It is interesting to see that $\varphi_f$ is independent of $\Lambda$. We point out that $x_4 \rightarrow 1$ and $\varphi_0 \rightarrow 0$ as $\Lambda \rightarrow \infty$. For various values of $\Lambda=50,100,500,1000$, we have $\left(\varphi _0,\varphi _f,x_4\right)=\{ (0.2644,13.11,0.9805),(0.1353,13.11,0.9891),\\(0.0273,13.11,0.9977),(0.0136,13.11,0.9988)\}$.

\begin{figure}[!htb]
	\centering
	\subfigure[]
	{ \includegraphics[width=0.49\linewidth]{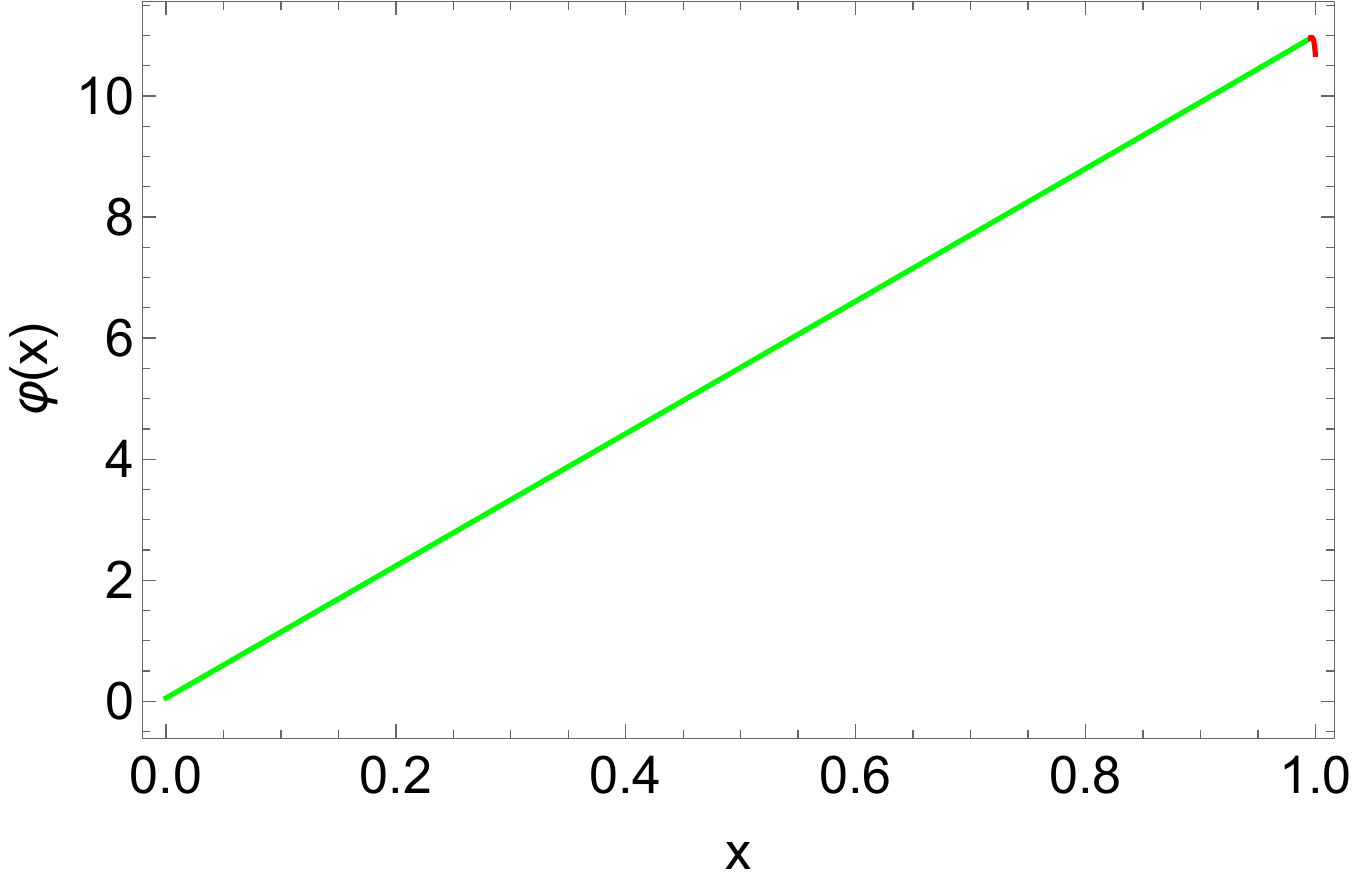}}
	\subfigure[]
	{ \includegraphics[width=0.49\linewidth]{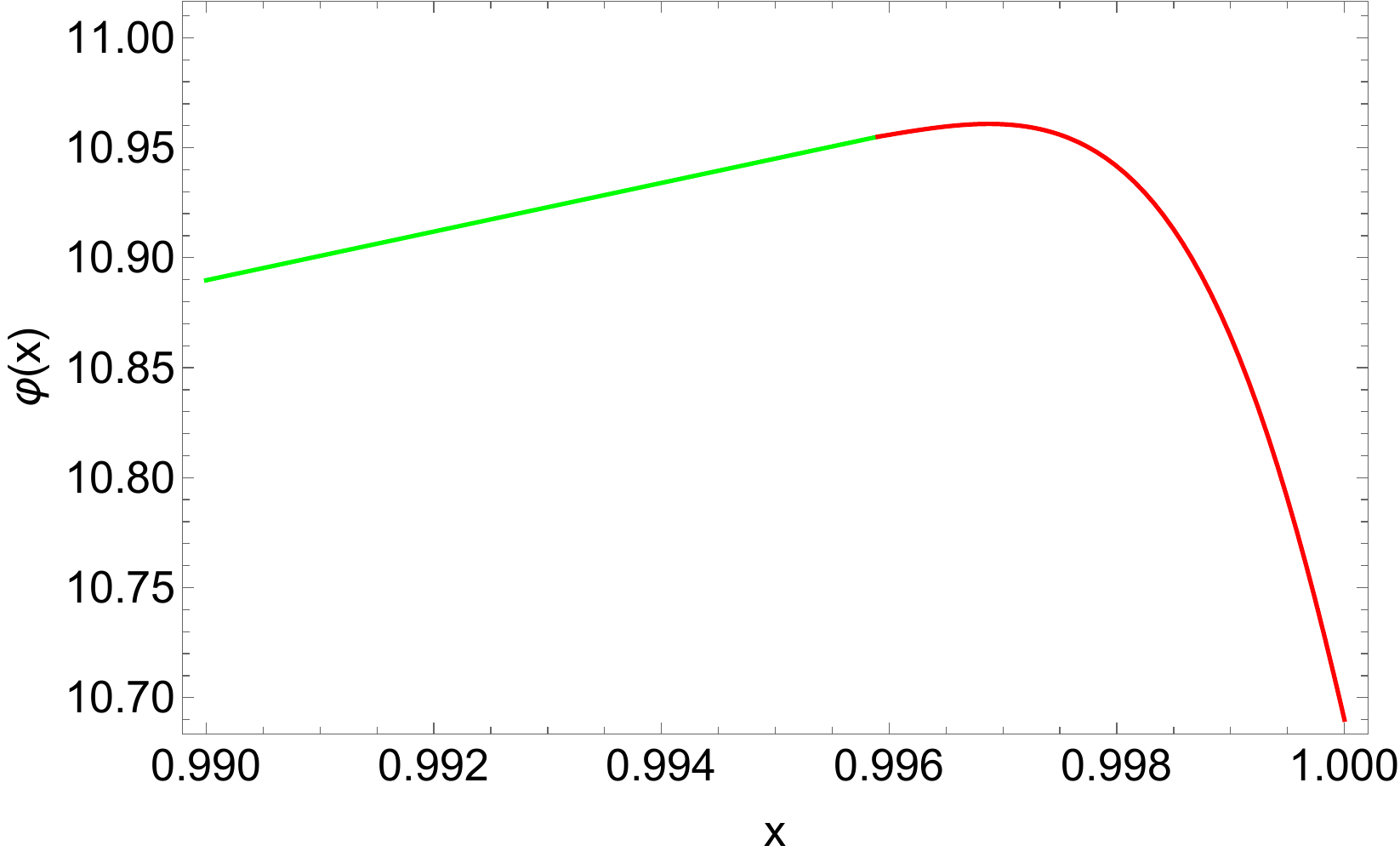}}
	
	\caption{ RG flow of the renormalized chiral condensate $\varphi(x)$. (a) $\varphi(x)$ vs. $x$ overall.	(b) $\varphi(x)$ vs. $x$ zoomed. }
	\label{phiplot}
\end{figure}

Figure \ref{phiplot} (a) and (b) are based on Fig.~\ref{phicont} (d): $\Lambda =1000, ~ \lambda _0=2.1, ~ \lambda _f=2.1, ~ w_r=1.2, ~ v_m=1$ with $\left(\varphi _0,\varphi _f,x_4\right)= (0.0136,13.11,0.9988)$. Green (red) curves in Fig.~\ref{phiplot} (a) and (b) are based on Eq. (\ref{eq:43}) (Eq. (\ref{eq:31})). The matching point is obtained by Eq. (\ref{eq:73}), Eq. (\ref{eq:74}), and Eq. (\ref{eq:75}).

\subsection{Introduction of $\delta \varphi(x,z)$ fluctuations near the fixed point}

Now, we are in the position to discuss how to find correlation functions, considering perturbations around the previously discussed vacuum solution. For example, we consider $\delta \varphi(x,z)$ fluctuations up to the Gaussian order around the vacuum fixed point as follows
\bqa && Z = Z_{vac.} \int D \delta \varphi(x,z) \exp\Bigg[ - N \int d^{2} x \Bigg\{ \frac{\lambda_{\chi}^{2}(z_{f})}{2} \delta \varphi(x,z_{f}) \int d^{2} x' \Pi(x-x',z_{f}) \delta \varphi(x',z_{f}) + \frac{\lambda_{\chi}(0)}{2} \delta \varphi^{2}(x,0) \Bigg\} \nn && - N \int_{0}^{z_{f}} d z \int d^{2} x \Bigg\{ \frac{\lambda_{\chi}(z)}{2} \Bigg( \partial_{z} \delta \varphi(x,z) + \frac{\lambda_{\chi}^{2}(z)}{2} \int d^{2} x' \Pi(x-x',z) \delta \varphi(x',z) \Bigg)^{2} \nn && + \frac{\lambda_{\chi}^{2}(z)}{2} \delta \varphi(x,z) \int d^{2} x' \Pi(x-x',z) \delta \varphi(x',z) \Bigg\} \Bigg] , \eqa
where the vacuum partition function is given by
\bqa && Z_{vac.} = \exp\Bigg[ - \beta L N \Bigg( - \frac{2}{\beta} \ln \Big\{ 2 w_{\psi}(z_{f}) \cosh \Big( \frac{\beta}{2 w_{\psi}(z_{f})} \sqrt{v_{\psi}^{2}(z_{f}) [\Lambda - z_{f}]^{2} - \lambda_{\chi}^{2}(z_{f}) \varphi^{2}(z_{f})} \Big) \Big\}  + \frac{\lambda_{\chi}(0)}{2} \varphi^{2}(0) \Bigg) \nn && - \beta L N \int_{0}^{z_{f}} d z \Bigg\{ \frac{\lambda_{\chi}(z)}{2} \Bigg( \partial_{z} \varphi(z) + \frac{1}{w_{\psi}(z)} \frac{ \lambda_{\chi}^{2}(z) \varphi(z)}{\sqrt{v_{\psi}^{2}(z) [\Lambda - z]^{2} - \lambda_{\chi}^{2}(z) \varphi^{2}(z)}} \tanh \Big( \frac{\beta}{2 w_{\psi}(z)} \sqrt{v_{\psi}^{2}(z) [\Lambda - z]^{2} - \lambda_{\chi}^{2}(z) \varphi^{2}(z)} \Big) \Bigg)^{2} \nn && - \frac{2}{\beta} \ln \Big\{ 2 w_{\psi}(z) \cosh \Big( \frac{\beta}{2 w_{\psi}(z)} \sqrt{v_{\psi}^{2}(z) [\Lambda - z]^{2} - \lambda_{\chi}^{2}(z) \varphi^{2}(z)} \Big) \Big\} \Bigg\} \Bigg] , \eqa
and discussed before. Here, we focus on $\delta \varphi(x,z)$ fluctuations only just for simplicity. In principle, we have to take into account all possible fluctuations of dynamical fields, more precisely, $\delta w_{\psi}(x,z)$, $\delta v_{\psi}(x,z)$, and $\delta \lambda_{\chi}(x,z)$ in addition to $\delta \varphi(x,z)$.

$\Pi(x-x',z)$ is the polarization function, given by
\bqa && N \Pi(x-x',z) \equiv \Big\langle \bar{\psi}_{\sigma}(x) \psi_{\sigma}(x) \bar{\psi}_{\sigma'}(x') \psi_{\sigma'}(x') \Big\rangle_{c} \equiv \frac{1}{Z(z)} \int D \psi_{\sigma}(y) ~ \bar{\psi}_{\sigma}(x) \psi_{\sigma}(x) \bar{\psi}_{\sigma'}(x') \psi_{\sigma'}(x') \nn && \exp\Big[ - \int d^{2} y \Big\{ \bar{\psi}_{\sigma}(y) \Big(w_{\psi}(z) \gamma^{\tau} \partial_{\tau} - i v_{\psi}(z) \gamma^{x} \partial_{y}\Big) \psi_{\sigma}(y) - i \lambda_{\chi}(z) \varphi(z) \bar{\psi}_{\sigma}(y) \psi_{\sigma}(y) \Big\} \Big] , \eqa
where the corresponding partition function for this correlation function is
\bqa && Z(z) = \int D \psi_{\sigma}(x) \exp\Big[ - \int d^{2} x \Big\{ \bar{\psi}_{\sigma}(x) \Big(w_{\psi}(z) \gamma^{\tau} \partial_{\tau} - i v_{\psi}(z) \gamma^{x} \partial_{x}\Big) \psi_{\sigma}(x) - i \lambda_{\chi}(z) \varphi(z) \bar{\psi}_{\sigma}(x) \psi_{\sigma}(x) \Big\} \Big] . \eqa
The last term of $\frac{\lambda_{\chi}^{2}(z)}{2} \delta \varphi(x,z) \int d^{2} x' \Pi(x-x',z) \delta \varphi(x',z)$ ($\frac{\lambda_{\chi}^{2}(z)}{2} \int d^{2} x' \Pi(x-x',z) \delta \varphi(x',z)$) results from the expansion of the effective potential $\mathcal{V}_{eff}[\varphi(x,z),w_{\psi}(x,z),v_{\psi}(x,z),\lambda_{\chi}(x,z)]$ (RG $\beta-$function $\beta_{\varphi}[\varphi(x,z),w_{\psi}(x,z),v_{\psi}(x,z),\lambda_{\chi}(x,z)]$) with respect to $\varphi(x,z) = \varphi(z) + \delta \varphi(x,z)$ up to the Gaussian (first) order.

The equation of motion for $\delta \varphi(x,z)$ is given by
\bqa && \partial_{z} \Big\{ \lambda_{\chi}(z) \Big( \partial_{z} \delta \varphi(x,z) + \frac{\lambda_{\chi}^{2}(z)}{2} \int d^{2} x' \Pi(x-x',z) \delta \varphi(x',z) \Big) \Big\} \nn && - \lambda_{\chi}(z) \frac{\lambda_{\chi}^{2}(z)}{2} \Pi(0,z) \Big( \partial_{z} \delta \varphi(x,z) + \frac{\lambda_{\chi}^{2}(z)}{2} \int d^{2} x' \Pi(x-x',z) \delta \varphi(x',z) \Big) \nn && - \lambda_{\chi}^{2}(z) \int d^{2} x' \Pi(x-x',z) \delta \varphi(x',z) = 0 \eqa
in the large $N$ limit. Taking $\lambda_{\chi}(z) = \lambda_{\chi}$ near the IR fixed point discussed in the previous subsection and performing the Fourier transformation, we obtain
\bqa && \partial_{z}^{2} \delta \varphi(i \Omega_{n}, \bm{q}, z) + \frac{\lambda_{\chi}^{2}}{2} \Big( \Pi(i \Omega_{n}, \bm{q}, z) - \frac{1}{\beta} \sum_{i \Omega_{n}} \int \frac{d^{D-1} \bm{q}}{(2\pi)^{D-1}} \Pi(i\Omega_{n},\bm{q},z) \Big) \partial_{z} \delta \varphi(i \Omega_{n}, \bm{q}, z) \nn && + \Big\{ \frac{\lambda_{\chi}^{2}}{2} \partial_{z} \Pi(i \Omega_{n}, \bm{q}, z) - \Big(\frac{\lambda_{\chi}^{2}}{2}\Big)^{2} \Big( \frac{1}{\beta} \sum_{i \Omega_{n}} \int \frac{d^{D-1} \bm{q}}{(2\pi)^{D-1}} \Pi(i\Omega_{n},\bm{q},z) \Big) \Pi(i\Omega_{n},\bm{q},z) - \lambda_{\chi} \Pi(i\Omega_{n},\bm{q},z) \Big\} \delta \varphi(i \Omega_{n}, \bm{q}, z) = 0 . \nn \eqa

Both IR and UV boundary conditions for these fluctuations can be found as follows
\bqa && \partial_{z_{f}} \delta \varphi(i\Omega_{n},\bm{q},z_{f}) = - \Big( 1 + \frac{1}{\lambda_{\chi}}\Big) \frac{\lambda_{\chi}^{2}}{2} \Pi(i\Omega_{n},\bm{q},z_{f}) \delta \varphi(i\Omega_{n},\bm{q},z_{f}) , \\ && \partial_{z} \delta \varphi(i\Omega_{n},\bm{q},z) \Big|_{z = 0} = \delta \varphi(i\Omega_{n},\bm{q},0) - \frac{\lambda_{\chi}^{2}}{2} \Pi(i\Omega_{n},\bm{q},0) \delta \varphi(i\Omega_{n},\bm{q},0) . \eqa
Here, the polarization function is
\bqa && \Pi(i \Omega_{n}, \bm{q}, z) = - \frac{1}{\beta} \sum_{i \omega_{n}} \int \frac{d k}{2 \pi} \mbox{tr}\Big\{G(i\omega_{n}+i\Omega_{n},k+q) G(i\omega_{n},k)\Big\} , \eqa
where the translational invariant vacuum Green's function is
\bqa && G(i\omega_{n},k) = \frac{1}{w_{\psi}(z) \gamma^{\tau} (i \omega_{n}) + v_{\psi}(z) \gamma^{x} k - i \lambda_{\chi} \varphi(z)} . \eqa
It is straightforward to calculate this polarization function for the vacuum state, not performed explicitly here.

Solving the equation of motion with two boundary conditions, we would find the `meson' spectrum in the chiral symmetry broken vacuum state. Although we do not perform this calculation here, the present discussion shows how the nonperturbative RG-MFT takes the essential features of the holographic dual effective field theory.

\section{Curvature and minimal surface on the three-dimensional curved spacetime}\label{HologarphicSpacetime}

By using the identification Eq. (\ref{ivb}) between the vierbein and the renormalized coupling functions, one can construct a curved spacetime metric containing information of the RG-MFT. In this section, we investigate properties of the curved spacetime constructed from the asymptotic solutions of the RG flow equations. In particular, we calculate the Ricci scalar curvature and area of the minimal surface.

\subsection{Three-dimensional Ricci scalar}

Let us consider the following three-dimensional spacetime
\begin{align}
d s^{2} = d z^{2} + g_{\mu\nu}d x^{\mu} d x^{\nu}=d z^{2} + g_{\tau\tau}d \tau d \tau + g_{yy}d y d y,\label{tcstm}
\end{align}
where $0\le z \le \Lambda$, $\tau$ is a time component, and $y$ is a spatial component. With the identification Eq. (\ref{ivb}), the metric $g_{\mu\nu}$ is expressed by
\begin{align}
g_{\tau\tau}=v_\psi^2, \;\;\; g_{yy}=w_\psi^2.\label{gmunu}
\end{align}
When $v_\psi$ and $w_\psi$ do not depend on $\tau$ and $y$ as studied in section \ref{Mean_Field_Theory}, the Ricci scalar $R$ of the three-dimensional spacetime Eq. (\ref{tcstm}) is given by
\begin{align}
R=-2\frac{(\partial_z v_\psi) (\partial_z w_\psi)+(\partial_z^2v_\psi) w_\psi+v_\psi (\partial_z^2w_\psi)}{v_\psi w_\psi}.
\end{align}
In section \ref{Mean_Field_Theory}, we obtained asymptotic IR solutions $w_{ir}(x)$ Eq. (\ref{eq:30}) and $v_{ir}(x)$ Eq. (\ref{eq:33}) near $x=1$ and asymptotic UV solutions $w_{uv}(x)$ Eq. (\ref{eq:51}) and $v_{uv}(x)$ Eq. (\ref{eq:50}) near $x=0$, where $x:=z/\Lambda$. By using them, we define two Ricci scalars as
\begin{align}
R_{uv}(x):=R|_{w_\psi=w_{uv}(x), v_\psi=v_{uv}(x)}, \;\;\; R_{ir}(x):=R|_{w_\psi=w_{ir}(x), v_\psi=v_{ir}(x)}.
\end{align}

\begin{figure}[!htb]
	\centering
	\subfigure[]
	{ \includegraphics[width=0.49\linewidth]{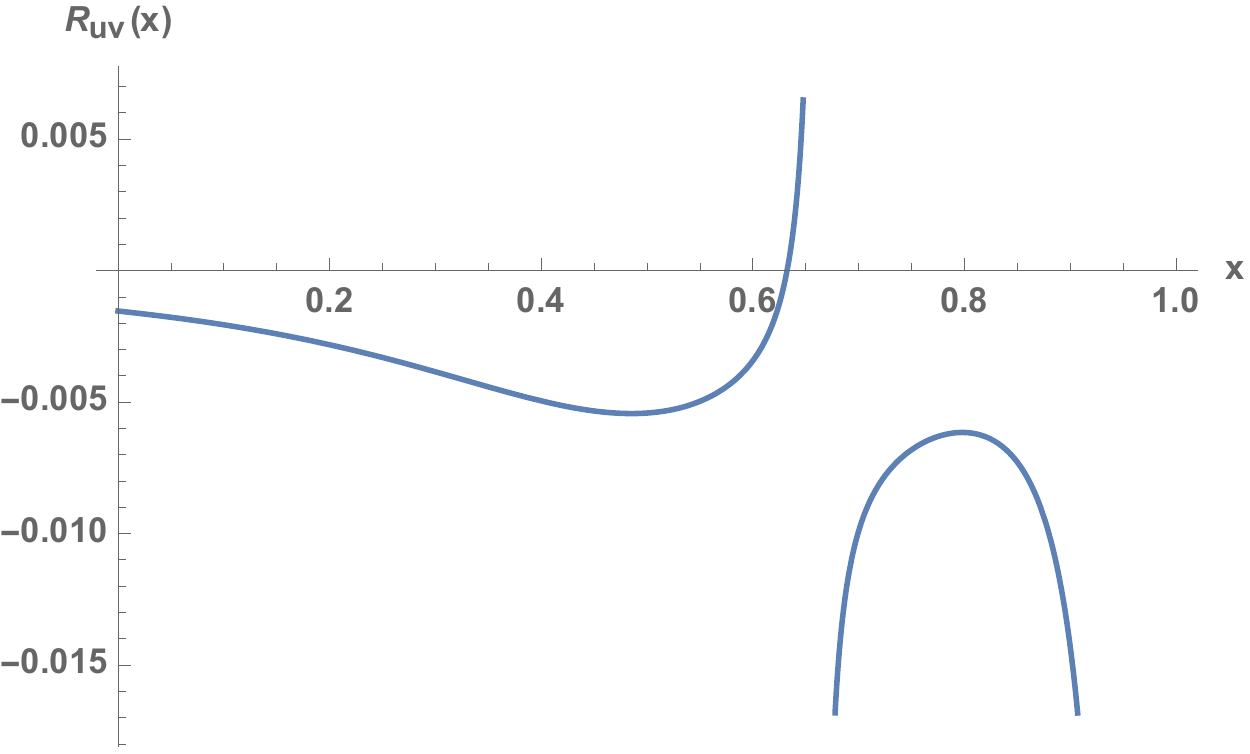}}
	\subfigure[]
	{ \includegraphics[width=0.49\linewidth]{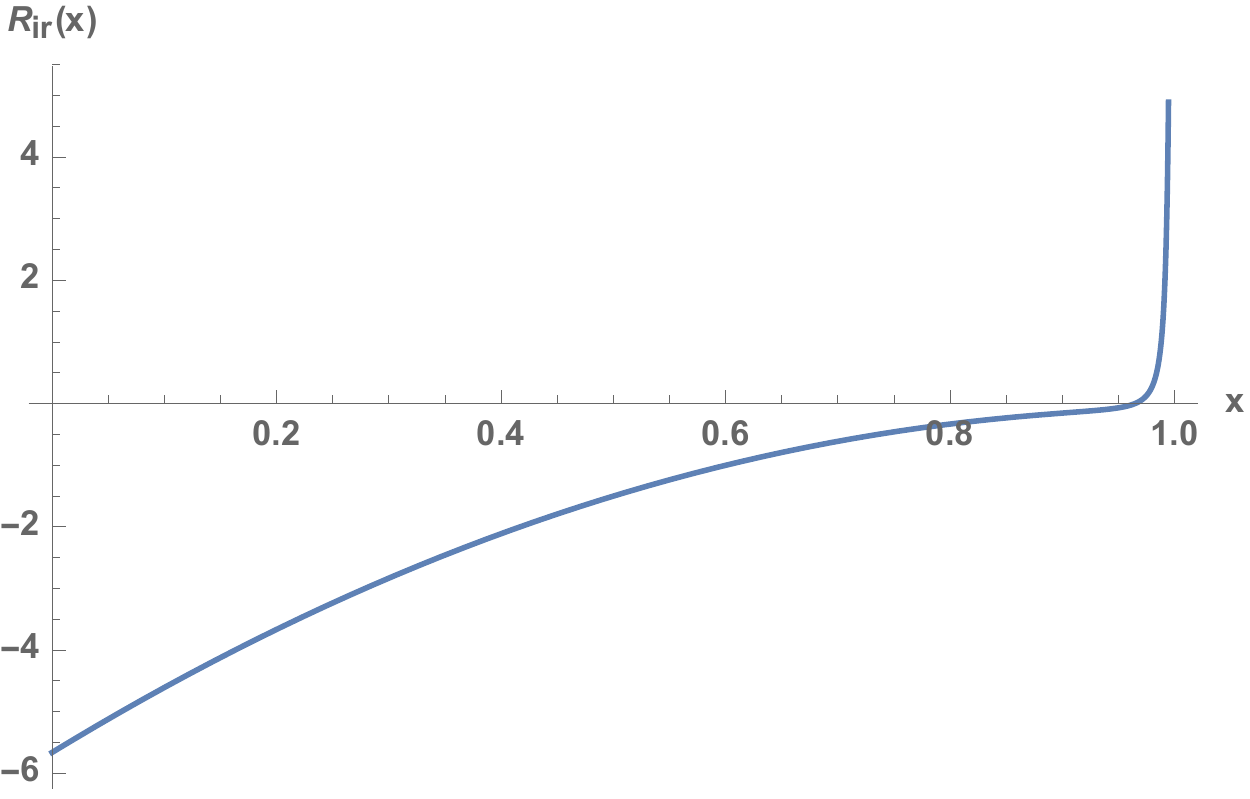}}
	
	\caption{Two Ricci scalars constructed from the asymptotic UV and IR solutions. (a) $R_{uv}(x)$ with $\Lambda =50, w_r=0.9,  v_r=\frac{v_m}{\Lambda }+w_r^3,  v_m=1$. We observe that it is given by a negative constant near the UV boundary. The behavior near the IR boundary is not relevant. (b) $R_{ir}(x)$ with $\Lambda =50, \lambda_f=5, w_f=0.5, v_f=0.0065, \varphi_f=1.007$. It diverges as $x \rightarrow 1$. The behavior near the UV boundary is not relevant.}
	\label{Rplot}
\end{figure}

Figure \ref{Rplot} shows plots of $R_{uv}(x)$ and $R_{ir}(x)$ with $\Lambda =50,   \lambda_f=5,  w_r=0.9,  w_f=0.5, v_r=\frac{v_m}{\Lambda }+w_r^3,  v_m=1,  v_f=0.0065,  \varphi_f=1.007$. With the above parameters, these two Ricci scalars have singular behaviors in certain regions. Near $x=0.65$, $R_{uv}(x)$ is singular because $v_{uv}(x)$ vanishes. Near $x=1$, $R_{uv}(x)$ is not well-defined since $w_0^3-\Lambda ^2 v_0 x$ in Eq. ($\ref{eq:50}$) is negative, where $v_0=v_r\Lambda$ and $w_0=w_r\Lambda$. Of course, these features are artifacts of the asymptotic UV solution extended to the IR region. A physically relevant point is that it is given by a negative constant near the UV boundary. The divergence of $R_{ir}(x)$ near $x=1$ is actually physical, which results from the singular behavior of the derivative of $w_{ir}(x)$. On the other hand, the asymptotic UV behavior of $R_{ir}(x)$ is not relevant.

Let us comment on the finiteness of the metric with finite $\Lambda$.
In Fig.~\ref{Rplot}, $R_{uv}(x)$ at $x=0$ is negative, which is similar to asymptotic AdS geometries. However, when $v_0$ and $w_0$ are finite, the metric Eq. (\ref{gmunu}) is finite at $x=0$, which is different from the asymptotic AdS geometries. Rather, its finiteness is similar to a property of the asymptotic AdS geometries with a finite cutoff radius.

\subsection{Minimal surface}

The authors of Refs. \cite{Ryu:2006bv,Ryu:2006ef} proposed the holographic entanglement entropy as the holographic dual of entanglement entropy in boundary QFTs. They gave a prescription to calculate the entanglement entropy using the minimal surface that anchored on the entangling surface. This celebrated proposal is the pioneering work to study quantum information from the viewpoint of holography. In this subsection, we evaluate the minimal surface for a single interval on the three-dimensional spacetime Eq. (\ref{tcstm}) with Eq. (\ref{gmunu}).

\begin{figure}[!htb]
	\centering
	\subfigure[]
	{ \includegraphics[width=0.3\linewidth]{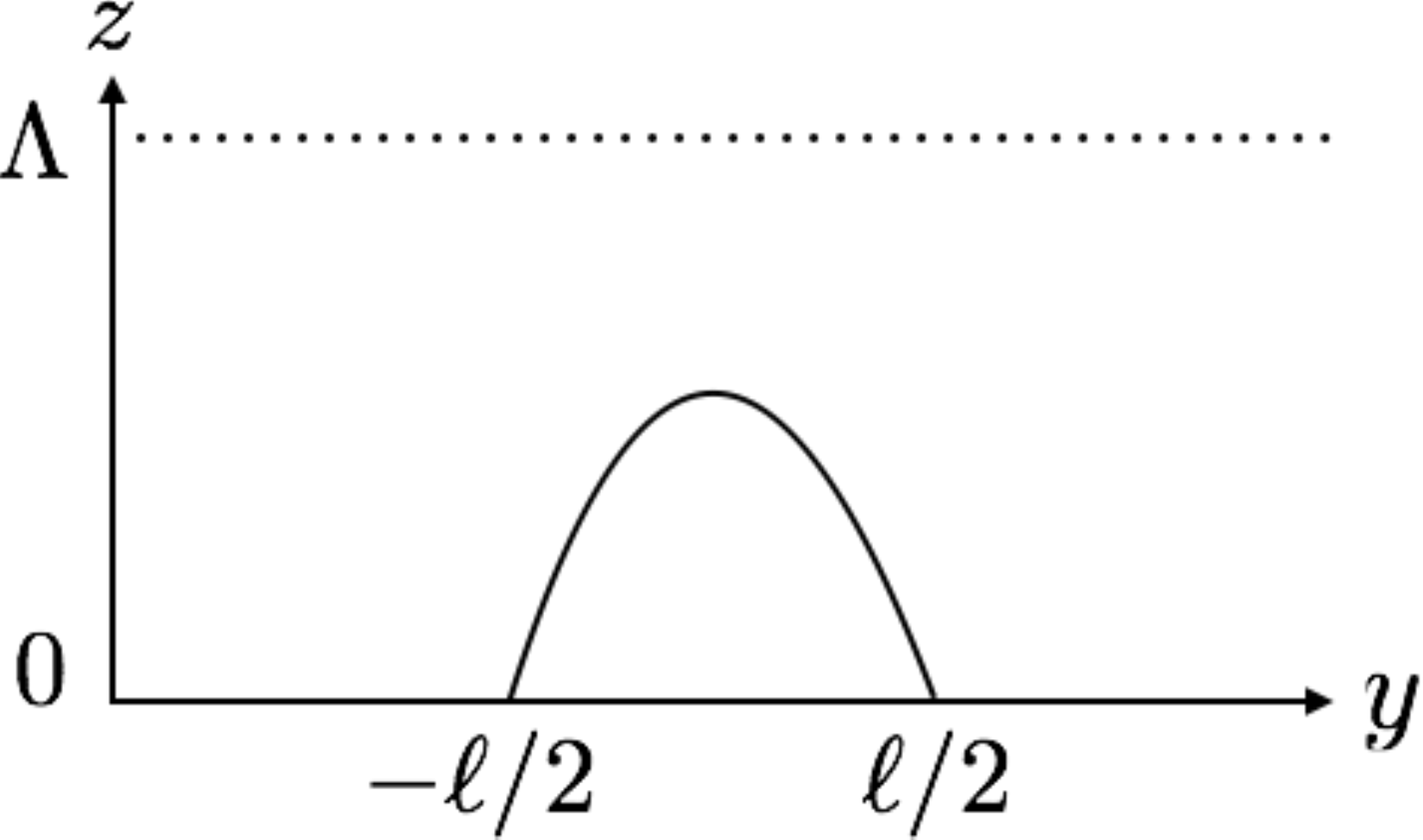}}
	\subfigure[]
	{ \includegraphics[width=0.3\linewidth]{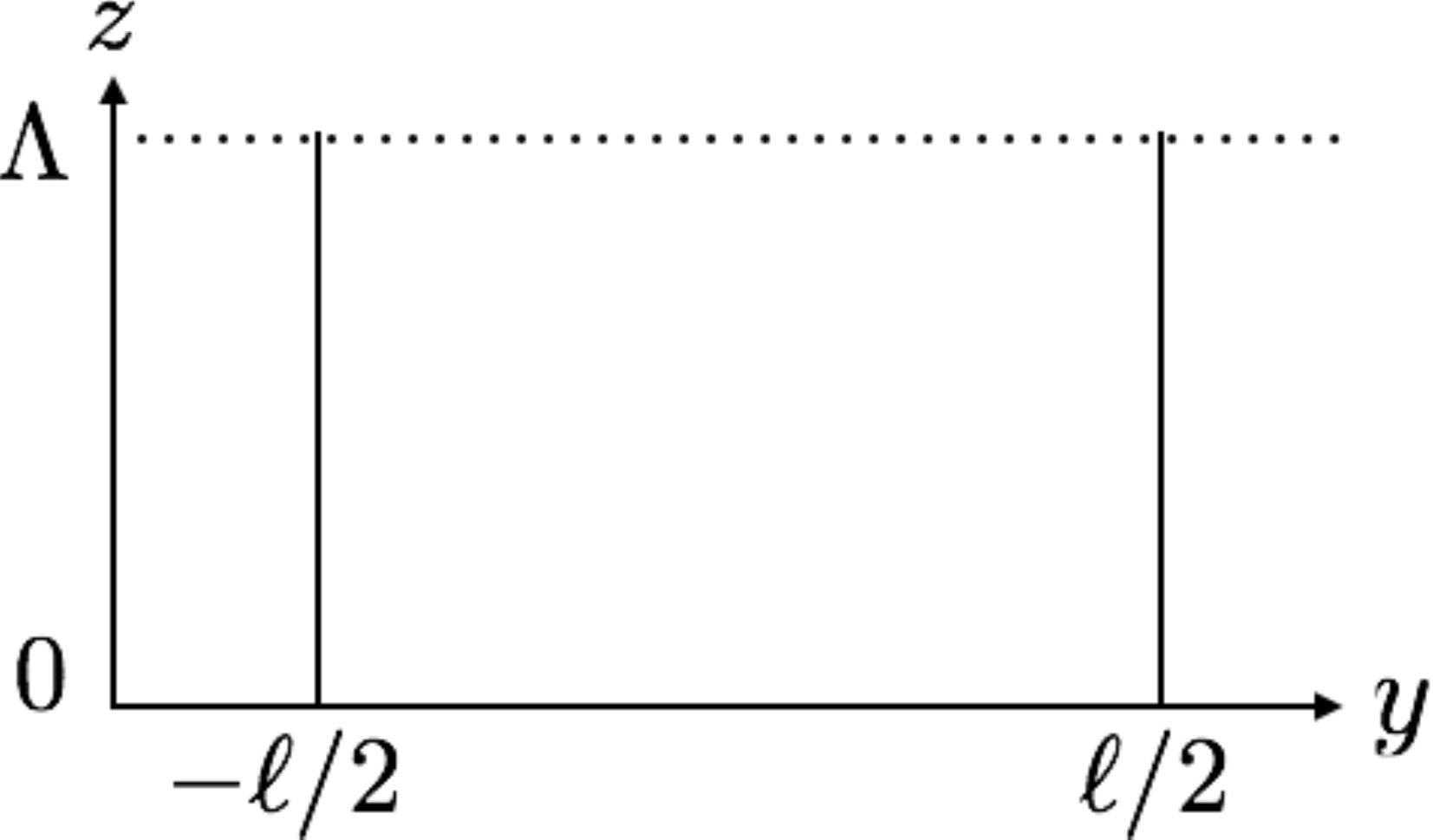}}

	\caption{Schematic pictures of the minimal surface for a single interval at $z=0$, where $\ell$ is the size of the interval. (a) Connected surface for $\ell\le\ell_c$. (b) Disconnected  surface for $\ell\ge\ell_c$. There is a transition of the minimal surface at $\ell=\ell_c$.}
	\label{spms}
\end{figure}

For a given one-dimensional surface $\mathcal{E}$ on a fixed time-slice, we define its area $A$ by
\begin{align}
A:=\int_{\mathcal{E}} \sqrt{dz^2+g_{yy}(z)dy^2}.\label{defarea}
\end{align}
The minimal surface anchored on the boundary of a single interval at $z=0$ is defined by the surface that minimizes the area $A$.
Suppose that $g_{yy}(z)=w_\psi^2(z/\Lambda)$ is a decreasing function of $z$ like Fig.~\ref{wf} (b). Then a spatial distance $g_{yy}(z)d y^2$ is minimal at $z=\Lambda$. Since the metric of Eq. (\ref{tcstm}) is finite at $z=\Lambda$, the minimal surface can reach $z=\Lambda$ if $\Lambda$ is finite. Such holographic geometries have been studied for gapped systems, e.g., \cite{Holographic_Duality_III, Girardello:1999hj, Erlich:2005qh}, and the shape of the minimal surface would be as follows. When the size of the interval is small, the minimal surface is a connected surface as shown in Fig.~\ref{spms} (a), where $\ell$ is the size of the interval. As $\ell$ increases, a disconnected surface becomes the minimal surface at $\ell=\ell_c$ as shown in Fig.~\ref{spms} (b). This behavior is known as the confinement/deconfinement transition at the critical length $\ell=\ell_c$ \cite{Nishioka:2006gr,  Klebanov:2007ws}.

First, we examine the connected surface (Fig.~\ref{spms} (a)), whose area $A_c(\ell)$ is
\begin{align}
A_c(\ell)=\int_{-\ell/2}^{\ell/2} dy \sqrt{\left(\frac{dz(y)}{dy}\right)^2+g_{yy}(z(y))},\label{mssa}
\end{align}
where $z(y)$ is a function that represents the shape of the surface.
To determine the  minimal connected surface, we need to find $z(y)$ that minimizes the area $A_c(\ell)$, and a differential equation of $z(y)$ for the minimal connected surface is \begin{align}
&\frac{d}{dy}\frac{\partial L\left(z, \frac{dz}{dy}\right)}{\partial \frac{dz}{dy}}-\frac{\partial L\left(z, \frac{dz}{dy}\right)}{\partial z}=0, \;\;\; L\left(z, \frac{dz}{dy}\right):=\sqrt{\left(\frac{dz}{dy}\right)^2+g_{yy}(z)},\\
\Longrightarrow&2g_{yy}(z)\frac{d^2z}{dy^2}-2\frac{d g_{yy}(z)}{dz} \left(\frac{dz}{dy}\right)^2-g_{yy}(z)\frac{d g_{yy}(z)}{dz}=0.\label{de}
\end{align}
The minimal connected surface is given by a solution of Eq. (\ref{de}) with two boundary conditions
\begin{align}
z(\ell/2)=0, \;\;\; \frac{dz(y)}{dy}\Big|_{y=0}=0.\label{bc}
\end{align}
We define $A^{min}_c(\ell)$ by the area of the minimal connected surface, which is Eq. (\ref{mssa}) with $z(y)$ for the minimal connected surface.

Next, the area $A_d$ of the disconnected surface (Fig.~\ref{spms} (b)) is given by
\begin{align}
A_d=2\int_0^\Lambda dz=2\Lambda,\label{madcs}
\end{align}
which does not depend on $\ell$. If $A^{min}_c(\ell)<A_d$, the connected surface is the minimal surface.  If $A^{min}_c(\ell)>A_d$, the disconnected surface is the minimal surface.

\begin{figure}[!htb]
	\centering
	\subfigure[]
	{ \includegraphics[width=0.4\linewidth]{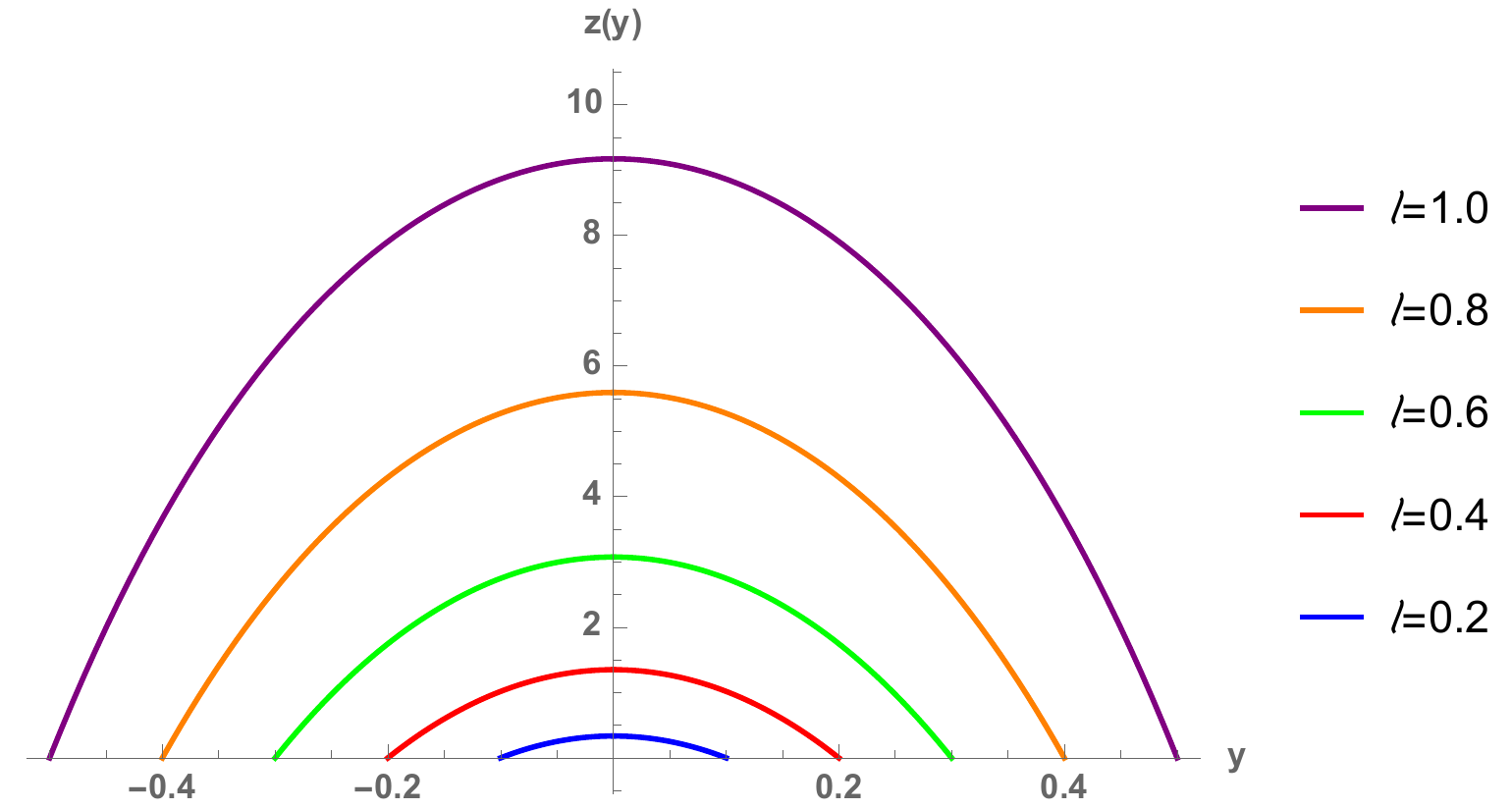}}
	\subfigure[]
	{ \includegraphics[width=0.4\linewidth]{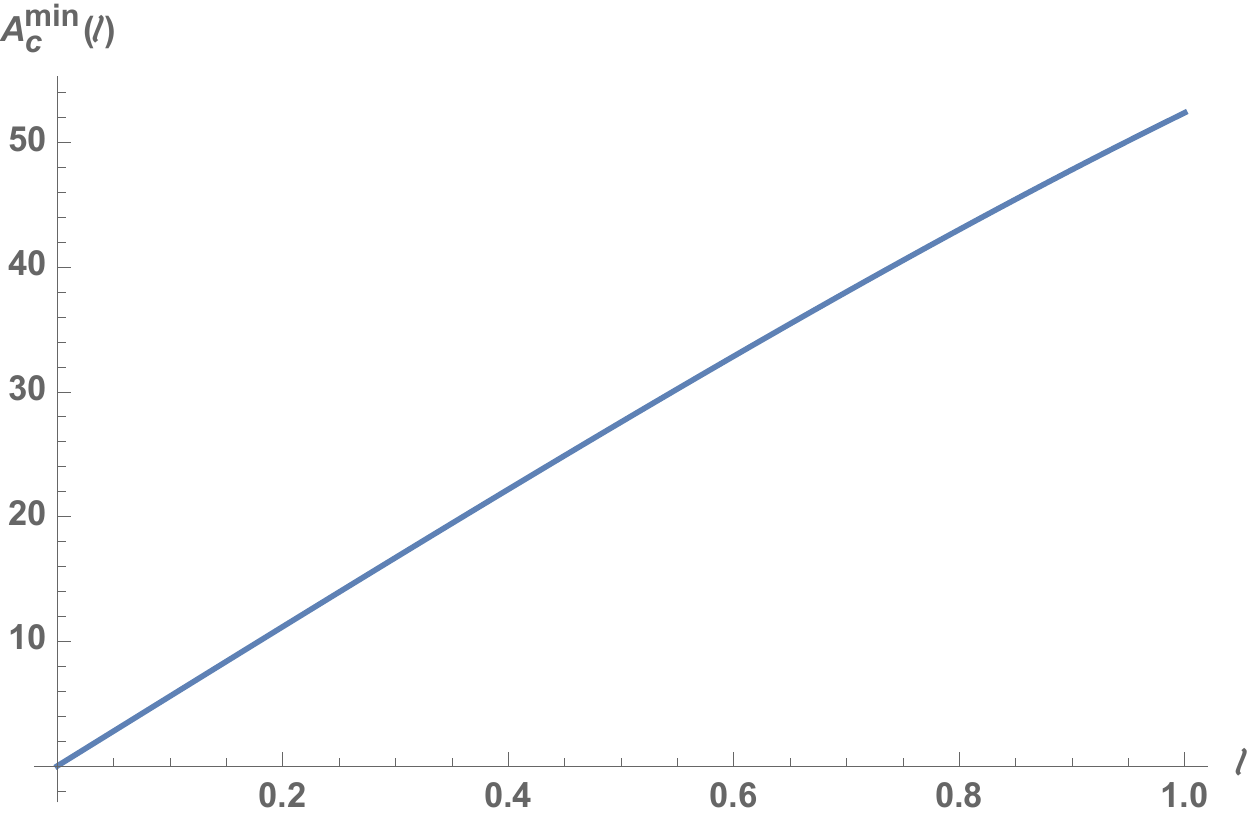}}
	
	\caption{(a) Shape of the minimal connected surface with $g_{yy}(z)=w^2_{uv}(z/\Lambda), \Lambda =50,  w_r=1.12, v_r=1.5$. (b) $\ell$-dependence of the minimal area $A^{min}_c(\ell)$.}
	\label{mss}
\end{figure}

Fig.~\ref{mss} (a) shows $z(y)$ for the minimal connected surface with $g_{yy}(z)=w^2_{uv}(z/\Lambda), \Lambda =50,  w_r=1.12, v_r=1.5$. We numerically solve Eq. (\ref{de}) with Eq. (\ref{bc}) and plot $z(y)$ with the size of the interval $\ell=0.2, 0.4, 0.6, 0.8, 1.0$. Fig.~\ref{mss} (b) shows the $\ell$-dependence of the minimal area $A_c^{min}(\ell)$. Note that $w_{uv}(z/\Lambda)$ is the asymptotic solution near $z=0$, and the approximation using $w_{uv}(z/\Lambda)$ would be reasonable when $\ell$ is small. Near $\ell=0$, the plot of the minimal area in Fig.~\ref{mss} (b) can be approximated by
\begin{align}
A^{min}_s(\ell)\sim\int_{-\ell/2}^{\ell/2} dy \sqrt{g_{yy}(0)}=w_{uv}(0)\ell \;\;\;\;\;\; (\ell\sim0).\label{appas}
\end{align}
This linear behavior is different from the logarithmic  behavior for the asymptotic AdS geometries, and the reason is that $w_{uv}(z/\Lambda)$ Eq. (\ref{eq:51}) does not decay exponentially with respect to $z$.
Due to $dz^2$ in Eq. (\ref{defarea}) and the $z$-dependence of $w_{uv}(z/\Lambda)$, the approximation Eq. (\ref{appas}) is not valid except near $\ell=0$.

\begin{figure}[!htb]
	\centering
	\subfigure[]
	{ \includegraphics[width=0.4\linewidth]{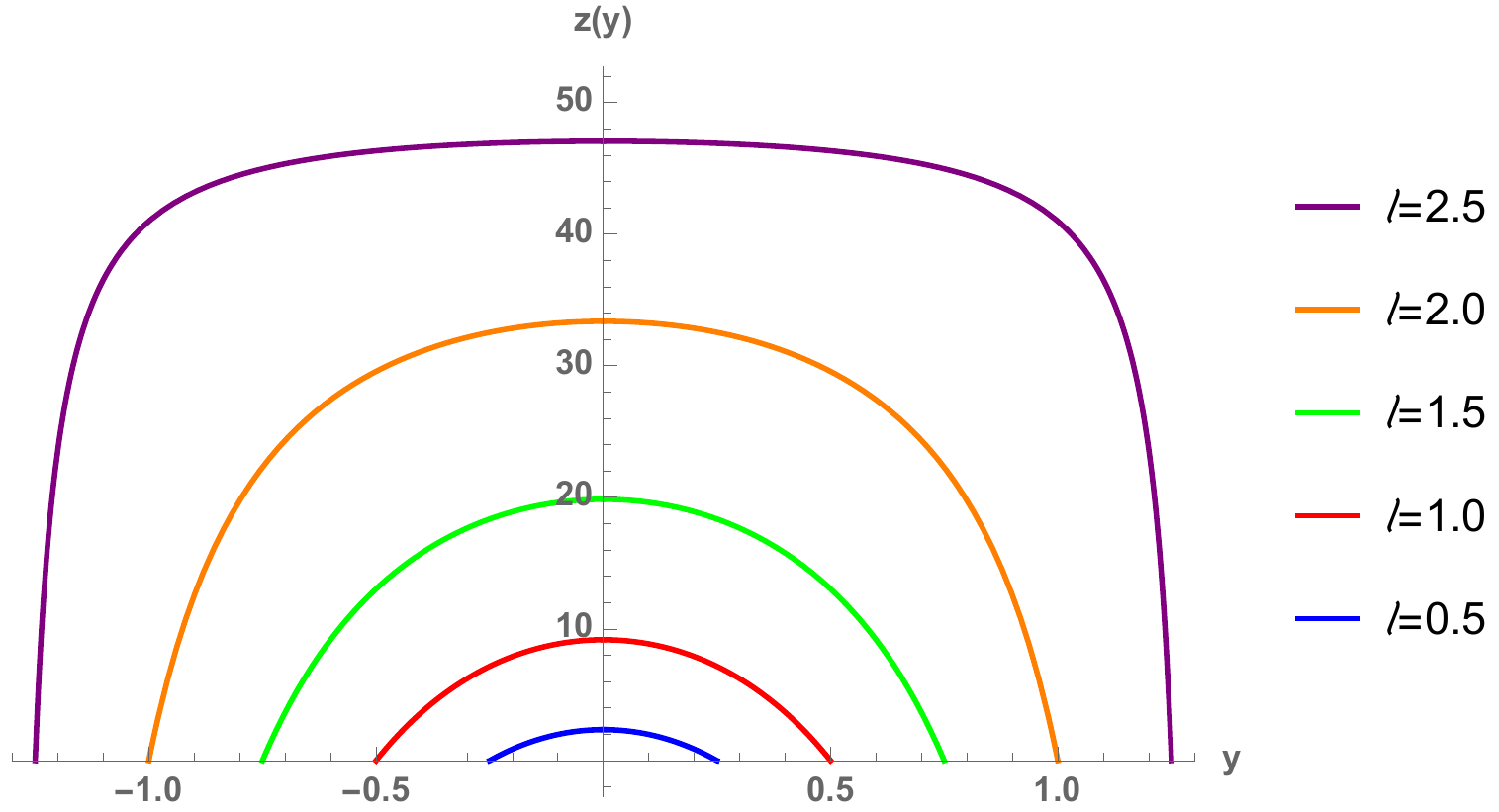}}
	\subfigure[]
	{ \includegraphics[width=0.4\linewidth]{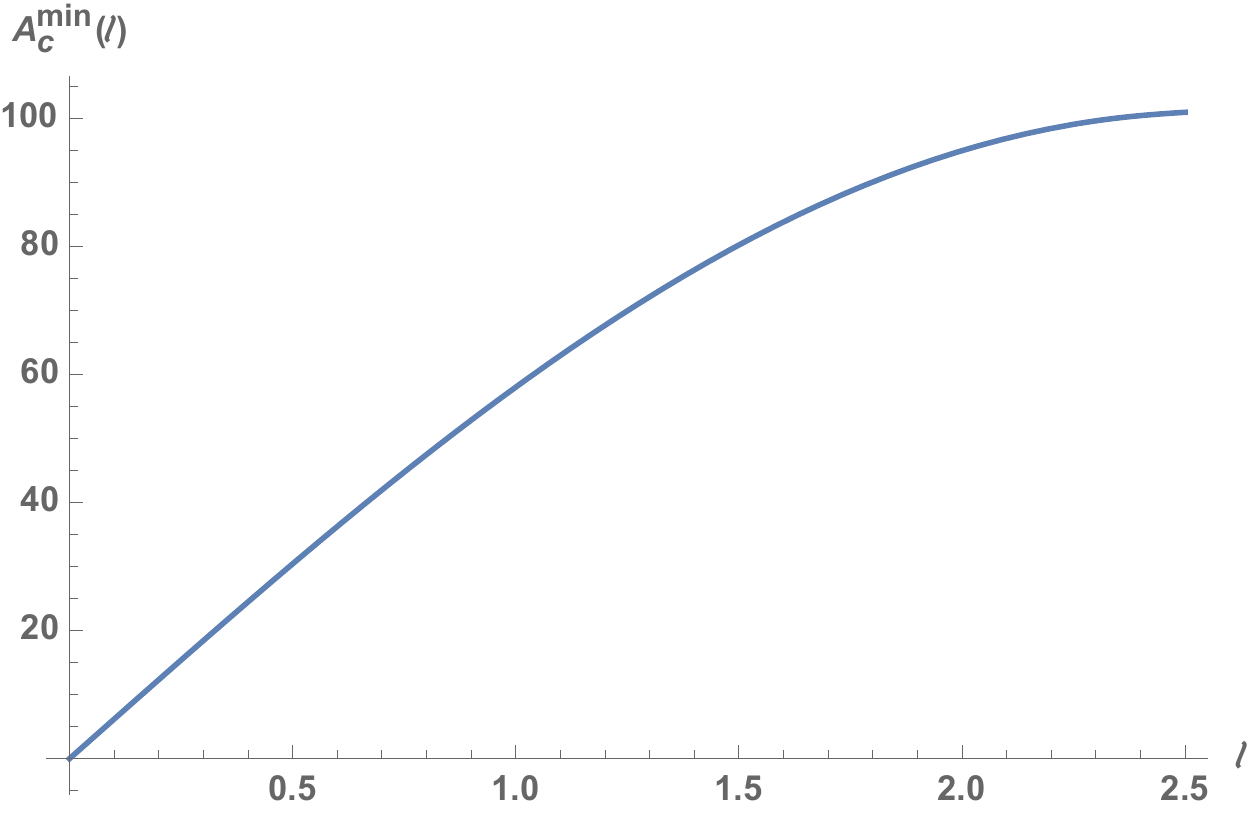}}
	
	\caption{(a) Shape of the minimal connected surface with $g_{yy}(z)=w^2_{ir}(z/\Lambda), \Lambda =50,  w_f=2, \lambda_f=2, \varphi_f=0.0001$. (b) $\ell$-dependence of the minimal area $A^{min}_c(\ell)$. At $\ell=\ell_c\sim2.4$, the disconnected surface becomes the minimal surface due to $A^{min}_c(\ell)>A_d=2\Lambda=100$.}
	\label{mssir}
\end{figure}

Fig.~\ref{mssir} (a) shows $z(y)$ with $\ell=0.5, 1.0, 1.5, 2.0, 2.5$ for the minimal connected surface with $g_{yy}(z)=w^2_{ir}(z/\Lambda), \Lambda =50,  w_f=2, \lambda_f=2, \varphi_f=0.0001$, and we plot the $\ell$-dependence of the minimal area $A_c^{min}(\ell)$ in Fig.~\ref{mssir} (b). At $\ell=\ell_c\sim2.4$, the area of the minimal connected surface $A^{min}_c(\ell)$ is larger than the area of the disconnected surface $A_d=2\Lambda=100$, which means the transition from the connected surface to the disconnected surface.
Although $w_{ir}(z/\Lambda)$ is the asymptotic solution, this result implies the existence of the critical length $\ell_c$ for the minimal surface in the three-dimensional spacetime Eq. (\ref{tcstm}) with finite $\Lambda$ constructed from information of the RG-MFT with Eq. (\ref{gmunu}). We note that $z(y)$ with $\ell=2.5$ in Fig.~\ref{mssir} (a) is similar to the minimal surface in a black hole type geometry, which gives the volume law of the holographic entanglement entropy.

Finally, we discuss the holographic geometry in the limit $\Lambda\to\infty$. If we choose the UV boundary values as $v_0=v_r\Lambda$ and $w_0=w_r\Lambda$, the metric Eq. (\ref{gmunu}) diverges in the limit $\Lambda\to\infty$. In this case, the minimal area is not well-defined, which may be interpreted as the entanglement entropy of QFTs is not well-defined in the continuum limit. If $v_\psi$ and $w_\psi$ are finite in the limit $\Lambda\to\infty$, there is no transition to the disconnected surface because Eq. (\ref{madcs}) diverges. Furthermore,  if $v_\psi\to0$ and $w_{\psi}\to w_f\ne0$ at $z=\Lambda\to\infty$, the metric Eq. (\ref{gmunu}) behaves as
\begin{align}
g_{\tau\tau}\to0, \;\;\; g_{yy}\to w_f^2\ne0 \;\;\;(z=\Lambda\to\infty),
\end{align}
which is similar to black hole geometries if $z=\Lambda$ is regarded as the horizon. Note that one can construct black holes with zero temperature by introducing scalar or $U(1)$ gauge field sources \cite{Andrade:2013gsa}.

Appearance of a black hole type geometry is rather unexpected since spontaneous chiral symmetry breaking occurs to cause a single-particle excitation gap. In this resect the only physical picture that we can have in our mind at present is that the single-particle excitation gap has to be filled with multi-particle excitations, here dual collective fields involved with $w_{\psi}$ and $v_{\psi}$ fluctuations. As discussed in the previous section, this claim is certainly verifiable, calculating correlation functions of $\delta w_{\psi}$, $\delta v_{\psi}$, and $\delta \varphi$ fluctuations around this vacuum solution. Frankly speaking, it is not clear at all whether the present effective UV field theory allows such a strongly coupled local fixed point or not. It is possible that the present theoretical framework, that is, our nonperturbative RG-MFT may overtake quantum fluctuations than expected. However, it is surprising that the nonperturbative RG-MFT can access the strong-coupling IR fixed point from the Gaussian UV fixed point beyond the perturbation framework, interpreted as an emergent black hole type geometry.

\section{Summary and discussions} \label{Discussion_Section}

In this study, we proposed a nonperturbative Wilsonian-RG improved mean-field theory in the large $N$ limit, where RG flows of the coupling functions manifest in the level of an effective action and the equation of motion for the order-parameter fields correspond to mean-field equations with self-consistently renormalized coupling functions at a given energy scale of the emergent extradimension. We argued that this nonperturbative generalization of the RG-improved mean-field theory takes into account essential features of the holographic dual effective field theory, where the dynamics of metric fluctuations describe the RG flows of the coupling functions and the equation of dual scalar and Maxwell fields correspond to the mean-field equations of order-parameter fields. Here, two modifications have been introduced into the holographic dual effective field theory. One is the appearance of RG $\beta-$functions in the bulk equation of motion for all the coupling functions, describing the RG flow from UV to IR. The other is that all IR boundary conditions are given by matching all the RG flows in the bulk to the renormalized coupling functions and order-parameter fields in the IR effective action, which can be identified with an effective on-shell action as the solution of the Hamilton-Jacobi equation in the holographic dual effective field theory.

However, the coupling functions were not dynamical in the nonperturbative Wilsonian-RG improved mean-field theory while metric fluctuations were in the holographic dual effective field theory. To promote all the coupling functions dynamically, we may introduce energy terms for random coupling fluctuations as follows
\bqa && Z = \int D \psi_{\sigma}(x) D \varphi(x) D w_{\psi}(x) D v_{\psi}(x) D \lambda_{\chi}(x) \nn && \exp\Big[ - \int d^{D} x \Big\{ \bar{\psi}_{\sigma}(x) \Big(w_{\psi}(x) \gamma^{\tau} \partial_{\tau} - i v_{\psi}(x) \gamma^{i} \partial_{i}\Big) \psi_{\sigma}(x) - i \lambda_{\chi}(x) \varphi(x) \bar{\psi}_{\sigma}(x) \psi_{\sigma}(x) \nn && + \frac{N \lambda_{\chi}(x)}{2} \varphi^{2}(x) + \frac{N \lambda_{w_{\psi}}}{2} \Big(w_{\psi}(x) - \bar{w}_{\psi}\Big)^{2} + \frac{N \lambda_{v_{\psi}}}{2} \Big(v_{\psi}(x) - \bar{v}_{\psi}\Big)^{2} + \frac{N \lambda_{\lambda_{\chi}}}{2} \Big(\lambda_{\chi}(x) - \bar{\lambda}_{\chi}\Big)^{2} \Big\} \Big] . \eqa
Taking $\lambda_{w_{\psi}} \rightarrow \infty$, $\lambda_{v_{\psi}} \rightarrow \infty$, and $\lambda_{\lambda_{\chi}} \rightarrow \infty$, such random fluctuations become frozen, where this effective UV partition function reduces into the original one that we discussed before. Based on this effective field theory, we obtain the following IR effective partition function
\bqa && Z = \int D \psi_{\sigma}(x) D \varphi(x,z) D \pi_{\varphi}(x,z) D w_{\psi}(x,z) D \pi_{w_{\psi}}(x,z) D v_{\psi}(x,z) D \pi_{v_{\psi}}(x,z) D \lambda_{\chi}(x,z) D \pi_{\lambda_{\chi}}(x,z) \nn && \exp\Big[ - \int d^{D} x \Big\{ \bar{\psi}_{\sigma}(x) \Big(w_{\psi}(x,z_{f}) \gamma^{\tau} \partial_{\tau} - i v_{\psi}(x,z_{f}) \gamma^{i} \partial_{i}\Big) \psi_{\sigma}(x) - i \lambda_{\chi}(x,z_{f}) \varphi(x,z_{f}) \bar{\psi}_{\sigma}(x) \psi_{\sigma}(x) \nn && + \frac{N \lambda_{\chi}(x,0)}{2} \varphi^{2}(x,0) + \frac{N \lambda_{w_{\psi}}}{2} \Big(w_{\psi}(x,0) - \bar{w}_{\psi}\Big)^{2} + \frac{N \lambda_{v_{\psi}}}{2} \Big(v_{\psi}(x,0) - \bar{v}_{\psi}\Big)^{2} + \frac{N \lambda_{\lambda_{\chi}}}{2} \Big(\lambda_{\chi}(x,0) - \bar{\lambda}_{\chi}\Big)^{2} \Big\} \nn && - N \int_{0}^{z_{f}} d z \int d^{D} x \Big\{ \pi_{\varphi}(x,z) \Big( \partial_{z} \varphi(x,z) - \beta_{\varphi}[\varphi(x,z),w_{\psi}(x,z),v_{\psi}(x,z),\lambda_{\chi}(x,z)] \Big) - \frac{1}{2 \lambda_{\chi}(x,z)} \pi_{\varphi}^{2}(x,z) \nn && + \pi_{w_{\psi}}(x,z) \Big( \partial_{z} w_{\psi}(x,z) - \beta_{w_{\psi}}[\varphi(x,z),w_{\psi}(x,z),v_{\psi}(x,z),\lambda_{\chi}(x,z)] \Big) - \frac{1}{2 \lambda_{w_{\psi}}} \pi_{w_{\psi}}^{2}(x,z) \nn && + \pi_{v_{\psi}}(x,z) \Big( \partial_{z} v_{\psi}(x,z) - \beta_{v_{\psi}}[\varphi(x,z),w_{\psi}(x,z),v_{\psi}(x,z),\lambda_{\chi}(x,z)] \Big) - \frac{1}{2 \lambda_{v_{\psi}}} \pi_{v_{\psi}}^{2}(x,z) \nn && + \pi_{\lambda_{\chi}}(x,z) \Big( \partial_{z} \lambda_{\chi}(x,z) - \beta_{\lambda_{\chi}}[\varphi(x,z),w_{\psi}(x,z),v_{\psi}(x,z),\lambda_{\chi}(x,z)] \Big) - \frac{1}{2 \lambda_{\lambda_{\chi}}} \pi_{\lambda_{\chi}}^{2}(x,z) \nn && + \mathcal{V}_{eff}[\varphi(x,z),w_{\psi}(x,z),v_{\psi}(x,z),\lambda_{\chi}(x,z)] \Big\} \Big] . \label{RG_MFT_TTbar_Deform} \eqa
As clearly seen in this IR effective field theory, all the coupling functions are promoted to be dynamical, where their RG flows are given by all coupled second-order differential equations instead of the first order. One may regard that this generalization is analogous to the $T \bar{T}$ deformation, as discussed before.

As mentioned above, the key ingredient is the effective potential $\mathcal{V}_{eff}[\varphi(x,z),w_{\psi}(x,z),v_{\psi}(x,z),\lambda_{\chi}(x,z)]$ at a given energy scale of the extradimension, which occurs from the path integral of matter fields, here Dirac fermions. It is quite natural to take the RG $\beta-$functions as functional derivatives of the effective potential with respect to the corresponding coupling function and order-parameter field, respectively, completely consistent with the quantum field theoretical framework.

Taking the large $N$ limit, one finds all coupled second-order differential equations in the presence of their RG $\beta-$functions, where their IR boundary conditions are given by matching all the RG flows in the bulk to the renormalized coupling functions and order-parameter fields in the IR effective action, as discussed above. To solve these coupled differential equations, we made several assumptions. First, we took $\lambda_{w_{\psi}} \rightarrow \infty$, $\lambda_{v_{\psi}} \rightarrow \infty$, and $\lambda_{\lambda_{\chi}} \rightarrow \infty$, resulting in the first-order differential equations for the coupling functions. Second, we considered translational invariance as a vacuum solution, where $D-$dimensional spacetime dependencies are all neglected. Third, we focused on the zero temperature limit, simplifying the manifold $S^{1} \times R^{D-1}$ at finite temperatures to $R^{D}$ at zero temperature. Even this simplification did not allow us to solve such coupled differential equations.

To solve these complicatedly intertwined differential equations, we applied the matching method to this problem. First, we solved such coupled differential equations near both UV and IR boundary regions independently, where these equations become simplified. Second, we applied the UV (IR) boundary condition to the UV-regional (IR-) solution. Since the number of boundary conditions would be less than that of integration constants, some of the integration constants remain undetermined in both UV- and IR-regional solutions. Third, we required that the UV-regional solution should be smoothly connected to the IR-regional solution at one point in the extradimensional space. Of course, there must be a certain condition for the existence of this matching solution. Based on this delicately working matching method, we found an RG flow from a weakly-coupled chiral-symmetric UV fixed point to a strongly-correlated chiral-symmetry broken IR fixed point, where the renormalized velocity of Dirac fermions vanishes most rapidly and effective quantum mechanics appears at IR. It is a feature of the nonperturbative RG-MFT the appearance of this local strong-coupling fixed point.

From the renormalized coupling functions that are solutions of RG flow equations, we can build the three-dimensional curved spacetime metric. Geometrical quantities on this emergent holographic spacetime may capture characteristics of the RG-MFT. One important measure is the minimal surface for the holographic entanglement entropy, and we calculate the minimal surface of the holographic spacetime associated with the UV- and IR-regional solutions. Our interesting result is that the holographic entanglement entropy shows the volume law instead of the area one. This indicates that the emergent geometry in the infinite cutoff limit is a black hole type even at zero temperature. We speculated that the origin of this emergent black hole type geometry is appearance of gapless multi-particle spectra dual to quantum fluctuations of the coupling functions and order-parameter field, where such gapless multi-particle spectra would fill the single-particle excitation gap due to spontaneous chiral symmetry breaking. We leave it as an interesting future study to calculate the correlation functions of collective excitations around the black hole type geometry. 

%
%

Although we focused on the zero temperature limit, we may reinterpret the present RG flow as follows. First of all, there is an interesting crossover regime during this dimensional reduction. We recall the appearance of spontaneous chiral symmetry breaking in the intermediate regime of the extradimension, where the renormalized velocity of Dirac fermions remains to be finite. In other words, the present self-consistent matching solution not only describes the RG flow from a weakly-coupled chiral-symmetric UV fixed point to a strongly-correlated chiral-symmetry broken IR fixed point but also indicates the existence of a weakly-coupled chiral-symmetry broken intermediate regime.

%
%

Let us close the present paper, asking the following question: What happens above two spacetime dimensions. Does the locally quantum critical strong coupling fixed point appear ubiquitously in the vicinity of spontaneous chiral symmetry breaking? Then, does the AdS$_{2}$ black hole appear as a signature of the RG flow to a strong coupling fixed point?

\begin{acknowledgments}
K.-S. Kim was supported by the Ministry of Education, Science, and Technology (NRF-2021R1A2C1006453 and NRF-2021R1A4A3029839) of the National Research Foundation of Korea (NRF) and by TJ Park Science Fellowship of the POSCO TJ Park Foundation. M.~Nishida was supported by the Ministry of Education, Science, and Technology (NRF-2020R1I1A1A01072726) of the National Research Foundation of Korea (NRF). Y.-S. Choun was supported through NRF Grant No. 2020-R1A2C2-007930.
\end{acknowledgments}


\begin{thebibliography}{9}
\bibitem{QFT_Textbook} Michael E. Peskin and Daniel V. Schroeder, \textit{An Introduction To Quantum Field Theory}, (CRC Press. Taylor and Francis Group, New York, 1995).
\bibitem{SungSik_Holography_I} Sung-Sik Lee, J. High Energy Phys. \textbf{10}, 160 (2012).
\bibitem{SungSik_Holography_II} Sung-Sik Lee, J. High Energy Phys. \textbf{01}, 076 (2014).
\bibitem{SungSik_Holography_III} P. Lunts, S. Bhattacharjee, J. Miller, E. Schnetter, Yong Baek Kim, and Sung-Sik Lee, J. High Energy Phys. \textbf{08}, 107 (2015).
\bibitem{Nonperturbative_Wilson_RG} Ki-Seok Kim, Shinsei Ryu, and Kanghoon Lee, Phys. Rev. D \textbf{105}, 086019 (2022).
\bibitem{Einstein_Klein_Gordon_RG_Kim} Ki-Seok Kim and Shinsei Ryu, JHEP05(2021)260, https://doi.org/10.1007/JHEP05(2021)260.
\bibitem{Einstein_Dirac_RG_Kim} Ki-Seok Kim, Phys. Rev. D \textbf{102}, 086014 (2020).
\bibitem{RG_GR_Geometry_I_Kim} Ki-Seok Kim, Phys. Rev. D \textbf{102}, 026022 (2020).
\bibitem{RG_GR_Geometry_II_Kim} Ki-Seok Kim, Nucl. Phys. B \textbf{959}, 115144 (2020).
\bibitem{Kondo_Holography_Kim} Ki-Seok Kim, Suk Bum Chung, Chanyong Park, and Jae-Ho Han, Phys. Rev. D \textbf{99}, 105012 (2019).
\bibitem{Kitaev_Entanglement_Entropy_Kim} K.-S. Kim, M. Park, J. Cho, and C. Park, Phys. Rev. D \textbf{96}, 086015 (2017).
\bibitem{RG_Holography_First_Kim} K.-S. Kim and C. Park, Phys. Rev. D \textbf{93}, 121702 (2016).
\bibitem{Holographic_Duality_I} J. M. Maldacena, Int. J. Theor. Phys. \textbf{38}, 1113 (1999).
\bibitem{Holographic_Duality_II} S. S. Gubser, I. R. Klebanov, and A. M. Polyakov, Phys. Lett. B \textbf{428}, 105 (1998).
\bibitem{Holographic_Duality_III} E. Witten, Adv. Theor. Math. Phys. \textbf{2}, 253 (1998).
\bibitem{Holographic_Duality_IV} O. Aharony, S. S. Gubser, J. Maldacena, H. Ooguri, and Y. Oz, Phys. Rep. \textbf{323}, 183 (2000).
\bibitem{Holographic_Duality_V} M. Bianchi, D. Z. Freedman, and K. Skenderis, Nucl. Phys. B \textbf{631}, 159 (2002).
\bibitem{Holographic_Duality_VI} J. de Boer, E. P. Verlinde, and H. L. Verlinde, J. High Energy Phys. \textbf{08}, 003 (2000).
\bibitem{Holographic_Duality_VII} E. P. Verlinde and H. L. Verlinde, J. High Energy Phys. \textbf{05}, 034 (2000).
\bibitem{Vasudev_Shyam_I} Vasudev Shyam, Phys. Rev. D \textbf{95}, 066003 (2017).
\bibitem{Vasudev_Shyam_II} Vasudev Shyam, JHEP10(2017)108, https://doi.org/10.1007/JHEP10(2017)108.
\bibitem{Vasudev_Shyam_III} Vasudev Shyam, JHEP03(2018)171, https://doi.org/10.1007/JHEP03(2018)171.
\bibitem{Vasudev_Shyam_IV} Vasudev Shyam, Ph. D. thesis (2020), \textit{Quantum Gravity Renormalization Group}, http://hdl.handle.net/10012/16081.
\bibitem{Holography_Matching_Method} Hong Liu, John McGreevy, and David Vegh, Phys. Rev. D \textbf{83}, 065029 (2011).
\bibitem{Ryu:2006bv} Shinsei Ryu and Tadashi Takayanagi, Phys. Rev. Lett. \textbf{96}, 181602 (2006).
\bibitem{Ryu:2006ef} Shinsei Ryu and Tadashi Takayanagi, J. High Energy Phys. \textbf{08} (2006) 045.
\bibitem{QFT_Finite_T} Alexander Altland and Ben D. Simons, \textit{Condensed Matter Field Theory 2nd Edition}, (Cambridge University Press, New York, 2010), ISBN-13: 978-0521769754.
\bibitem{GR_Textbook} Matthias Blau, \textit{Lecture Notes on General Relativity}, http://www.blau.itp.unibe.ch/GRLecturenotes.html (``unpublished").
\bibitem{ADM_Hamiltonian_Formulation} R. Arnowitt, S. Deser, and C. W. Misner, Phys. Rev. \textbf{116}, 1322 (1959).
\bibitem{Gradient_Expansion_Gravity_I} A. Sakharov, Sov. Phys. Dokl. \textbf{12}, 1040 (1968), and reprinted in Gen. Rel. Grav. \textbf{32}, 365 (2000).
\bibitem{Gradient_Expansion_Gravity_II} M. Visser, Mod. Phys. Lett. \textbf{A 17}, 977 (2002), and references therein [arXiv:gr-qc/0204062v1].
\bibitem{Higher_Curvature_GR} Pablo A. Cano, PhD thesis, arXiv:1912.07035v1 [hep-th].
\bibitem{DeWitt_Metric} Bryce S. DeWitt, Phys. Rev. \textbf{160}, 1113 (1967).
\bibitem{TTbar_Deformation} J. Cardy, J. High Energy Phys. \textbf{10} 186 (2018).
\bibitem{Holographic_Liquid_Son_I} G. Policastro, D. T. Son, and A. O. Starinets, J. High Energy Phys. \textbf{09} (2002) 043.
\bibitem{Holographic_Liquid_Son_II} G. Policastro, D. T. Son, and A. O. Starinets, J. High Energy Phys. \textbf{12} (2002) 054.
\bibitem{Holographic_Liquid_Son_III} Pavel Kovtun, Dam T. Son, and Andrei O. Starinets, J. High Energy Phys. \textbf{10} (2003) 064.
\bibitem{Holographic_Liquid_Son_IV} P. Kovtun, D. T. Son, and A. O. Starinets, Phys. Rev. Lett. \textbf{94}, 111601 (2005).
%
%
\bibitem{Girardello:1999hj} L. Girardello, M. Petrini, M. Porrati, and A. Zaffaroni, J. High Energy Phys. \textbf{05} (1999) 026.
\bibitem{Erlich:2005qh} Joshua Erlich, Emanuel Katz, Dam T. Son, and Mikhail A. Stephanov Phys. Rev. Lett. \textbf{95}, 261602 (2005).
\bibitem{Nishioka:2006gr} Tatsuma Nishioka and Tadashi Takayanagi, J. High Energy Phys. \textbf{01} (2007) 090.
\bibitem{Klebanov:2007ws} Igor R.~Klebanov, David Kutasov, and Arvind Murugan, Nucl. Phys. B \textbf{796}, 274293 (2008).
\bibitem{Andrade:2013gsa} Tomas Andrade and Benjamin Withers, J. High Energy Phys. \textbf{05} (2014) 101.
%
%
\end{thebibliography}
\end{document}